\documentclass{ieeeaccess}
\usepackage{cite}
\usepackage{amsmath,amssymb,amsfonts}
\usepackage{algorithmic}
\usepackage{graphicx}
\usepackage{textcomp}

\usepackage{array}
\usepackage{subcaption}
\usepackage{enumitem}
\usepackage{pifont}% http://ctan.org/pkg/pifont
\usepackage{xkvltxp}
\usepackage[binary-units]{siunitx}
\usepackage[hyphens]{url}
\usepackage{makecell}
\usepackage{lipsum}
\usepackage{tablefootnote}
\usepackage[dvipsnames]{xcolor}

\DeclareSIUnit \dBm {dBm}
\DeclareSIUnit \dB {dB} 
\DeclareSIUnit \dBi {dBi} 
\DeclareSIUnit \Kbps {Kbps}
\DeclareSIUnit \Mbps {Mbps}
\DeclareSIUnit \Gbps {Gbps}
\DeclareSIUnit \kBps {kBps}
\DeclareSIUnit \MBps {MBps}
\DeclareSIUnit \GBps {GBps}
\graphicspath{ {./Images/} }

\newcommand{\xmark}{\ding{55}}
\newcolumntype{N}{>{\centering\arraybackslash}m{3cm}}
\newcolumntype{K}{>{\centering\arraybackslash}m{0.8cm}}
\newcolumntype{P}[1]{>{\centering\arraybackslash}p{#1}}
\newcolumntype{M}[1]{>{\centering\arraybackslash}m{#1}}

\def\BibTeX{{\rm B\kern-.05em{\sc i\kern-.025em b}\kern-.08em
    T\kern-.1667em\lower.7ex\hbox{E}\kern-.125emX}}

\begin{document}
\bstctlcite{IEEEexample:BSTcontrol}
\history{Date of publication xxxx 00, 0000, date of current version xxxx 00, 0000.}
\doi{10.1109/ACCESS.2023.0322000}

\title{UMBRELLA: A One-stop Shop Bridging the Gap from Lab to Real-World IoT Experimentation}
% \title{UMBRELLA: A One-stop Shop Bridging the Gap Between Lab and Real-World IoT Experimentation}
% \title{UMBRELLA: A System-of-Systems Bridging the Gap Between Lab and Real-World IoT Experiments}
\author{\uppercase{Ioannis Mavromatis}\authorrefmark{1}$^*$,
\uppercase{Yichao Jin}\authorrefmark{2}, \uppercase{Aleksandar Stanoev}\authorrefmark{3}$^*$, \uppercase{Anthony Portelli}\authorrefmark{4}$^*$, \uppercase{Ingram Weeks}\authorrefmark{2}, \uppercase{Ben Holden}\authorrefmark{5}$^*$, \uppercase{Eliot Glasspole}\authorrefmark{6}$^*$, \uppercase{Tim Farnham}\authorrefmark{2}, \uppercase{Aftab Khan}\authorrefmark{2}, \uppercase{Usman Raza}\authorrefmark{7}$^*$, \uppercase{Adnan Aijaz}\authorrefmark{2}, \uppercase{Thomas Bierton}\authorrefmark{2}, \uppercase{Ichiro Seto}\authorrefmark{8}, \uppercase{Nita Patel}\authorrefmark{9}, \uppercase{Mahesh Sooriyabandara}\authorrefmark{2}}

\address[1]{Digital Catapult, NW1 2RA, London, U.K.}
\address[2]{Bristol Research \& Innovation Laboratory, Toshiba Europe Ltd., Avon, BS1 4ND, Bristol, UK}
\address[3]{Nordic Semiconductor, Bristol, Avon, BS1 4EX, UK}
\address[4]{Turtle Beach, Basingstoke, Hampshire, RG24 8GU, UK}
\address[5]{Q5D Technologies Ltd
, Portishead, North Somerset, S20 7GF UK}
\address[6]{Amiosec Ltd, Tewkesbury, Gloucestershire, GL20 8DN, UK}
\address[7]{Waymap, London, EC4R 2SU, UK}
\address[8]{Corporate R\&D Centre, Toshiba Corporation, Saiwai-ku, Kawasaki-shi, 212-8582, Japan}
\address[9]{South Gloucestershire Council, Gloucester, Yate, Gloucestershire, BS37 7PN, UK}
\tfootnote{This work was funded in part by the West of England Local Enterprise Partnership (LEP) Local Growth Fund, administered by the West of England Combined Authority (WECA), and in part by Toshiba Europe Ltd. and Bristol Research Innovation Laboratory (BRIL).}
% \tfootnote{asdfasdfsad}

\markboth
{I. Mavromatis \headeretal: Preparation of Papers for IEEE TRANSACTIONS and JOURNALS}
{I. Mavromatis \headeretal: Preparation of Papers for IEEE TRANSACTIONS and JOURNALS}

\corresp{Corresponding authors: Ioannis.Mavromatis@digicatapult.org.uk, \{Yichao.Jin, Tim.Farnham, 
Mahesh\}@toshiba-bril.com, and Nita.Patel@southglos.gov.uk.\\$^*$The co-author was affiliated with Toshiba Europe Ltd. during this project.}

\begin{abstract}
UMBRELLA\footnote{UMBRELLA - A living lab: \url{https://www.umbrellaiot.com/}} is an open, large-scale IoT ecosystem deployed across South Gloucestershire, UK. It is intended to accelerate innovation across multiple technology domains. UMBRELLA is built to bridge the gap between existing specialised testbeds and address holistically real-world technological challenges in a System-of-Systems (SoS) fashion. UMBRELLA provides open access to real-world devices and infrastructure, enabling researchers and the industry to evaluate solutions for Smart Cities, Robotics, Wireless Communications, Edge Intelligence, and more. Key features include over 200 multi-sensor nodes installed on public infrastructure, a robotics arena with 20 mobile robots, a 5G network-in-a-box solution, and a unified backend platform for management, control and secure user access. The heterogeneity of hardware components, including diverse sensors, communication interfaces, and GPU-enabled edge devices, coupled with tools like digital twins, allows for comprehensive experimentation and benchmarking of innovative solutions not viable in lab environments. This paper provides a comprehensive overview of UMBRELLA’s multi-domain architecture and capabilities, making it an ideal playground for Internet of Things (IoT) and Industrial IoT (IIoT) innovation. It discusses the challenges in designing, developing and operating UMBRELLA as an open, sustainable testbed and shares lessons learned to guide similar future initiatives. With its unique openness, heterogeneity, realism and tools, UMBRELLA aims to continue accelerating cutting-edge technology research, development and translation into real-world progress.
\end{abstract}


\titlepgskip=-21pt
\maketitle

\section{Introduction}\label{sec:introduction}
% For inspiration read these papers
% https://hal.inria.fr/hal-01147346/document

\PARstart{C}{yber-Physical} Systems (CPSs) are built from, and depend upon, the seamless integration of computational algorithms and physical components~\cite{physicalDigitalAssets}. These CPSs often operate within larger, more complex environments, where the concept of a System-of-Systems (SoS) comes into play~\cite{systemOfSystems}. A SoS is an ensemble of Constituent Systems (CSs) that collaboratively achieve a common goal. This goal usually extends beyond the reach of any individual system. The Internet of Things (IoT) becomes particularly pertinent in this context. Individual IoT devices or entire IoT networks can be viewed as ``systems'' and, integrated, can materialise complex use cases and applications.

Modern IoT systems inherently support multi-system integration due to their interconnected nature. It is estimated that over 80 billion devices will be active by 2030~\cite{iotBillionDevices}. By that time, each IoT device will be within reach (i.e., have the ability to communicate) by more than 20 other nearby devices~\cite{devicesNearby}. That enables collaboration among different systems but also significantly strains every aspect of an IoT ecosystem, e.g., networking, decision-making, data analytics, etc.

An IoT ecosystem can be described in the form of both physical and digital assets~\cite{physicalDigitalAssets}. The basis of the transformational nature of CPS/IoT systems relies on this tight logical-physical linkage. The similarities, differences and importance of such an interaction are described by the National Institute of Standards and Technology (NIST) in~\cite{nistDoc}. Due to this interrelationship, when developing IoT SoS, it is of paramount importance to evaluate different CS components and algorithms, not only independently but also in a holistic platform-wide fashion~\cite{iotPlatformEvaluation}

This is the focus of Urban Multi Wireless Broadband and IoT Testing for Local Authority and Industrial Applications (UMBRELLA). UMBRELLA is an IoT SoS provided in an ``as-a-service'' fashion. UMBRELLA has been developed as a ``living lab'' where users can emulate real-world scenarios, prototype, and evaluate their IoT solutions. UMBRELLA provides the underlying capability, tools and infrastructure for testing both physical and digital solutions across various use cases such as Smart City, IoT, Industrial IoT (IIoT) applications, etc. 

All CSs in an SoS should retain their independent ownership, management, geographic distribution and individual focus~\cite{systemOfSystems}. However, in collaboration with other CSs, a global SoS goal can be achieved. This is the principle behind UMBRELLA. Each existing system could be independently built, managed, and extended, but the interoperation of various CSs can enable more complex scenario demonstration. For example, the South Gloucestershire Council already utilises the facilities for air quality monitoring and street light maintenance, two use cases that make use of multiple subsystems within UMBRELLA. UMBRELLA also manages and facilitates multiple external systems developed by Small and Medium-sized Enterprises (SMEs). For example, Altered Carbon and Awaretag directly integrated their hardware solutions into the UMBRELLA ecosystem. They currently use the provided Low-Power Wide-Area Network (LPWAN) interfaces to collect air quality data. Similarly, CyberHive tested the roaming capabilities of their cyber security solution on UMBRELLA's roadside infrastructure. UMBRELLA is unique in its nature, as it provides access to multiple nodes, sensors, network interfaces, and AI-capable edge devices, something not common in other existing testbeds. Moreover, the large scale of the infrastructure and network, currently stretching across a large geographical area, provides opportunities for building unique, scalable, interoperable, and flexible solutions for complex academic and industrial use cases.

% This paper provides an overview of the UMBRELLA ecosystem, describing the main functionality provided, the different systems and subsystems already provided ``as-services'' and how users can interact with the platform in the form of ``experiments''.  
This paper provides an overview of the UMBRELLA ecosystem, detailing its functionalities, systems, and user interactions. All the different components are accessible from a unified portal and via standardised interfaces. We will compare UMBRELLA with other existing solutions, describing the unique features provided and presenting existing use cases already tested on our infrastructure. Finally, we will briefly discuss the lessons learned from developing and operating UMBRELLA and how the existing ecosystem facilitated other projects.

The rest of the paper is organised as follows. Sec.~\ref{sec:relatedWork} provides an overview of other similar testbeds and platforms and compares UMBRELLA against them. The requirements and expected features for an IoT testbed are discussed in Sec.~\ref{sec:requirement_analysis}. The architecture of UMBRELLA is presented in Sec.~\ref{sec:architecture} along with an overview of the infrastructure, hardware and software components provided. The deployment of the testbed and the potential users are also described as part of this section. Sec.~\ref{sec:use_cases} provides a brief overview of several up-and-running key use cases, intending to provide the broad functionality of the UMBRELLA ecosystem and motivate researchers to demonstrate their solutions. Sec.~\ref{sec:spinoffs} briefly touches upon spinoff and UMBRELLA-enabled projects, followed by Sec.~\ref{sec:lessons_learned} that provides our critical thinking on the lessons learned from building and operating UMBRELLA. Lastly, the paper is concluded in Sec.~\ref{sec:conclusion}, also providing some details on future UMBRELLA extensions.

\begin{table*}[h!]
\renewcommand{\arraystretch}{1.15}
    \centering
    \caption{\textbf{Existing IoT testbeds and platforms and a comparison with UMBRELLA.}} 
    \begin{tabular}{|r|M{1.7cm}|M{0.9cm}|M{2.4cm}|M{2.3cm}|M{4.8cm}|}
        \hline
        \textbf{Testbed} & \textbf{Focus} & \textbf{Open Access} & \textbf{No. of Nodes} & \textbf{Wireless Experimentation}  & \textbf{Sensors} \\ 
        \hline \hline
        FitLab-IoT~\cite{fitlab_iot} & Wireless / Robotics & \checkmark &  1500+ Wireless & Bluetooth, LoRa & - \\ \hline
        FlockLab 2~\cite{flocklab2} & Wireless & \checkmark & 102 Wireless & Bluetooth, LoRa & - \\ \hline
        D-Cube~\cite{dcube} & Wireless & \checkmark & 50 Wireless & Bluetooth & - \\ \hline
        KU Leuven~\cite{leuven_iot}~\cite{leuven_5g} & Wireless & \xmark & \makecell{86 Wireless / \\ 46 USRPs} & Bluetooth, SDRs, LoRa & - \\ \hline
        w-ILab 1\&2~\cite{w_ilab} & Wireless / Robotics / Sensors &  \checkmark & \makecell{100 Wireless / \\ 16 Robots} & Wi-Fi, Bluetooth, SDRs, LTE & Temperature \\ \hline
        Orbit-Lab~\cite{orbit_lab} & Wireless / ML / Distrib. ML & \checkmark & \makecell{20 Wireless / \\ 3 ML} & Wi-Fi, Bluetooth, ZigBee, SDRs & - \\ \hline
        Arno~\cite{arno} & Wireless & \xmark & 10 LTE & USRPs & - \\ \hline
        NITOS~\cite{nitos} & Wireless & \checkmark & 100 Wireless & Wi-Fi, WiMAX, LTE, Bluetooth & - \\ \hline
        Cosmos Lab~\cite{cosmosLab} & Wireless & \checkmark & 100 Wireless & SDRs & - \\ \hline
        LinkLab~\cite{LinkLab} & Wireless & \xmark & 150 Wireless & Bluetooh, LoRa, Wi-Fi & - \\ \hline
        INDRIYA2~\cite{indriya} & Wireless / Sensors & \checkmark & 58 Wireless & Bluetooth &  Ambient light, Microphone, Magnetometer, Humidity, Pressure, Accelerometer, Gyroscope, Temperature \\ \hline
        LOG-a-TEC~\cite{log_a_tec} & Wireless / Sensors & \checkmark & 79 / 52 Wireless & ZigBee, LoRa, 6LoWPAN, Bluetooth &  Ambient light, Humidity, Pressure, Accelerometer, Temperature, VOC, CO, NO2, etc. \\ \hline
        Gradient~\cite{gradient} & ML & \xmark & - & - & - \\ \hline
        OpenAI~\cite{openai} & ML & \xmark & - & - & - \\ \hline 
        Colab~\cite{colab} & ML & \checkmark & - & - & - \\ \hline
        Robotarium~\cite{robotarium} & Robotics & \checkmark &  100 Robots & - & - \\ \hline
        IRIS~\cite{iris} & Robotics & \xmark & 10 Robots & - & - \\ \hline
        LivingLab~\cite{fraunhofer} & Robotics & \xmark & 50 Robots & - & - \\ \hline
        OpenCyberCity~\cite{openCityTestbed,openCyberCityTestbed} & Smart Buildings / Robotics & \xmark & Unknown & - & - \\ \hline \hline
        \textbf{UMBRELLA} & \textbf{All} & \checkmark & \makecell{\makecell{\textbf{200 Wireless /} \\ \textbf{75 ML / 20 Robots /}}  \\ \textbf{2 5G-in-a-box}} & \textbf{Bluetooth, LoRa, Wi-Fi} & \textbf{Ambient light, Microphone, Camera, Magnetometer, Humidity, Pressure, Accelerometer, Gyroscope, Temperature, VOC, CO, NO2, Laser TOF, Strain gauge etc.} \\
        \hline
    \end{tabular}
    \label{table:testbedcomparison}
\end{table*}

\section{Related Work and Testbeds}\label{sec:relatedWork}
IoT testbeds focusing on researchers and industrial collaborators have existed for many years. The available testbeds and platforms provide access to multiple use cases and present unique features and capabilities. Briefly, we can find testbeds related to wireless experimentation, robotics research and Artificial Intelligence (AI)-related activities, either publicly available, under subscription schemes, or available to personnel from specific organisations and/or labs. For our investigation, we focus on:
\begin{itemize}
    \item  Testbeds that are still operational (many testbeds have been decommissioned, thus will not be included in the comparison).
    \item Platforms that are remotely accessible. Testbed implementations that could be deployed ``in-house'' (using a provided list of hardware and software components and instructions from a repository) will not be considered.
\end{itemize} 
Tab.~\ref{table:testbedcomparison} summarises all available testbeds and platforms, presenting their functionality, the available sensors, wireless interfaces, no. of nodes and the use cases they focus on.

\vspace{0.2em}

\noindent \textbf{Wireless Experimentation}: With regards to wireless experimentation, some notable examples are the FIT IoT-Lab~\cite{fitlab_iot}, FlockLab 2 Testbed~\cite{flocklab2}, D-Cube~\cite{dcube}, w-ILab~\cite{w_ilab}, Cosmos Lab~\cite{cosmosLab}, etc. These testbeds are publicly accessible and can enable experimentation around the LPWAN area. They provide access to one or multiple testing locations with different network architectures and setups. They are primarily used for testing wireless networking protocols and algorithms developed by the research community. They primarily focus on Bluetooth and LoRaWAN. Some testbeds, e.g., w-ILab, Cosmos Lab, etc., provide access to several Software-Defined Radios (SDRs) for LTE experimentation. All but D-Cube allow the users to design their own scenarios. D-Cube provides access to specific application scenarios pre-defined by the testbed administrators. All the above testbeds are installed in indoor environments. This limits the number of use cases that could be demonstrated on them or the level of realism required in the real world where many ``systems'' interact with each other in an SoS fashion.

\vspace{0.2em}

\noindent \textbf{AI/ML Testbeds}: Furthermore, various platforms focus on AI/Machine Learning (ML) algorithms and pipelines. Some examples are the Gradient~\cite{gradient} and OpenAI~\cite{openai}. Users can upload Jupiter Notebooks and evaluate their AI/ML algorithms on pre-existing datasets and models or develop their pipelines from scratch. Both operate on a subscription scheme that can become costly for independent researchers and students. The Google Colaboratory (Colab)~\cite{colab} platform also provides similar functionality. Compared to OpenAI and Gradient, Colab is free of charge. All platforms provide access to powerful hardware, such as Graphics Processing Units (GPUs) and Tensor Processing Units (TPUs). However, none of these platforms provides access to real-world data generated in real-time.

\vspace{0.2em}

\noindent \textbf{Robotic Testbeds}: With regards to robotics and swarm activities, we find Robotarium~\cite{robotarium}, Intelligent Robotic IoT System (IRIS)~\cite{iris} and Fraunhofer's LivingLab robotic testbed~\cite{fraunhofer}. FIT IoT-Lab~\cite{fitlab_iot} and w-ILab~\cite{w_ilab} mentioned earlier provide access to robotic facilities as well. Robotarium and Iris are publicly available, while LivingLab is only available to the lab collaborators. Robotarium offers access to miniature custom-designed robots, which, even though it enhances the battery's autonomy and reduces the testbed's footprint, limits the robots' hardware capabilities. IRIS robots are equipped with fully programmable wireless IEEE 802.15.4 radio interfaces. However, the robots' position accuracy deteriorates for long experiments due to accumulated drift. Finally, FIT IoT-Lab and w-ILab functionality is still quite limited as both are still under development.

\vspace{0.2em}

\noindent \textbf{Wireless Sensor Networks}: INDRIYA2~\cite{indriya}, w-ILal~\cite{w_ilab}, and Log-a-TEC~\cite{log_a_tec} testbeds operate as wireless sensor networks providing access to various air quality monitoring sensors, e.g., humidity, temperature, pressure, etc. The idea is that the sensor data collected can be exchanged via an LPWAN interface (usually Bluetooth) and used for post-processing by the end-users. These sensors are installed in indoor environments (usually inside buildings and warehouses), making it difficult for end-users to demonstrate and emulate realistic smart-cities applications.

\vspace{0.2em}

\noindent \textbf{Smart Cities}: Finally, OpenCyberCity~\cite{openCityTestbed,openCyberCityTestbed} provides access to a miniaturised Smart City testbed, emulating a Smart Building and an Intelligent Transportation System (ITS) scenarios. The small scale of the testbed does not allow for realistic experimentation, but the architecture is similar to UMBRELLA. The developers provide ways to run experiments, interact with the sensors and actuators, reset the testbed between experiments, run ML-based optimisations and visualise the results. They also emphasise the security features provided.

Overall, all available testbeds are usually \textit{limited to one or a couple of very specific use cases and applications}. Despite their distinctive features, many existing platforms are \textit{constrained to indoor environments}, \textit{do not provide true-to-life facilities and scenarios}, equip \textit{limited hardware}, or \textit{lack integration} with multiple use cases. This limits their ability to emulate real-world Smart Cities scenarios where systems interact in a SoS manner. Moreover, as described, there is a lack of access to real-time, real-world data, particularly in the ML-enabled platforms. Finally, as described, some available \textit{testbeds are limited to a specific audience} or can be used under subscription-based schemes, which could become rather costly for independent researchers. These are the \textbf{gaps that UMBRELLA aims to fill}, offering users a \textbf{diverse}, \textbf{true-to-life playground} for experimentation that can \textbf{provide} researchers and industrial partners with \textbf{an open-access environment and foster innovation}.

\section{Requirement Analysis \& Design Reflections}\label{sec:requirement_analysis}
% For inspiration read these papers
% https://hal.inria.fr/hal-01147346/document
Consider a typical example of an IoT application, e.g., a city-wide air quality monitoring use-case, as described in~\cite{airquality}. This use case relies on several components. At first, we have the sensors installed on street furniture and building facades that collect the air quality readings. We later have the wireless interfaces responsible for transmitting the sensor data to one or multiple locations. The data collected are then post-processed and stored on local servers or a cloud system. All sensory data can be later visualised from a web portal. Based on the provided data, an IoT system can make recommendations to the decision maker in an automated or human-triggered fashion to optimise the system's behaviour (e.g., if air pollution is increased on the highway during peak time, the speed limit on a smart motorway could be set to a suitable number to reduce pollution). This scenario could be tackled holistically (e.g., a team builds the entire pipeline and evaluates it on a real-world system) or from the research point of view (e.g., a novel traffic signal control algorithm is designed based on historical data).

\subsection{Evaluation of a typical IoT ecosystem}\label{subsec:evalIoTEcosystem}
Designing and developing a software application or a hardware component typically starts by identifying the objective and formulating a hypothesis. From there, the idea is conceptualised and refined. Planning around the idea involves selecting appropriate tools and defining user experiences, interfaces, and key features. Next, the idea is materialised, implemented and validated by conducting tests and collecting data. The idea and the implementation are further refined through an iterative process until a working proof-of-concept is achieved. This follows the traditional Software Development Life-Cycle (SDLC) models~\cite{sdlc}.

For each new idea implemented within an IoT ecosystem, it is essential to consider the tight integration of all physical and digital components~\cite{sdlc,iotDeviceLifecycle}. Evaluating a new idea is a laborious task, with simulation tools providing a cost-effective way to assess a complex system or use case. However, simulations are built on assumptions about several parameters of the environment that cause uncertainties. As described in~\cite{agileCalibration}, results vary compared to the real world, even after thoroughly calibrating a simulation framework. 

As described in~\cite{gPapadopoulos}, the complementarity between simulation and real-world studies is particularly important for IoT application characterisation. It is also noted that high-quality assessment has a strong need for experimentation on real hardware and at scale. Particularly for complex IoT scenarios, cultivating, refining, or augmenting an entire ecosystem requires interaction with both the physical and digital components of the given infrastructure~\cite{plcm}. This is the aim of UMBRELLA. It provides ways for users to build complex use cases involving both new hardware and software components and allows for designing complex scenarios that are rather tricky to assess in existing testbed.

\subsection{Considerations for an IoT Experiment}\label{sub:exp_lifecycle}
\begin{figure}[ht]
    \centering
    \includegraphics[width=0.95\columnwidth]{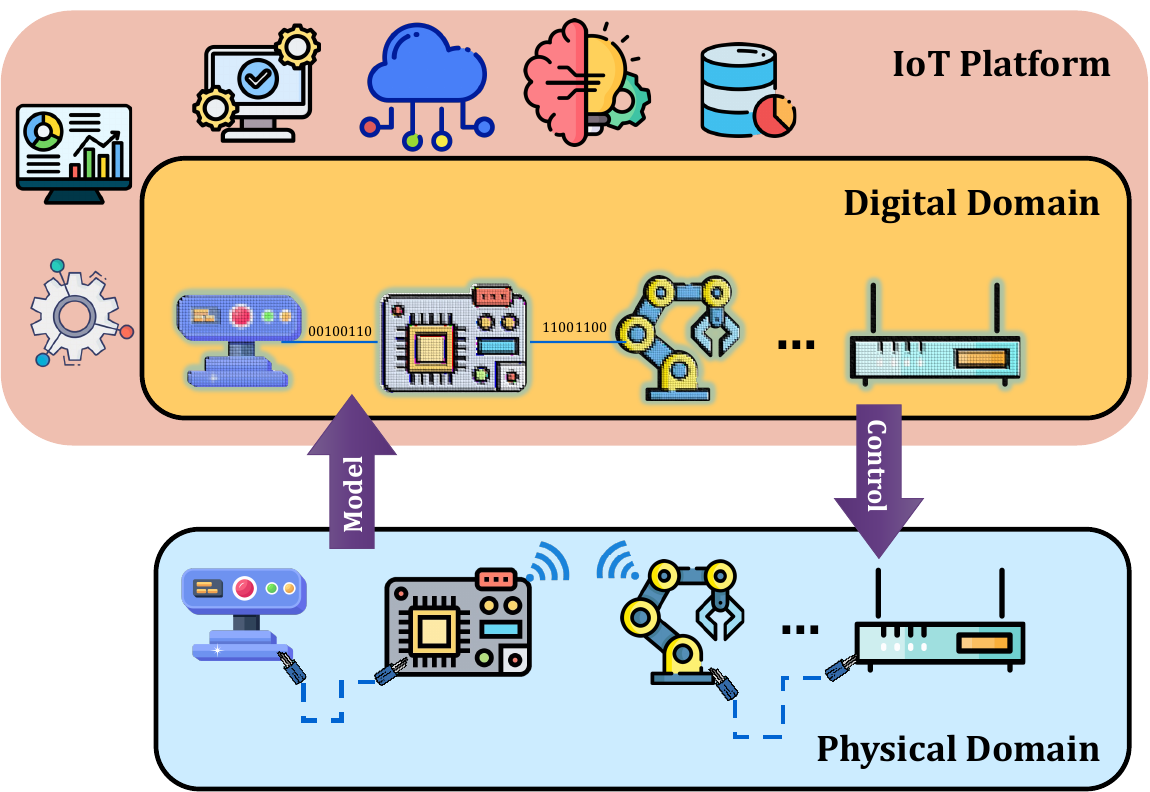}
    \caption{\textbf{Physical and digital assets in an IoT ecosystem. The IoT platform is the centre of the digital realm.}}
    \label{fig:physical_digital}
\end{figure}

An IoT ecosystem is characterised by its \textit{vast physical and digital assets}~\cite{physicalDigitalAssets}. In the \textit{physical domain}, we find tangible devices like sensors, actuators, microcontrollers, and communication interfaces, either wirelessly connected or hard-wired~\cite{iotDeviceTypes}. In the \textit{digital domain}, we have software applications consisting of reusable components, modules, and various building blocks which serve multiple applications and products~\cite{iotplatform} (Fig.~\ref{fig:physical_digital}). An \textit{IoT platform is the centre} of the digital realm and streamlines the deployment and orchestration of applications, manages sensing and actuating devices, and fosters a seamless integration of public and private infrastructure. This holistic digital solution is often termed a Platform as a Service (PaaS)~\cite{paas}.

When considering the life-cycle of an IoT experiment, IoT testbeds and platforms provide a playground that closely aligns with the real world in terms of scale, behaviour, functionality, environment and constraints. They also provide tools to run and manage the life-cycle of the experiment~\cite{experimentalLifecycle}. Chronologically, this life cycle could be broken down into the ``before'', ``during'', and ``after'' phases, each with unique requirements and designs.

\subsubsection{Before the Experiment - Design Phase}
During this phase, one defines the \textit{experimental specification}. This implies that the experimenter should be given ways to choose the resources, i.e., the number, type and properties required for the experiment and the interactions with these resources. A testbed, particularly when heterogeneous devices are involved, should offer ways for the users to \textit{develop their own client applications} leveraging existing services and \textit{exposed Application Programming Interfaces (APIs)} while utilising tools and programming languages they are familiar with.

Furthermore, the user should be given the \textit{flexibility to specify the data to be collected}, if and how they are visualised and have an easy way of validating the experimental configuration. Nevertheless, \textit{security} questions can be raised in that phase to protect other users' data and the system from malicious activities.

\subsubsection{During the Experiment - Execution Phase}
During this phase, it is necessary to \textit{provide interfaces for interacting and adjusting parameters} of the devices, sensors and actuators that take part in the experimentation. \textit{Debugging interfaces} and feedback loops are also crucial. Based on the experiment type, different tools could be provided that monitor network metrics, energy consumption, healthiness of the applications, etc. 

At this phase, the \textit{accessibility} to the testbed and the \textit{load balancing} between different experiments are also critical. The accessibility could be achieved in various ways, ranging from web servers and portals to direct SSH access to devices. The solutions implemented should again consider the \textit{security} requirements of a given system. Regarding load balancing, a system should handle this automatically to avoid the voracious use of resources by greedy users.

Finally, \textit{inter-experiment interference must be avoided}. For example, parallel radio experiments should run on different channels or be scheduled at different times. Moreover, utilising existing features (e.g., visualising past results) should not impact other ongoing experiments.

\subsubsection{After the Experiment - Analysis Phase}
After the end of the experiment, a testbed should provide ways to \textit{collect, store and analyse data}. Different ways of visualising the results can be provided, and different statistical models could be applied during the visualisation. Furthermore, accessing the experiment logs that could give valuable information or more in-depth results is also essential for post-processing and further analysis.

\qquad

\textbf{UMBRELLA ecosystem accommodates all the above requirements}. It provides ways to interact with the available devices before and during the experimentation and stores and visualises data and analytics on the fly while providing an easy and scalable way to access the hardware devices and deploy experiments in a scalable and flexible way. In the following sections, we will go into more detail on how all the above are achieved.

\subsection{Testbed Features}
An IoT testbed accommodating a spectrum of experiments across various use cases should provide broad functionality and features. Broadly, testbed requirements are formalisation, scheduling, replication and repetition, monitoring, cleanup, query and discovery, and adaptation~\cite{experimentalLifecycle}. As \textbf{UMBRELLA provides access to a multitude of systems and platforms}, in the following sections, we capture the features required across the different physical and digital domains.

\subsubsection{Hardware Features}

The goal of any experimental platform like UMBRELLA is uninterrupted access to real-world testing equipment. The hardware should operate in realistic conditions and constraints, matching the requirements for the envisioned use cases. When considering the hardware requirements of an IoT testbed and platform, we have:

\begin{enumerate}[wide, labelwidth=!, labelindent=9pt, itemindent=0pt, label=\roman*)]
% \begin{enumerate}[wide, labelwidth=!, labelindent=9pt, itemindent=0pt, label=\roman*)]
    \item \textbf{Scalability and Expandability}: An IoT testbed should support \textit{easy access to multiple devices} during an experiment. Adding new devices should not interfere with existing devices in the system. The expandability of the hardware should require minimal human intervention. Finally, the addition of new hardware should be in a cost-effective way. 
    
    \item \textbf{Replaceability and Modularity}: Components should be \textit{easily swappable}, catering to wear and tear (e.g., due to exposure to adverse weather conditions) and obsolescence (e.g.,  inadequate functionality for future use cases). The swappable capabilities should favour a plug-and-play approach. 

    \item \textbf{Heterogeneity}: The system should provide \textit{diverse sensors, wireless interfaces, and computational capabilities}. This will ensure flexibility and adaptability to new and existing challenges and experiments. Seamless communication between diverse devices necessitates \textit{well-defined interfaces} and access to \textit{communication buses}, communication libraries, etc. Finally, the different devices should be \textit{remotely programmable}; thus, the necessary tools should be provided, e.g., templates and examples, programming tools, bootloaders, etc. 

    \item \textbf{Federation}: \textit{Multi-site deployments} can augment the testbed's scale and functionality. The federation can distribute the resources across all available nodes and different sites. A unified abstraction layer should facilitate resource reservation and authentication across testbeds. Finally, common communication buses can be used to exchange information across the distributed applications.

    \item \textbf{Adaptability}: A flexible testbed should easily adapt to \textit{evolving technological trends} and experimental needs, ensuring the hardware remains relevant and capable of meeting emerging requirements.
\end{enumerate}

\subsubsection{Experimentation Features}
The experimental lifecycle described in Sec.~\ref{sub:exp_lifecycle} must be enabled by several features. A testbed should provide services and tools to interact with an experiment intuitively and easily. Some more important features to consider are as follows: 

\begin{enumerate}[wide, labelwidth=!, labelindent=9pt, itemindent=0pt, label=\roman*)]
% \begin{enumerate}[wide, labelwidth=!, labelindent=9pt, itemindent=0pt, label=\roman*)]

    \item \textbf{Scheduling}: The testbed should provide an \textit{efficient and effective way to schedule experiments}, allow multiple users to reserve the resources and avoid conflicts (or mitigate against them). The users should be able to schedule experiments on demand (when resources are available) or at a specific timeframe. Finally, the overutilisation of resources should be mitigated (e.g., by limiting the length of the experiments).

    \item \textbf{Repeatability}: Ensuring \textit{consistent configuration} across repeated experiments is crucial, especially when assessing varying parameters. A testbed should provide easy ways to repeat an experiment with the same or different configurations. Standardising the configuration parameters and copying them to the new experiment can ease the repeatability and simplify the data collection. 

    \item \textbf{Monitoring and Discovery}: A testbed should provide \textit{continuous real-time monitoring} capabilities for running experiments. This enables tracking progress or detecting anomalies/hardware faults. A user interface can be used to query the testbed for available resources, visualise past experiment results, configure experiments and monitor the status of scheduled jobs. This could be achieved with access to a Graphical User Interface (GUI). The GUI can display relevant information and warnings to guide users on fixing errors or facilitate informed decision-making.

    \item \textbf{Cleanup}: Post-experiment, the testbed should automatically \textit{reset and restore} the system to its original state. This enables the repeatability of the experiments, releases unutilised resources and improves the speed and performance of the system. Moreover, hanging processes could create noisy data. Thus, they must be automatically removed.
    
    \item \textbf{Simulations}: Sec.~\ref{subsec:evalIoTEcosystem} described the \textit{need for both real-world and simulated experiments}. A testbed should combine simulated, emulated and real-world-like environments under a common platform. The transition between simulated and real-world experiments should be seamless. This could be achieved by, e.g., allowing the same code to be deployed in both the simulated and real-world environment. Moreover, \textit{digital twining} can enhance the experimentation and the seamless integration of both worlds. Complex scenarios, e.g. injecting non-expected data or modifying the behaviour of a device during a real-time experiment, can significantly push the limits of algorithms implemented and evaluated.

    \item \textbf{Interactions and Software Interfaces}: A testbed should provide \textit{standardised interfaces} for interacting with various components and collecting data. Varied abstraction layers and API interfaces are necessary to cater for different experiment requirements. Command line tools could interface with the testbed and the running experiments. A Graphical User Interface (GUI) can be used to visualise results, configure experiments and monitor the status of scheduled jobs. Finally, messaging protocols can be used for standardising data transfers.

    \item \textbf{Data Management}: A testbed should provide various \textit{data storage solutions} to handle structured, semi-structured, and unstructured data. This includes traditional databases, time-series databases for sensor data or data lakes for raw data. Moreover, normalisation and transformation of the data must be considered. This may include converting units of measure, time-stamping, or aligning data to a common schema.
 
    \item \textbf{Open Access and Fine-grained Access Control}: A testbed should be broadly reached by the community. It must be \textit{remotely accessible} so users can remotely run experiments on the provided infrastructure. To enable greater reach, it could also be \textit{free-of-charge and open-access}. However, the testbed must \textit{ensure the security} of the user data and implement mechanisms for fine-grained control and user access. The users and their data must be protected from leaks and the system from malicious activities.

\end{enumerate}

\begin{figure*}[ht]
    \centering
    \includegraphics[width=0.85\textwidth]{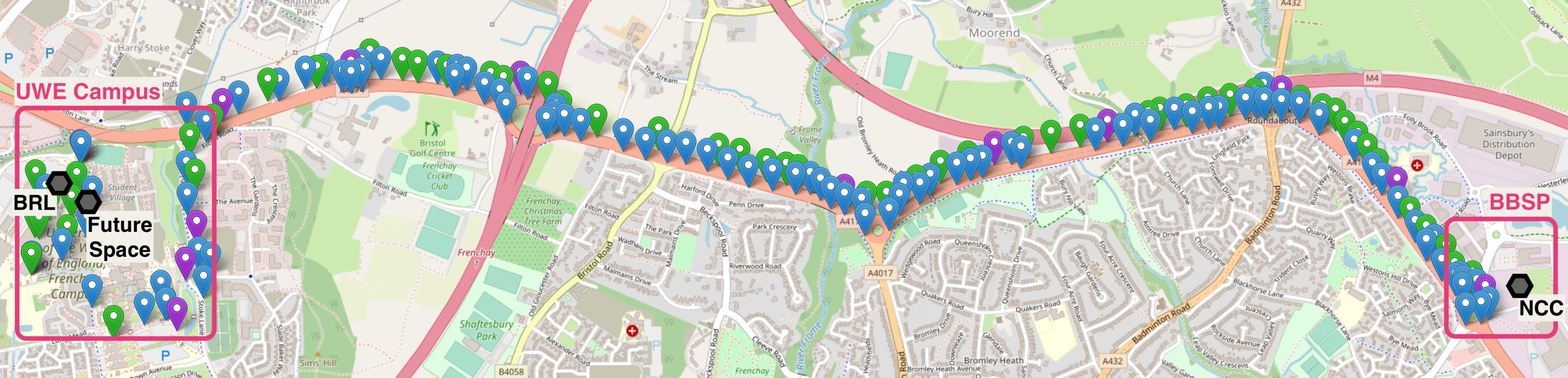}
    \caption{\textbf{The UMBRELLA network. All nodes are installed on public lampposts across a $\sim$\SI{7.2}{\kilo\meter} stretch of road and the UWE campus. The colours represent the nodes' connectivity, i.e., green is fibre-connected, blue is Wi-Fi-connected, and purple is fibre-connected and can act as a LoRa gateway.}}
    \label{fig:umbrella_network}
\end{figure*}

\subsubsection{Integration of Mobile and Remote Devices}
Given the growing interest in IoT applications in logistics, smart farming, predictive maintenance, etc., the testbed should not only focus on fixed infrastructure but also support mobile and remote nodes. With regards to mobile nodes, \textit{localisation and path planning assisting subsystems} can help accurately position mobile nodes in the 3D space and track their movement. Also, the hardware's safety, the experiments' continuity, and \textit{human-robot interactions} should be carefully considered for mobile nodes. 

Use cases like Smart Farming require\textit{ hierarchical device architecture and control} with layers of edge and remote embedded devices in a microgrid format~\cite{hierarchicalIoT}. As an IoT testbed cannot always provide all the necessary equipment for any given use case, it should enable the \textit{integration of remote nodes} that can enhance its capabilities. It should \textit{provide wireless interfaces and protocols} for enabling the connection of new devices, common \textit{communication buses} for the data exchange and \textit{standardised practices} for interacting with these remote nodes. \textit{Identity management} is critical for such as setup, and the testbed must provide ways to authenticate new devices.

Finally, a testbed should allow interaction and integration with other collaborative IoT ecosystems so that more complex use cases can be devised and tested there. Standardised communication gateways can enable protocol translation and data normalisation when common interfaces are unavailable.

\subsubsection{Security Considerations}
In the landscape of an IoT ecosystem, security remains paramount. Ensuring robust protection at every layer -- hardware, software, and network -- is crucial. An IoT testbed must address these concerns as detailed below:

\begin{enumerate}[wide, labelwidth=!, labelindent=9pt, itemindent=0pt, label=\roman*)]
% \begin{enumerate}[wide, labelwidth=!, labelindent=9pt, itemindent=0pt, label=\roman*)]
    \item \textbf{Hardware Security}: The IoT testbed must ensure that all devices only boot up with malware-free software, so \textit{safeguarding against malicious firmware} is essential. Nodes installed in public spaces should be physically secured or not easily accessible by the general public. Trusted Platform Modules (TPMs) can help prevent hardware tampering by storing used cryptographic keys. Finally, \textit{safeguarding data at rest} can ensure that even if the hardware is compromised, the data will remain inaccessible. 
    
    \item \textbf{Software Security}: The IoT ecosystem must be updated with \textit{frequent software and vulnerability patches}. All software applications running on the testbed (either user experiments or tools deployed by the administrators) should be authenticated and integrity checked against Common Vulnerabilities and Exposures (CVE) databases. Moreover, the devices connecting to the system should be authenticated and authorised. This could be done via standard Public Key Infrastructure (PKI) strategies. Finally, particularly for the user experiments, they must be executed within a \textit{sandbox with limited access rights}, so if compromised, they will not jeopardise the entire system. 
    
    \item \textbf{Network Security}: All \textit{data in transit must be encrypted}. An IoT platform should incorporate Intrusion Detection Systems (IDS) and Intrusion Prevention Systems (IPS) to monitor the network traffic for malicious activity and block potential threats from penetrating the network. Furthermore, network segmentation is important in isolating different parts of the network and ensuring breaches will not propagate. Finally, external access to the network, particularly to systems, portals, and components not meant to be accessible via the public Internet, must be through an encrypted tunnel via a Virtual Private Network (VPN).
    
    \item \textbf{Monitoring and Incident Response}: \textit{Continuous monitoring} of the system can allow immediate detection of any anomalies or unauthorised access. In case of security breaches, a predefined \textit{incident response plan} should be in place to ensure rapid containment and mitigation of threats.

    \item \textbf{User Access Control}: \textit{Fine-grained access control} can allow for customisable user access levels, minimising access to internal resources based on assigned roles. Finally, user authentication must be considered by verifying email addresses, ensuring their validity, and identifying users before accessing the experimentation platform. 
\end{enumerate}

\begin{figure*}[ht]
    \centering
    \includegraphics[width=0.85\textwidth]{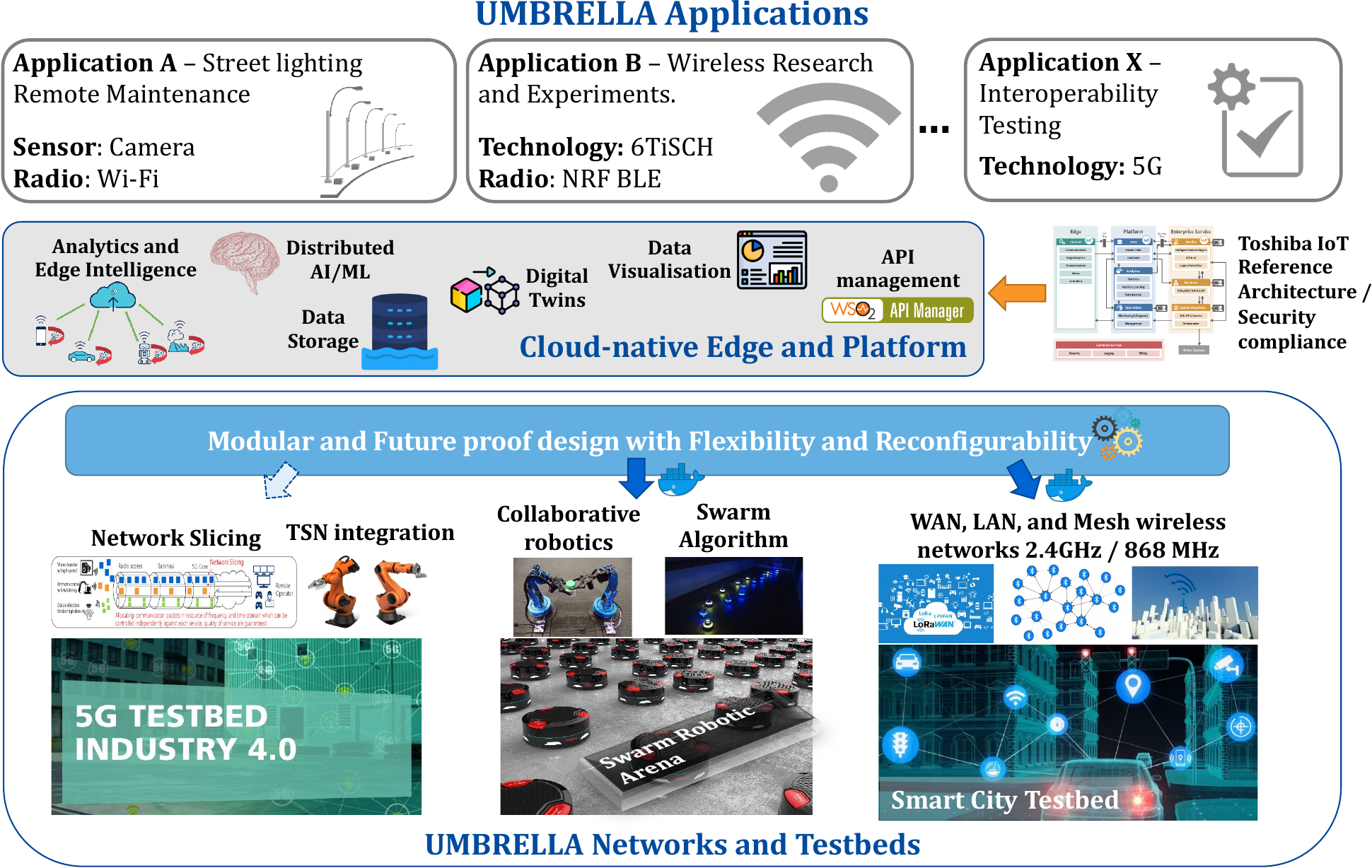}
    \caption{\textbf{UMBRELLA SoS architecture overview with support of multiple sub-system testbeds.}}
    \label{fig:testbeds}
\end{figure*}

\subsection{UMBRELLA: Open Access System-of-Systems}
As seen in Sec.~\ref{sec:relatedWork},  existing IoT testbeds predominantly target specific technologies or a handful of use cases. That overlooks the multifaceted nature of IoT implementations. While such specialised testbeds have their merits, there is a gap in addressing holistic, real-world-like systems. Such systems, encompassing diverse devices, communication protocols, and scenarios, are vital for comprehensively validating IoT solutions.

This is the gap that UMBRELLA tries to bridge. \textbf{UMBRELLA provides a platform where multiple technologies and use cases converge, enabling more representative experimentation and fostering innovation that aligns closely with real-world demands.} UMBRELLA platform and the nodes installed are located around the South Gloucestershire region in the UK, across a $\sim$\SI{7.2}{\kilo\meter} stretch of road and connecting five innovation hubs in the area, the Bristol \& Bath Science Park (BBSP), National Composites Centre (NCC), the University of the West of England (UWE), Bristol Robotics Laboratory (BRL), and Future Space (Fig.~\ref{fig:umbrella_network}). Over 200 multi-wireless, multi-sensor, edge computing devices and robotics nodes have been deployed. The platform aims to accelerate the deployment of innovative technologies and solutions and significantly increase efficiency in creating new services, leading to inward investment, job growth, sustainable transport, low carbon emissions, and improved regional health and well-being. 

A user can interact with UMBRELLA in the form of ``\textit{experiments}''. An online portal enables easy deployment and monitoring capabilities. Designing and running an experiment on the testbed gives \textit{access to a number of wireless interfaces and sensors}, \textit{data generated} in real-time and \textit{diverse environments} that can help emulate real-world scenarios. More details about the interfaces currently provided for experimentation can be found in Secs.~\ref{subsubsection:endpoint_boards} and~\ref{subsec:wireless_testbed}. The sensors available are detailed in Sec.~\ref{subsubsec:sensingpod}. The \textit{AI-enabled edge devices} allow the end-users to train ML or Federated Learning (FL) models in real time based on collected data from their experiments or test the inference of their pre-trained models in a real-world environment. In the following sections, we go into more detail on how UMBRELLA has been implemented and its core functionality. We also present existing use cases that run as part of the ecosystem deployed. A high-level overview of the currently deployed ecosystem can be seen in Fig.~\ref{fig:testbeds}.

\section{UMBRELLA Testbed Architecture}\label{sec:architecture}
UMBRELLA offers a mature and comprehensive platform for researching and developing IoT technologies and applications. As outlined in the subsequent sections, UMBRELLA's design and architecture were driven by creating tangible benefits for diverse stakeholders. For end users and the industry, the testbed enables rapid prototyping and evaluation of IoT solutions, saving costs and accelerating time-to-market. For governmental bodies, it provides a platform to trial Smart City technologies, enhance municipal services and support the central Digital and Industrial strategic vision. Industry players can commercialise proven innovations and use UMBRELLA as a playground for demonstrations. Researchers in academia gain access to real-world data at scale for their studies. 

Overall, UMBRELLA aims to foster an ecosystem of collaboration and innovation between users, industry, government and academia to create both economic and societal value. This overarching vision influenced the design choices and architecture of UMBRELLA and enabled a rapid prototyping platform for evaluating IoT solutions across hardware, software, and communication domains. 

\subsection{System Design and requirement analysis} 

\begin{table}[ht]
\renewcommand{\arraystretch}{1.1}
\caption{\textbf{High-level requirements summary.}}
\label{table:requirements}
\begin{tabular}{rl}
\textbf{Domain} &
  \textbf{Requirement} \\ \hline \hline
\textbf{Security} &
  \begin{tabular}[c]{@{}l@{}}Secure user access to testbed\\ Security monitoring and vulnerability scanning\\ Service and experiment access control\\ Application/user authorisation\\ Endpoint identity verification\\ Secure endpoint communication\\ Endpoint integrity protection\\ Endpoint authentication\\ User Identity and Authentication\\ Role-based access control to testbed resources\\ Storage and backup of testbed data\\ Privacy protection of  personal data\\ High reliability and availability\end{tabular} \\ \hline
\textbf{Safety} &
  \begin{tabular}[c]{@{}l@{}}Safety for robot systems - human/environment protection\\ Public safety when deploying nodes on street furniture\\ Experiment supervision and control\end{tabular} \\ \hline
\textbf{Testbed} &
  \begin{tabular}[c]{@{}l@{}}Experiment scheduling\\ Collect performance metrics \\ Inject external data to wireless \\ Evaluate software\\ Protocol software or firmware\\ Wireless firmware update\\ Evaluation performance metrics\\ Security software solution\\ Evaluate hardware\\ Retrieve experiments\\ Run wireless connectivity tests\\ Store experiment configurations and results\\ Visualisation of power consumption\\ View completed experiments\\ Archive experiments\\ Delete experiments and user data\\ Support collaborative experimentation via shared projects\\ Visualisation of KPIs through graphs/dashboards\\ Access to sensor data from external applications/ services\\ Wiki and guides for users and developers\end{tabular} \\ \hline
\textbf{Robotics} &
  \begin{tabular}[c]{@{}l@{}}Multi-collaborative robotic intralogistics experiments\\ Monitor experiment execution and Groundtruth\\ Wireless connectivity support and evaluation/expansion\\ Digital twin validation prior to arena deployment\\ Changing of warehouse layouts\\ Support experiment KPIs for performance evaluation\\ Remote visualisation of experiments (arena/digital twin)\end{tabular}  \\ \hline
\end{tabular}
\end{table}

The key requirements obtained from all stakeholders cover a broad area across different aspects, such as security, functionality, and operation. They are summarised in Tab.~\ref{table:requirements}. These requirements formed the basis of our system architecture and infrastructure design. UMBRELLA focused on being an open, generic and future-proof implementation that can cover multiple technologies and use cases, benefiting businesses, academia and society. Some of our key design considerations were the following:

\begin{figure}[t]
    \centering
    \begin{minipage}[t]{\columnwidth} % Adjust width as needed
        \begin{subfigure}[t]{0.413\columnwidth}
            \includegraphics[width=\columnwidth]{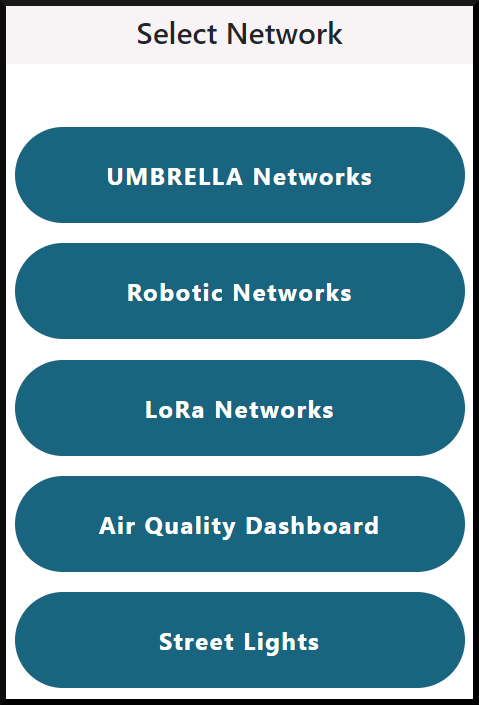}
            \caption{\textbf{Choosing a network.}}
            \label{fig:networks_to_choose}
        \end{subfigure}
        \vspace{3mm}
        \begin{subfigure}[t]{0.557\columnwidth}
            \includegraphics[width=\columnwidth]{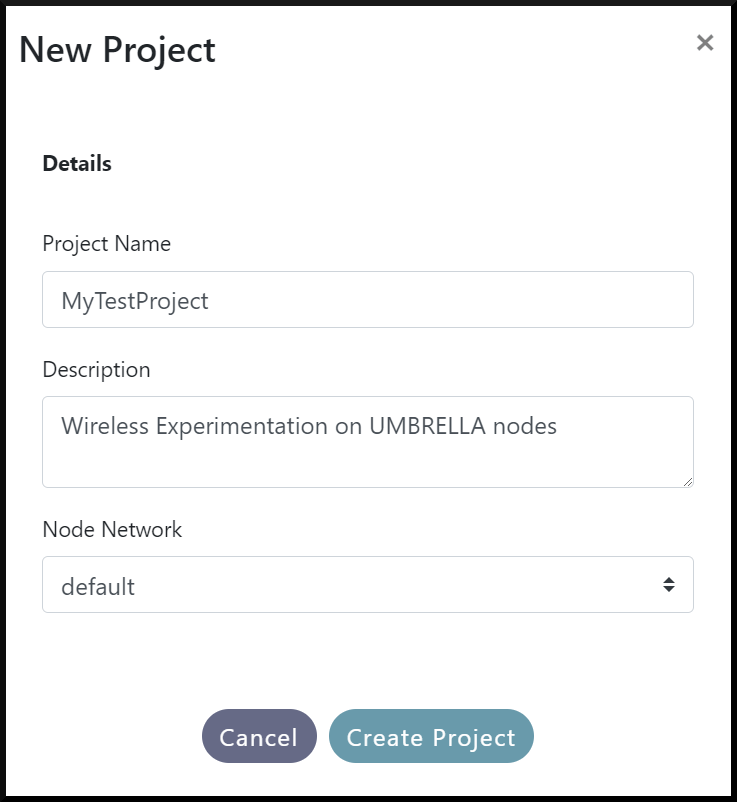}
            \caption{\textbf{Creating a new project.}}
            \label{fig:newproject}
        \end{subfigure}
    \end{minipage}%
    \hfill
    \begin{minipage}[b]{\columnwidth} % Adjust width as needed
        \begin{subfigure}[b]{\columnwidth}
            \includegraphics[width=\columnwidth,height=0.9\textheight,keepaspectratio]{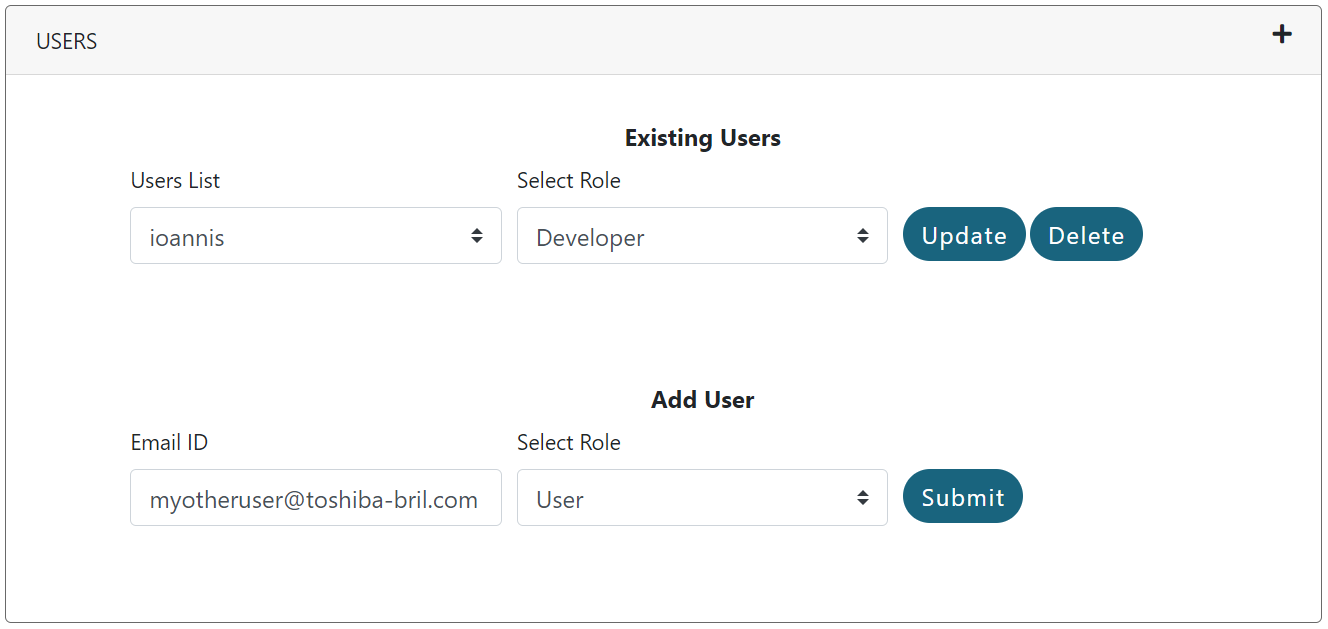} % Adjust height if needed
            \caption{\textbf{Adding users to existing projects and updating roles.}}
            \label{fig:project_update}
        \end{subfigure}
    \end{minipage}

    \caption{\textbf{The interfaces for choosing an available network and creating or updating projects.}}
    \label{fig:combined_projects}
\end{figure}

\begin{figure*}[h!]
    \centering
    \begin{minipage}[b]{0.52\textwidth} % Adjust width as needed
        \begin{subfigure}[b]{\textwidth}
            \includegraphics[width=\textwidth]{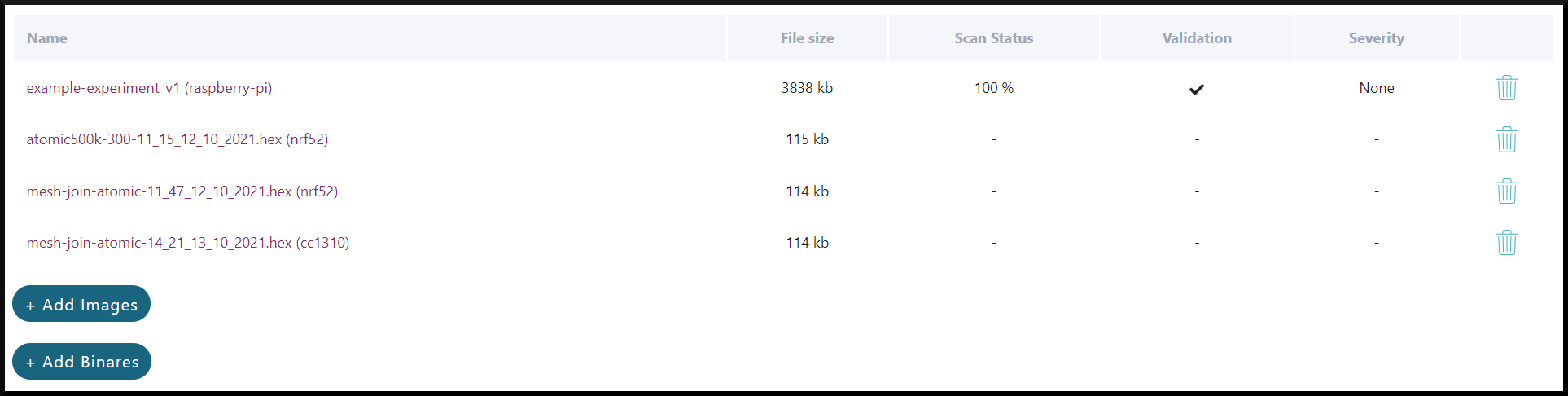}
            \caption{\textbf{Container images and the firmware binaries uploaded by the user.}}
            \label{fig:image_upload}
        \end{subfigure}

        \begin{subfigure}[b]{\textwidth}
            \includegraphics[width=\textwidth]{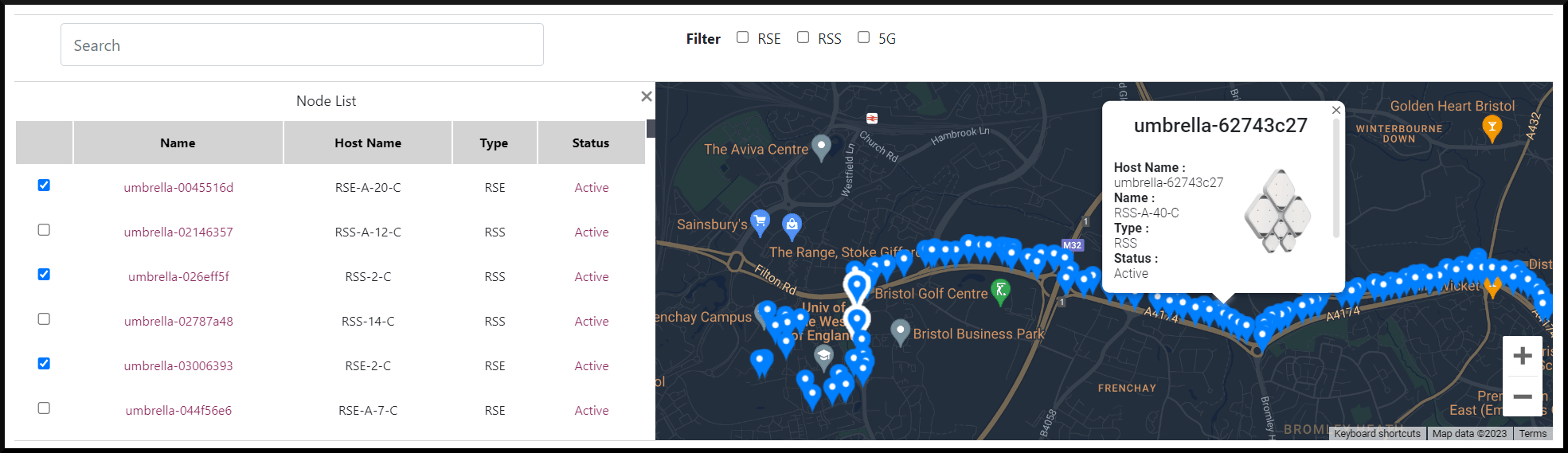}
            \caption{\textbf{A user chooses a number of nodes for an experiment -- the nodes' availability is based on the experiment time, date, and length.}}
            \label{fig:node_selection}
        \end{subfigure}
    \end{minipage}%
    \hfill
    \begin{minipage}[t]{0.223\textwidth} % Adjust width as needed
        \begin{subfigure}[b]{\textwidth}
            \includegraphics[width=\textwidth,height=1.1\textheight,keepaspectratio]{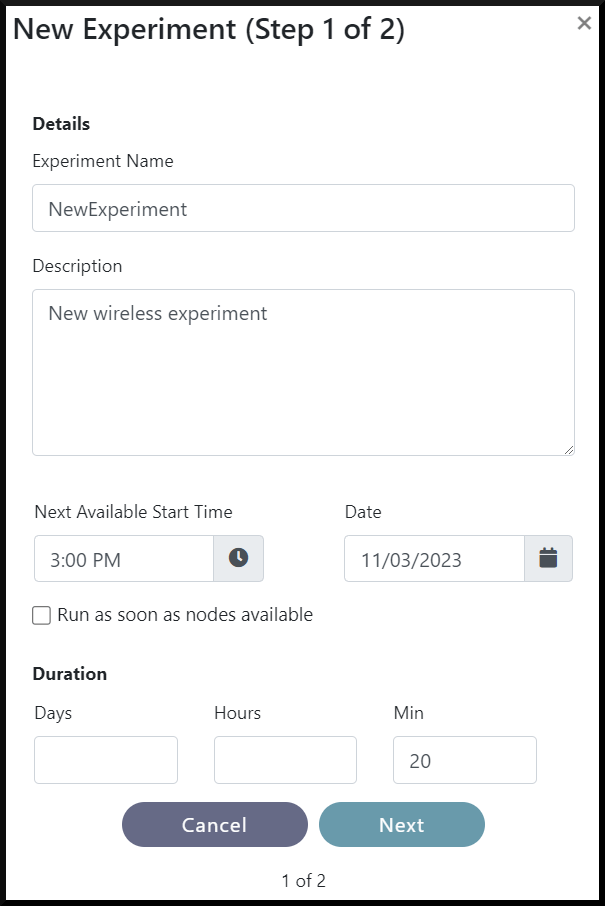} % Adjust height if needed
            \caption{\textbf{The experiment screen.}}
            \label{fig:experiment_screen}
        \end{subfigure}
    \end{minipage}
    \hfill
    \begin{minipage}[t]{0.23\textwidth} % Adjust width as needed
        \begin{subfigure}[b]{\textwidth}
            \includegraphics[width=\textwidth,height=0.9\textheight,keepaspectratio]{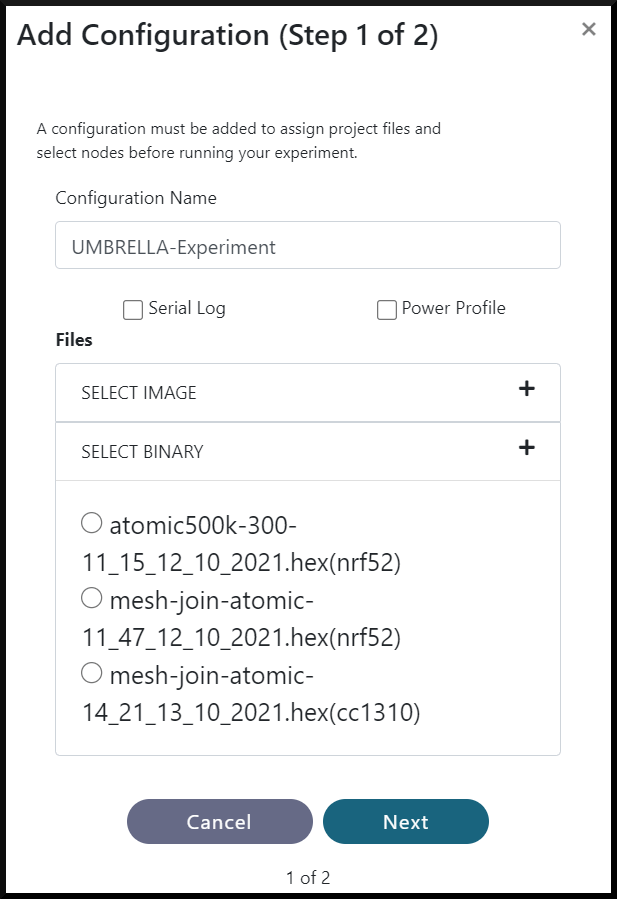} % Adjust height if needed
            \caption{\textbf{The configuration screen.}}
            \label{fig:configuration_screen}
        \end{subfigure}
    \end{minipage}

    \caption{\textbf{The interfaces for creating an experiment. A user can upload all required files that are scanned and validated by the backend and choose a number of nodes for experimentation.}}
    \label{fig:combined}
\end{figure*}

\begin{enumerate}[wide, labelwidth=!, labelindent=9pt, itemindent=0pt]
    \item \textbf{Modularity}: The system was designed modularly to allow flexibility in adding/removing hardware and software components. New hardware can be added in a Hardware-Attached-on-Top (HAT) fashion, allowing for expansion with new sensors, devices, and capabilities without major architecture changes. Software components run as isolated containers either on the backend or the nodes' side, interacting with well-defined interfaces. 
    \item \textbf{Interoperability}: Standard interfaces and common data formats like JSON enable easy integration of new components from different vendors. Common communication protocols like Message Queuing Telemetry Transport (MQTT) and well-defined REST APIs ensure compatibility across different applications. Containerisation and microservices architecture enabled independent deployment of components across the same platform.
    \item \textbf{Scalability}: The system was built cloud-natively to enable scaling and supporting many concurrent users and experiments. User experiments are stateless, allowing for easy recovery and avoiding bottlenecks. Load balancing and auto-scaling techniques were implemented in the cloud-native backend. The network was designed to handle increased traffic. Stateful applications, such as databases, are provided in high availability to avoid system failures and data corruption.
    \item \textbf{Security}: Authentication, authorisation, encrypted communication, sandboxing of experiments, and other security measures were implemented to protect the platforms and the different testbeds. Role-based access control was used to restrict user privileges. Network segmentation protects from attack propagation across different network domains.
    \item \textbf{Usability}: The user interface and APIs were designed for ease of use. Detailed documentation, code samples, and tutorials are provided. Scheduling systems and tools for monitoring and repeating experiments were added. Containerisation provided environment consistency and enabled portability across different devices. Components could be migrated easily in case of hardware failure.
    \item \textbf{Automation}: Processes were automated where possible, e.g. for scheduling, deployment, monitoring, and cleanup. This improved efficiency and reduced human effort. Job schedulers, within the cloud-native environments or running natively on the operating system, automate testing and deployment tasks while ensuring synchronisation across the different nodes and environments.
    \item \textbf{Analytics}: Data analytics modules were included for aggregating data, identifying trends/patterns, and generating insights from experiments. Users can analyse results using their own preferred tools, either in real-time or as a postprocessing step, downloading the data from the portal.
    \item \textbf{Sim-to-real}: A digital twin simulation environment was added to allow prototyping and sim-to-real transfer of experiments.
    \item \textbf{Real-world}: The hardware components are easily adaptable to a range of applications and both indoor and outdoor placements. Outdoor nodes are weatherproof, robust, and have a long service life. 
\end{enumerate}

In the following sections, we go into more detail about the different components of UMBRELLA, discuss the technologies chosen for fulfilling the requirements captured and discuss existing use cases currently operational across the different testbeds provided.

\subsection{Terminology and Experiment Description}\label{subsec:terminology}
As described earlier, users can interact with UMBRELLA through ``\textit{experiments}''. An \textit{experiment} is considered an application executed at a specific date and time for a pre-defined duration. Users can choose one or many ``\textit{nodes}'' to participate in an experiment. A \textit{node} is defined as an \textit{enclosed collection of hardware components} (sensors, actuators, processing units, etc.). For example, currently, UMBRELLA provides three types of \textit{node}, i.e., a robotic node, an outdoor-installed node (referred to as \textit{UMBRELLA node}), and a 5G node. In the following sections, we describe them in more detail.

\noindent \textbf{Experimentation networks}: After creating an account on the UMBRELLA portal\footnote{UMBRELLA portal: \url{https://portal.umbrellaiot.com/}}, users can run experiments over the different ``\textit{networks}'' provided (Fig.~\ref{fig:testbeds}). A \textit{network} is a collection of \textit{nodes} with common characteristics (Fig.~\ref{fig:networks_to_choose}). Access to different \textit{networks} is role-based. By default, users have access to the ``UMBRELLA networks'', i.e., all the networks with outdoor installed nodes; the ``Robotic networks'', i.e., all robotic testbeds; and the ``LoRa networks'', i.e., all networks with LoRaWAN enabled nodes. Access to other networks and use cases can be granted upon request from the UMBRELLA administrators. 

\noindent \textbf{Projects and users definition}: By choosing a \textit{network}, a user can later create ``\textit{projects}''. A \textit{project} requires a name, a description, and an available \textit{network} (Fig.~\ref{fig:newproject}). Within a \textit{project}, users can upload all the required application files for their experiments, schedule experiments, and visualise results. A \textit{project} can also be shared with other users, enabling collaboration (Fig.~\ref{fig:project_update}). A ``developer'' is a collaborator who can run experiments and modify the uploaded files, whereas a ``user'' is a collaborator who can only visualise the results of completed experiments. 

\noindent \textbf{Experiment creation}: In Fig.~\ref{fig:combined}, we see the interfaces for creating an experiment. A user initially uploads all firmware binaries and containerised applications on the portal (Fig.~\ref{fig:image_upload}). For each application, a user chooses a ``type''. The type could be either the wireless interface the binary will be uploaded to (e.g., Nordic nRF52840) or the ARM architecture family the application belongs to (e.g., ARM64). Each uploaded file is checked for vulnerabilities. If an application is considered vulnerable, a user cannot use it for experimentation and needs prior authorisation by the UMBRELLA administrators. 

\begin{figure*}[t]
    \centering
    \includegraphics[width=0.8\textwidth]{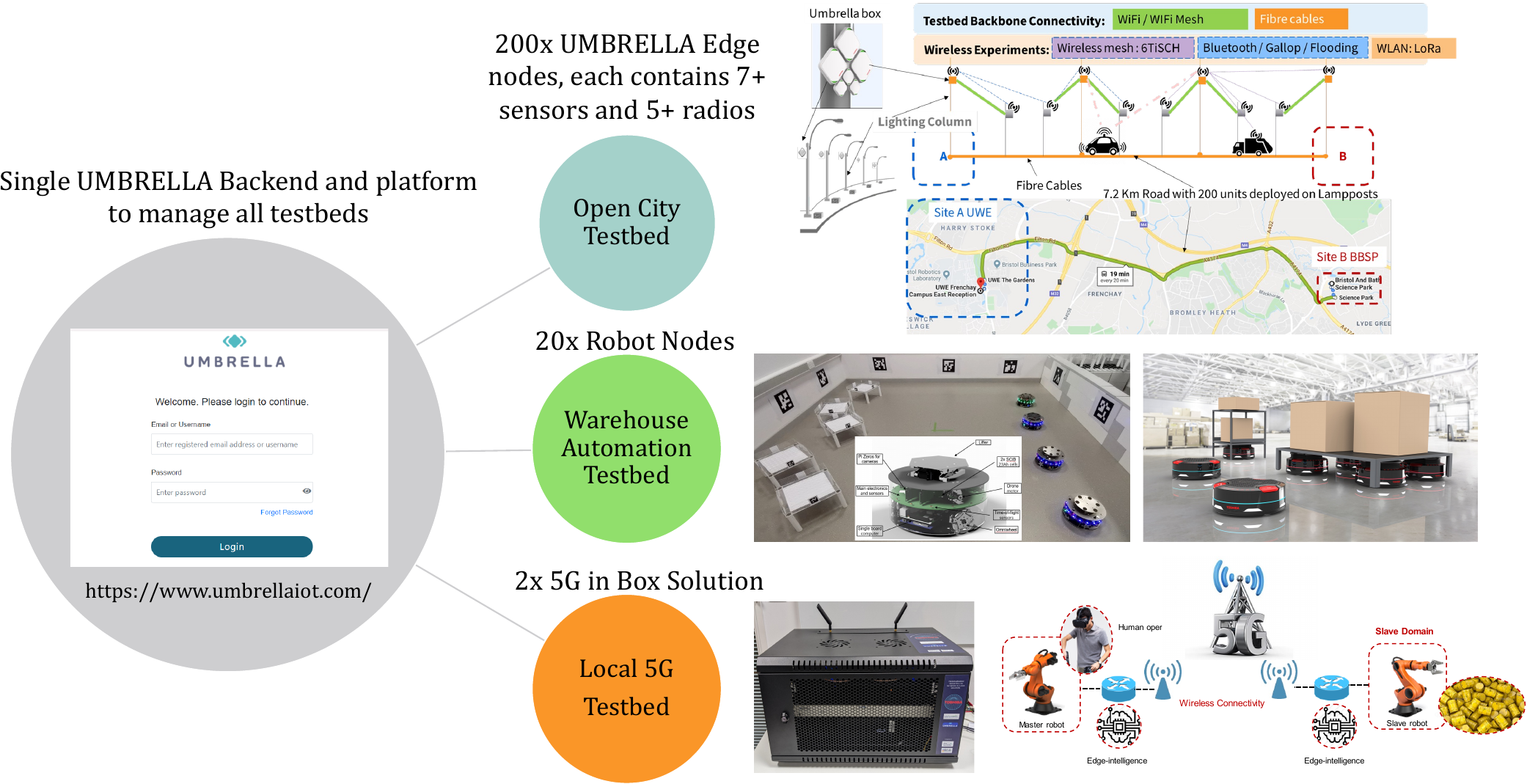}
    \caption{\textbf{UMBRELLA Infrastructure Overview.}}
    \label{fig:infrastructure}
\end{figure*}

\noindent \textbf{Experiment configuration}: A user can later create new experiments defining their duration, starting date, and time (Fig.~\ref{fig:experiment_screen}). Depending on the nodes' availability, a user can choose one or many nodes to participate in the experiment (Fig.~\ref{fig:node_selection}). The portal groups nodes depending on their connectivity. In the following sections, we will define what each group implies for each testbed. Finally, a user can define ``\textit{configurations}'' for each experiment. A \textit{configuration} controls the container images and firmware that will be deployed on a subset or all the \textit{nodes}. For an experiment, a user can provide multiple \textit{configurations}, so complex scenarios are designed and investigated.

\noindent \textbf{After experimentation}: After an experiment, all participating nodes are reset to their initial state and are prepared for a new experiment. This includes removing any user applications running on the nodes, resetting the wireless interfaces with dummy binary files, and deleting all user data after they are stored on the backend databases. This ensures that a new experiment can be executed with interference from previous executions and all resources are again available for the user.

The above experimental pipeline applies to all available networks, with minor modifications, e.g., the robotic experiments need to be validated in the simulated environment before being deployed on the real robots. Our Wiki page\footnote{UMBRELLA Wiki page: \url{https://wiki.umbrellaiot.com/}} provides more details for each individual network and use case.

\subsection {Infrastructure overview} 

UMBRELLA is a live testbed and network consisting of:  
\begin{itemize}
\item 200 UMBRELLA nodes 
\item \SI{22}{\kilo\meter} of fibre cables
\item 1000+ wireless radio devices
\item 1500+ IoT sensors
\item 20 robotic nodes
\item Digital twins
\item Two 5G network-in-a-box solutions
\item 75 edge GPUs
\end{itemize}
The deployed infrastructure is managed by a containerised, hybrid edge-cloud IIoT collaboration platform. As shown in Fig.~\ref{fig:infrastructure}, UMBRELLA contains three main testbeds with remotely accessible and re-programmable devices. These testbeds are: an open city testbed deployed along a $\sim$\SI{7.2}{\kilo\meter} stretch of the road and UWE campus in the region of South Gloucestershire in the West of England; a robotic testbed for warehouse automation applications; and a private 5G testbed for low latency and high reliability industrial wireless applications. 

\begin{figure}[t]
    \centering
    \begin{subfigure}{.34\columnwidth}
        \centering
        \includegraphics[width=\linewidth]{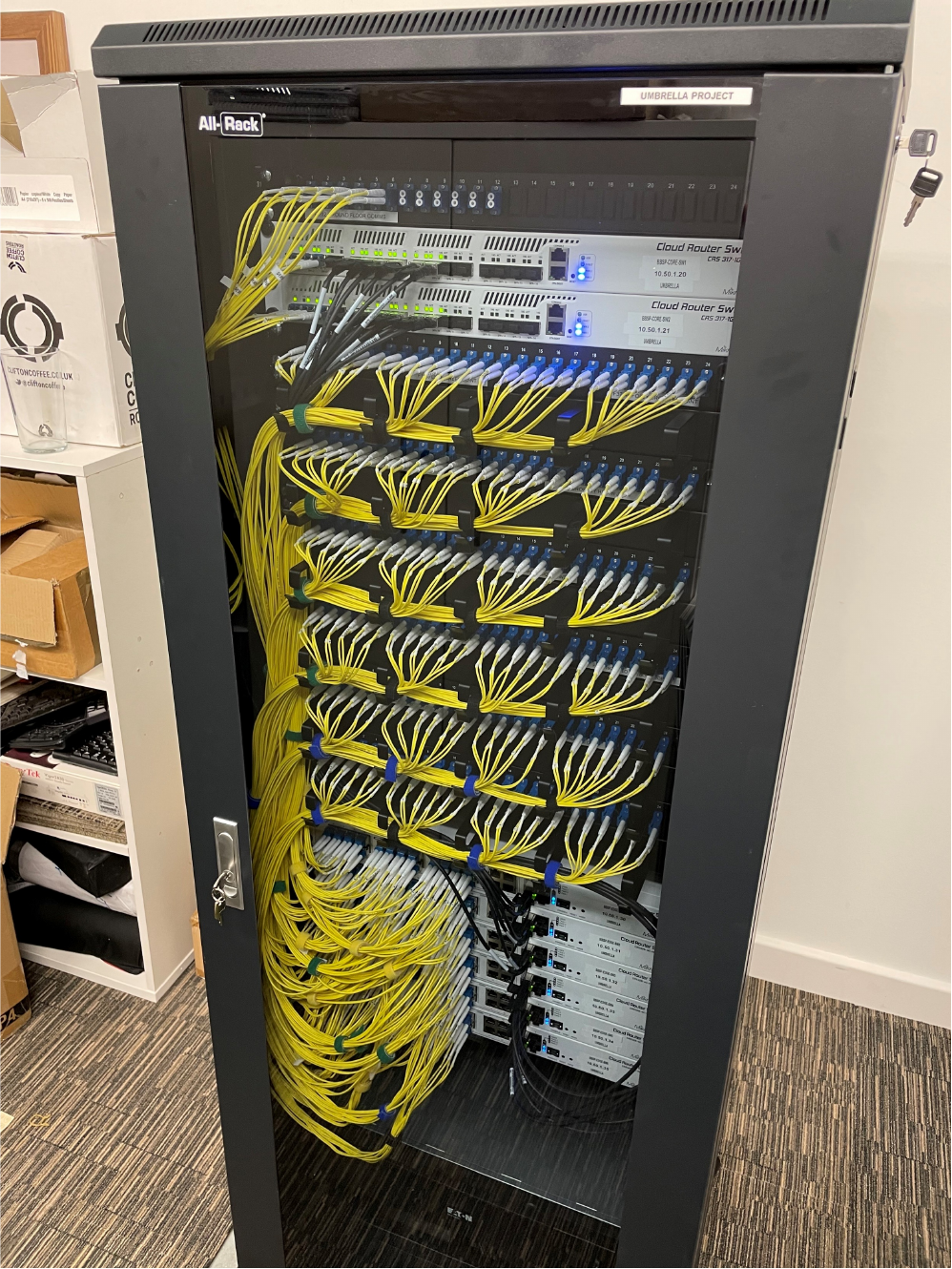}
        \caption{\textbf{Fibre switches.}}
        \label{fig:switch}
    \end{subfigure}
    % \hfill
    \begin{subfigure}{.6\columnwidth}
        \centering
        \includegraphics[width=\linewidth]{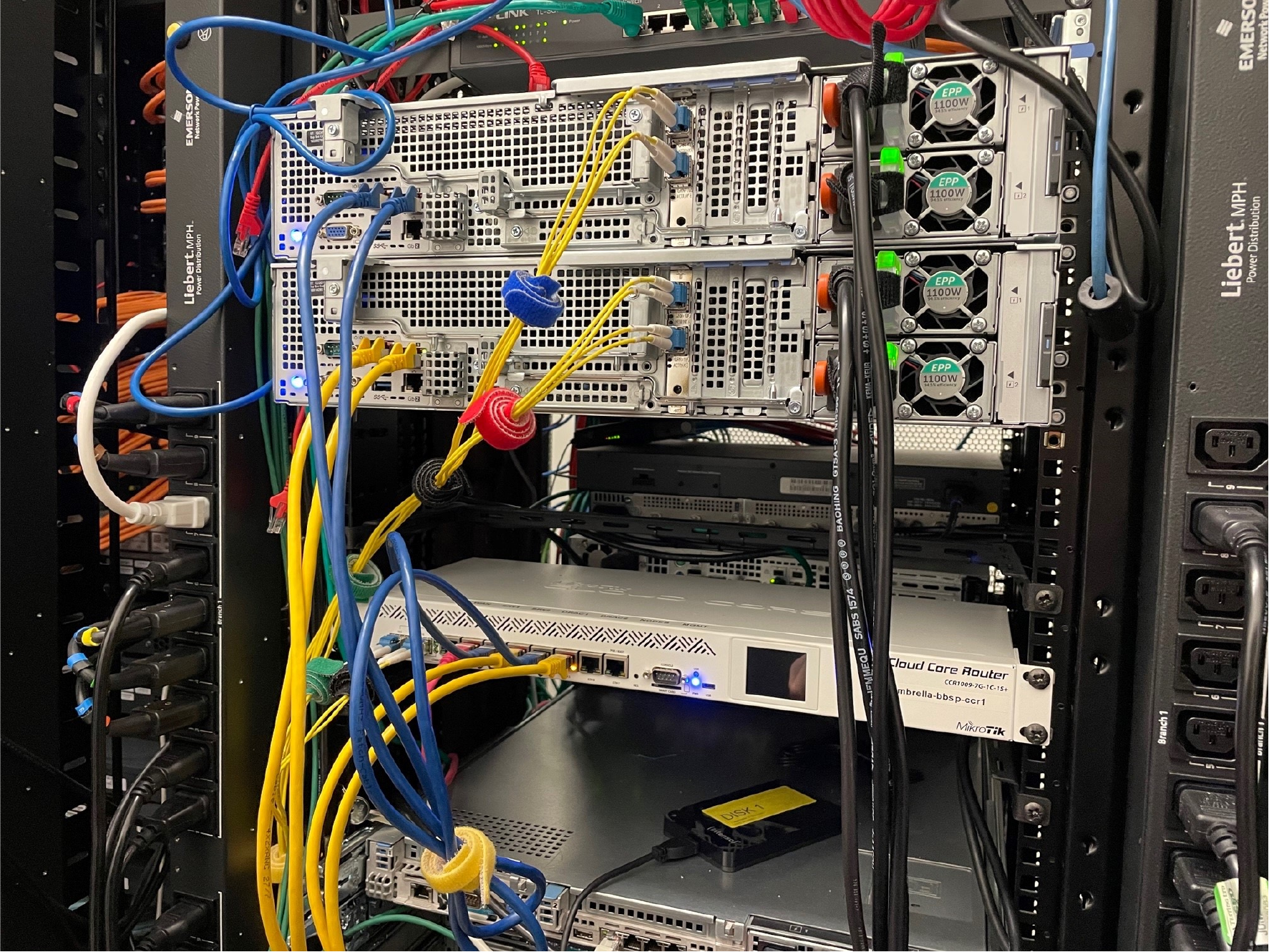}
        \caption{\textbf{Core router and servers.}}
        \label{fig:server}
    \end{subfigure}
    \caption{\textbf{UMBRELLA backend hardware and network infrastructure deployment.}}
    \label{fig:Deployment1}
\end{figure}  

The testbed devices are connected via \textit{two forms of backbone connectivity}. The first one is a fibre connection between every other UMBRELLA node. All fibres are patched at switches located at BBSP (Fig.~\ref{fig:switch}). The second form of backbone connectivity is a Wi-Fi network. All fibre nodes serve as Wi-Fi gateways, forming an enterprise Wi-Fi network. The wireless nodes connect as Wi-Fi clients to the nearest gateway. Such a hybrid approach cuts the fibre implementation cost by half but achieves high speed and high bandwidth across all nodes for efficient testbed operation. A \textit{VPN connection} enables nodes not on the same physical connection to reach UMBRELLA via the Internet. This allows the encryption of the data in transit and enables remote nodes to be used in a LAN-like system architecture, providing a seamless operation to the end user. 

The different testbeds are controlled from a \textit{unified backend} developed in a \textit{hybrid way}. Some components live on our on-prem servers (Fig.~\ref{fig:server}) installed as part of the UMBRELLA infrastructure, while additional components are hosted on Amazon Web Services (AWS). The robotics testbed ingress point is the arena server, which is the main entry point for the testbed services and digital twin environments (connected over a VPN tunnel). More information about the platform developed can be found in Sec.~\ref{subsec:platform}. The hybrid implementation reduces the operational cost but enables future expandability of the platform. Also, by utilising pre-existing services from AWS (e.g., AWS's Simple Email Service (SES)), we can enhance UMBRELLA's operation with minor configurations.

\subsection{Node Hardware Overview}
The UMBRELLA project developed three types of nodes, i.e., the UMBRELLA node, the robot node, and the 5G node. Each node comes with its unique features and capabilities. The different nodes are deployed across the different testbeds. This section will briefly describe the design paradigm behind the UMBRELLA node. The decisions taken also motivated the design of the other two nodes. More details about the robot node can be found in Sec.~\ref{subsec:robotarena} and for the 5G node in Sec.~\ref{subsec:private5G}.

\begin{figure}[t]
    \centering
    \includegraphics[width=0.97\columnwidth]{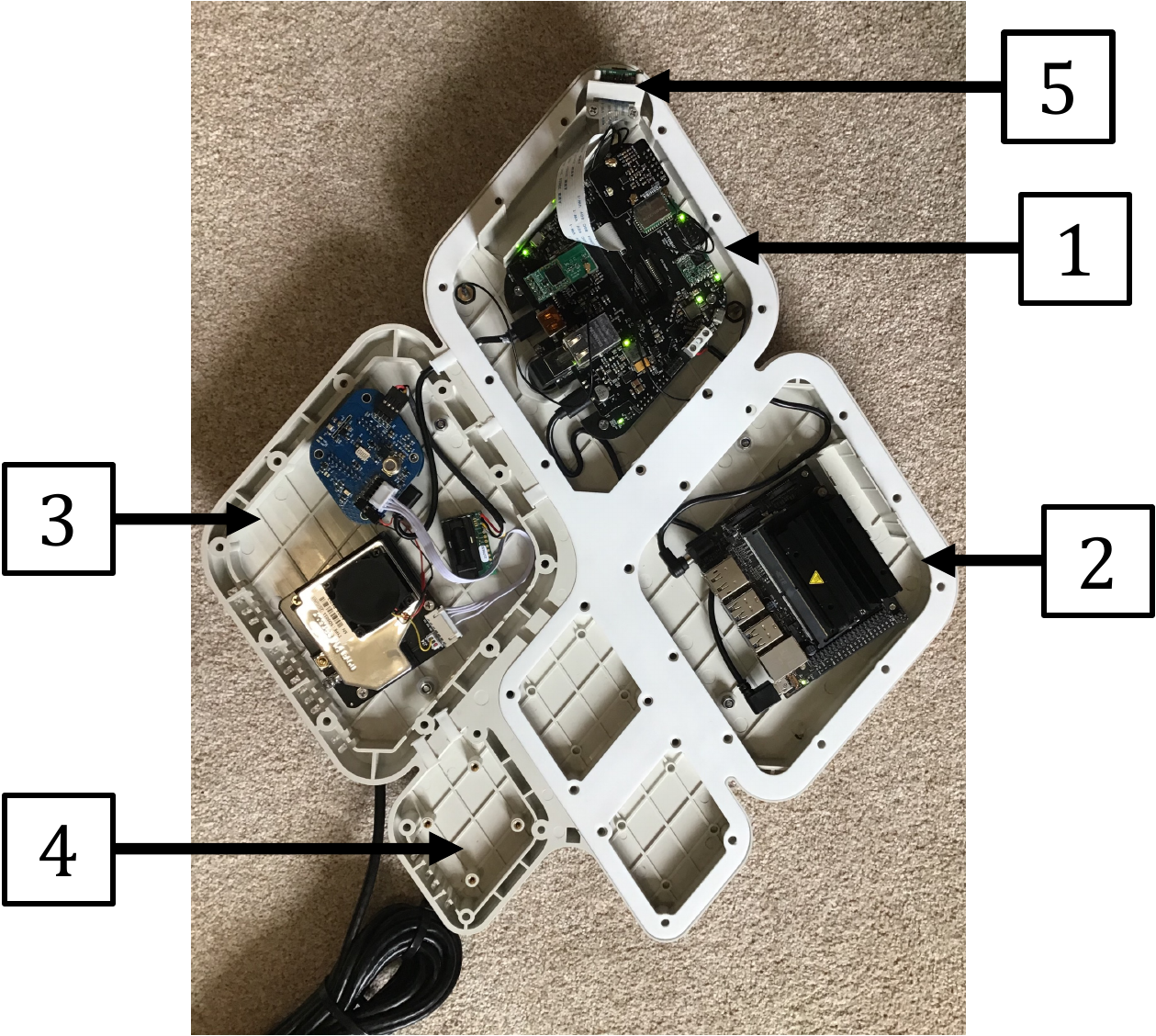}
    \caption{\textbf{An UMBRELLA node with the enclosure open. 1) Shows the Mothership module. 2) Shows the Edge module. 3) Shows the Ambient sensing module. 4) Shows a vacant ambient sensing expansion pod. 5) The RPi Camera Module.}}
    \label{fig:node_open}
\end{figure}

\begin{figure}[t]
    \centering
    \includegraphics[width=\columnwidth]{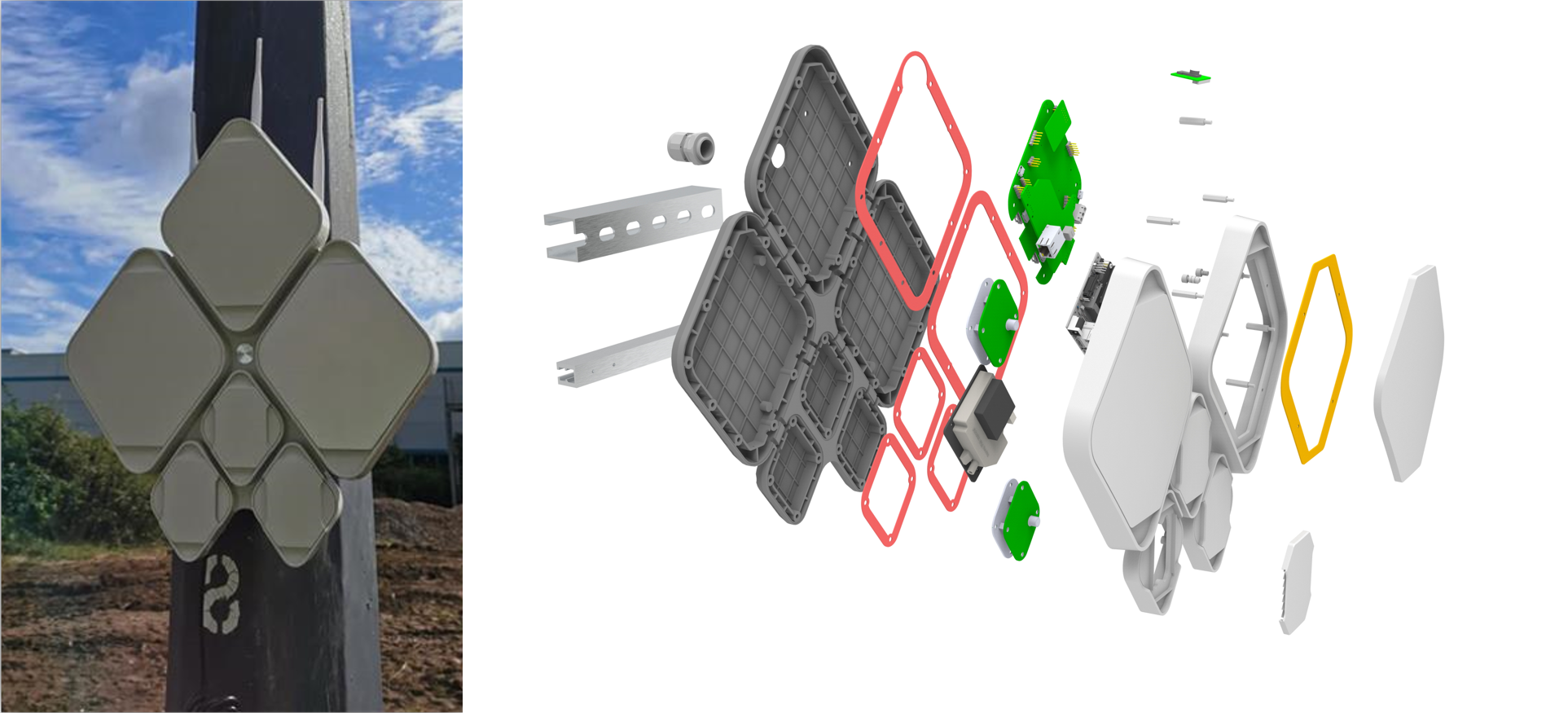}
    \caption{\textbf{UMBRELLA node hardware design with its exploded view.}}
    \label{fig:cad}
\end{figure}

The UMBRELLA node, shown in Fig.~\ref{fig:node_open}, is composed of separate hardware modules, centred around a Raspberry Pi Compute Module 3+ single board computer~\cite{RPI_3_COMPUTE}. The hardware has been designed to be flexible and expandable. Additional modules can be added, providing further functionality or sensing capabilities. The modules themselves are mounted in an injection moulded casing separated into ``pods'' broadly by functionality, as shown in Fig.~\ref{fig:node_open}. The contents of these pods are described in the following sections. Finally, UMBRELLA nodes are installed on public streetlights and are designed to be weatherproof, shielded with a rubber seal (Fig.~\ref{fig:cad}). The sensor pods require external airflow and are designed for enhanced air venting without the ingress of water (Sec.~\ref{subsec:airquality}). 

The testbed deployed consists of different node configurations, with differences in the available wireless interfaces, backbone connectivity, edge processing capabilities, and sensor availability. A user deploying an experiment via the portal (Sec.~\ref{subsec:terminology} and Fig.~\ref{fig:node_selection}) should keep in mind that ``RSE-*'' nodes are fibre-connected and enclose an edge compute module, ``RSE-L-*'' nodes can act as LoRaWAN gateways, and finally, ``RSS-*'' nodes are Wi-Fi connected without an edge processing unit. A 3-digit unique identifier is given to each node, e.g., ``RSE-045'' for identification purposes. More information about the different node configurations can be found on our Wiki page.

\subsubsection{Mothership Pod}
\begin{figure}[t]
    \centering
    \includegraphics[width=0.70\columnwidth]{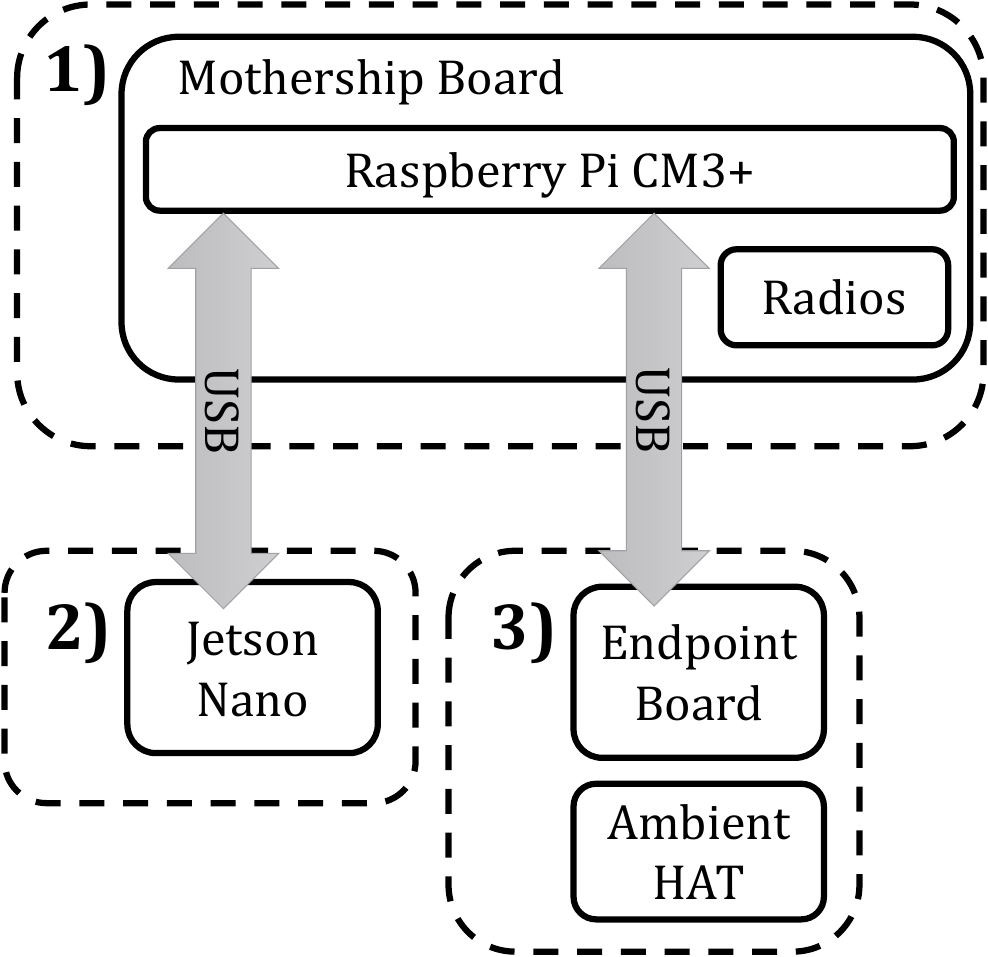}
    \caption{\textbf{Diagram of pods in the UMBRELLA node. The dashed lines indicate the physical boundaries of the pods themselves. 1) Mothership pod: houses a custom carrier board for the Raspberry Pi CM3+ and experimental radios. 2) Edge computing pod: houses the Jetson Nano. 3) Ambient sensing pod: houses an Endpoint board with an Ambient sensing HAT.}}
    \label{fig:node_module_diagram}
\end{figure}

The ``mothership pod'' houses the Raspberry Pi CM3+~\cite{RPI_3_COMPUTE} module, mounted on a custom carrier board (Fig.~\ref{fig:node_module_diagram}-1). This can be considered to be the operational core of the node. Facilitated by Raspbian GNU/Linux 10 (buster) 32-bit and a custom kernel based on ver. 4.19.95-v7+, the mothership delivers network connectivity and inter-module communication. Furthermore, the carrier board includes a range of radio hardware to provide both network connectivity and wireless testbed capabilities. 

\subsubsection{Edge Computing Pod}
\label{sec:edge_module}
The edge pod is optional and houses an NVIDIA Jetson Nano connected via USB (Fig.~\ref{fig:node_module_diagram}-2). The Jetson Nano runs the official Linux4Tegra 64-bit OS provided by NVIDIA (based on Ubuntu 18.04)~\cite{jetson_nano} with minor modifications. This module offers edge computing capabilities and can be used to perform more GPU-intensive activities, such as training local ML/FL models.

\subsubsection{Ambient Sensing Pod}\label{subsubsec:sensingpod}
The ambient sensing module comprises an array of environmental sensors (Fig.~\ref{fig:node_module_diagram}-3). A microcontroller board is responsible for relaying sensor values to the mothership. The module is exposed to the atmosphere through a vented enclosure, allowing the sensors to provide representative readings of the external environment. Sec.~\ref{subsec:airquality} provides more details on how these sensors are used in an air quality sensing use-case operating currently at UMBRELLA. Finally, a summary of the ambient sensors found in UMBRELLA can be seen in Tab.~\ref{table:nodeambientsensors}.

\begin{table}[t]
\renewcommand{\arraystretch}{1.10}
    \caption{\textbf{Node ambient sensors.}} 
    \centering
    \begin{tabular}{r||l}
        \textbf{Property Measured} & \textbf{Sensor}\\
        \hline
        \hline
        PM2.5 & Plantower\\
        PM2.5 & Nova PM\\
        PM10 & Plantower\\
        PM10 & Nova PM\\
        CO2 & Sensiron SCD41\\
        CO2 & Bosch BME680\\
        NO2 & AlphaSense NO2-B43F\\
        OX & AlphaSense OX-B431\\
        Multi-Gas & Amphenol Mics-6814\\
        Atmospheric Pressure & Bosch BME680\\
        Temperature & Bosch BME680\\
        Noise & Adafruit SPH0645LM4H\\
        
        \hline
    \end{tabular}\label{table:nodeambientsensors}
\end{table}

\subsubsection{Endpoint Boards}\label{subsubsection:endpoint_boards}
The endpoint boards are custom Printed Circuit Board Assembly (PCBA)  centred around a microcontroller (Fig.~\ref{fig:node_module_diagram}-3). The design provides a layer of abstraction between the modular sensors or radios and the rest of the node. The endpoint boards perform lower-level communication with the sensors using protocols such as I2C, SPI and I2S. This is also advantageous as it offloads time-critical operations from the main processing unit. To maximise the flexibility, the endpoint-provided interface is uniform across the entire range of sensor suites used by the testbed. The sensors are mounted on an additional HAT PCBA with a pre-programmed EEPROM.

Similarly, the radio interfaces in the UMBRELLA node are built on HATs that mount on top of the mothership carrier board. UMBRELLA nodes provide two Bluetooth and one LoRaWAN interface accessible by the end user, a Wi-Fi interface for backbone connectivity, and a cellular interface for fail-safe management tasks when a node is not connected to the main LAN. Tab.~\ref{table:noderadiospec} summarises the radio interfaces inside an UMBRELLA node. 

\begin{figure}[t]
    \centering
    \includegraphics[width=0.8\columnwidth]{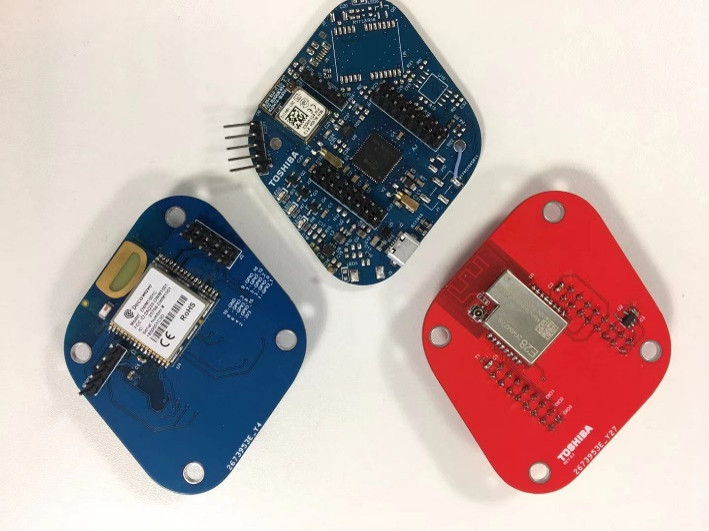}
    \caption{\textbf{Reference designs for plug-and-play UMBRELLA HATs including a Decawave UWB module and an EBYTE LoRa 2.4Ghz module.}}
    \label{fig:hat}
\end{figure}

Such modular design of UMBRELLA HATs enables the plug-and-play functionality. Several reference HAT designs were produced during the development phase, as shown in Fig.~\ref{fig:hat}, but they were not integrated into the final deployment due to space limitations. Users interested in designing new HATs for UMBRELLA can find more details on our Wiki page.

\begin{table}[t]
\renewcommand{\arraystretch}{1.10}
    \caption{\textbf{UMBRELLA node radio specifications. ``Experimental'' indicates the interfaces where users can upload their custom firmware.}} 
    \centering

    \begin{tabular}{r||c c}
        \textbf{Technology} & \textbf{Radio} & \textbf{Experimental}\\
        \hline
        \hline
        Bluetooth & Laird BL654 Module~\cite{bl654pa} & Yes\\
        \hline
        LoRa (EU 868) & Hope RFM95W~\cite{hoperf} & No\\
        \hline
        Sub-1 GHz & Texas Instruments CC1310~\cite{CC1310} & Yes \\
        \hline
        Wi-Fi & Atheros AR9271 [Generic] & No \\
        \hline
        Cellular & 4G LTE Dongle [Generic] & No \\
        \hline
        LoRaWAN & RakWireless RAK2247~\cite{rakwireless}\tablefootnote{Only included in base-station type nodes.} & Yes \\
        \hline
    \end{tabular}\label{table:noderadiospec}
\end{table}

\subsection{Node Software Overview}\label{subsec:node_software}
All nodes across the different testbeds handle the execution of the user experiments. Moreover, all nodes run several applications that enhance user experimentation and facilitate the testbed operations.

\begin{figure*}[t]
    \centering
    \includegraphics[width=0.85\textwidth]{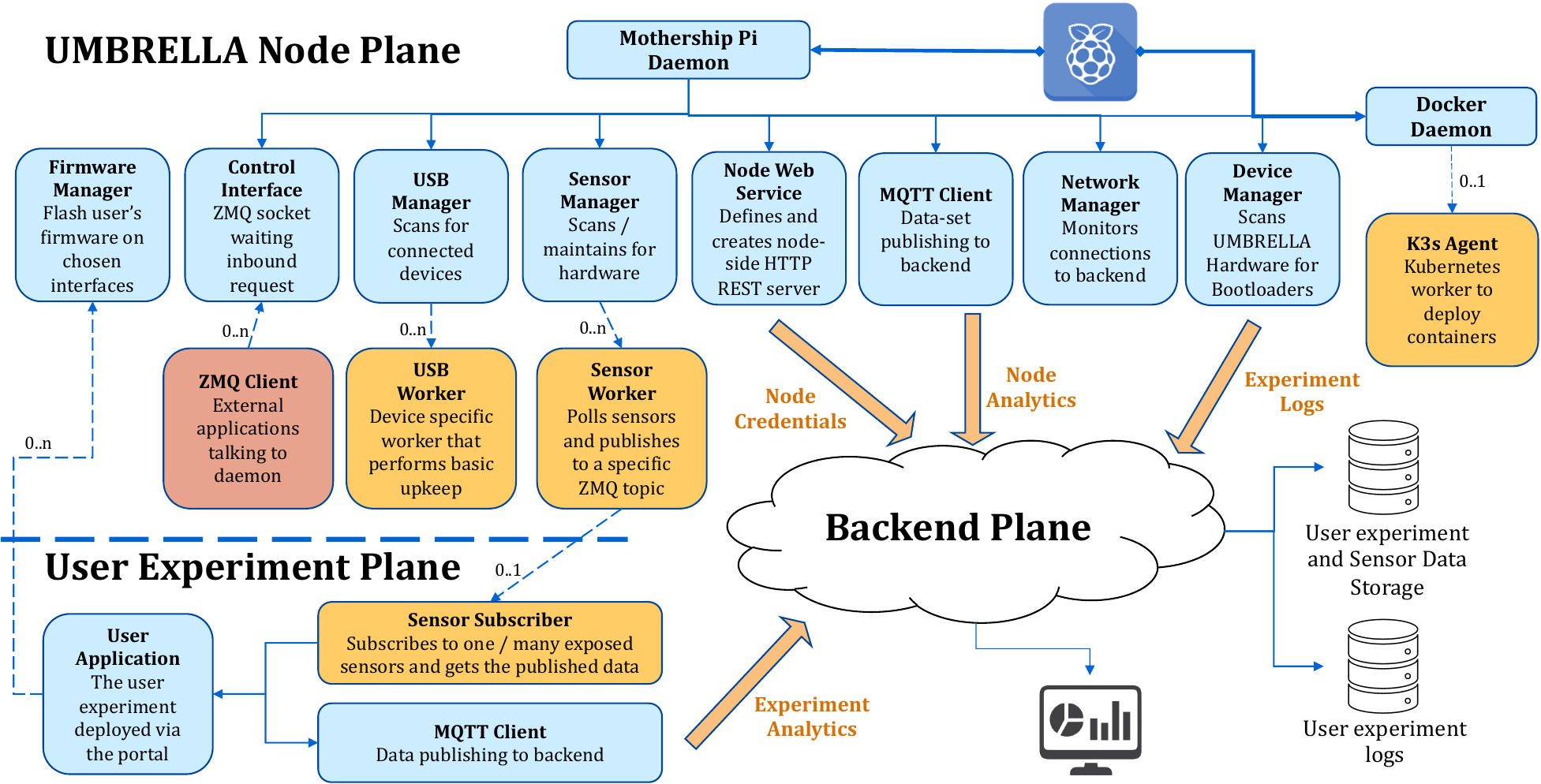}
    \caption{\textbf{UMBRELLA node main software blocks and interaction with backend.}}
    \label{fig:SW_Deployment}
\end{figure*}

\subsubsection{Container Deployment on Nodes}
All experiments and applications are deployed as containers on all available devices. The containers are orchestrated via a Kubernetes cluster\footnote{Kubernetes Container Orchestration: \url{https://kubernetes.io/}}. The chosen Kubernetes distribution is K3s\footnote{K3s - Lightweight Kubernetes: \url{https://k3s.io/}} due to its availability for all hardware architectures and its lightweight operation. Users can package their applications as Docker\footnote{Docker: \url{https://www.docker.com/}} images, upload them on the UMBRELLA portal and deploy them on the available nodes.

When an experiment is scheduled, it runs as a stateless instance on either the Raspberry Pi or the Jetson Nano. Upon completion (based on the duration of the experiment), the container is stopped and the image is removed. The experimental ideas are only limited by the user's imagination and the available resources on the nodes.

In parallel, the nodes run several stateful applications intended to expose interfaces to the end user, provide health checks, collect sensor data, and automate the nodes' functionality. The three applications of particular interest for the end users are the following.

\subsubsection{UMBRELLA Node Daemon}\label{subsubsec:node_daemon}
The ``\textit{UMBRELLA Daemon}'' acts as custom middleware between the UMBRELLA portal and the other pre-built applications on the node. The daemon also provides hardware abstraction from the sensing devices connected to the node, which will be discussed later. It is written in Go (Golang) due to its readily available web services libraries and the memory-safe and multi-threaded operations provided.

The daemon architecture can be seen in Fig.~\ref{fig:SW_Deployment}. The daemon facilitates the following main tasks:
\begin{itemize}
    \item Provides a REST interface such that HTTP requests from the UMBRELLA portal can be served.
    \item Manages the operation of UMBRELLA devices, including resetting and updating firmware.
    \item Polls and parses sensor data from UMBRELLA devices.
    \item Makes sensor data globally available to Zero-MQ subscribers on the node.
    \item Periodically publishes node analytics data to the backend.
    \item Interfaces with experiments on the node.
    \item Buffers and publishes high-speed power measurements from the radios directly to the UMBRELLA backend.  
\end{itemize}

Even though the architecture presented is specific for the UMBRELLA node, similar ``\textit{daemon}'' applications can also be found in other node types and testbeds as well. These daemon applications allow users to build and run applications without having to manage the underlying infrastructure and are integral components of all available testbeds.

\subsubsection{UMBRELLA Sensor Collector}
\begin{figure}[t]
    \centering
    \includegraphics[width=0.97\columnwidth]{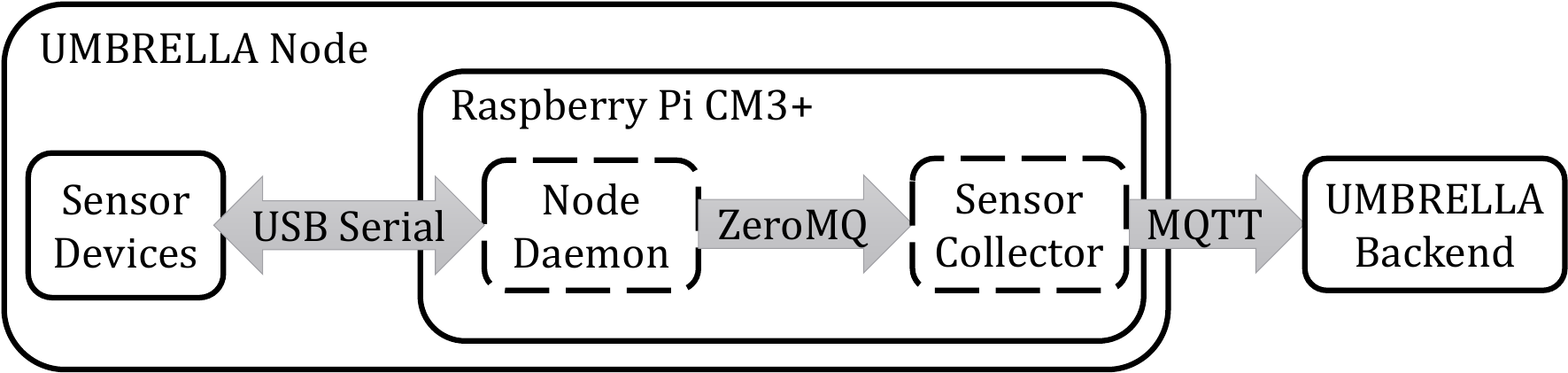}
    \caption{\textbf{A diagram showing the flow of sensor data from the UMBRELLA node to the backend services. Arrows indicate an interface through which data flows.}}
    \label{fig:environmental_data}
\end{figure}

The role of the sensor collector is to forward all UMBRELLA sensor data published by the UMBRELLA node. This application is developed as part of the air quality use case presented in Sec.~\ref{subsec:airquality}. Our application maintains an MQTT client connection with the backend, subscribes to all sensor data published by the UMBRELLA Daemon and then publishes it as it arrives. The flow of sensor data through the node is shown in Fig.~\ref{fig:environmental_data}. As neither the UMBRELLA daemon nor the sensor collector buffer environmental data, the rate at which sensor data is published is governed by the sensor polling rate.

Similar ``collector'' applications can be developed by the end users and deployed as ``experiments''. These applications could, for example, provide a processing step on the raw data (e.g., a drift detection analysis~\cite{le3d}) before being sent to the backend for visualisation and storage.

\begin{figure}[t]
    \centering
    \includegraphics[width=0.97\columnwidth]{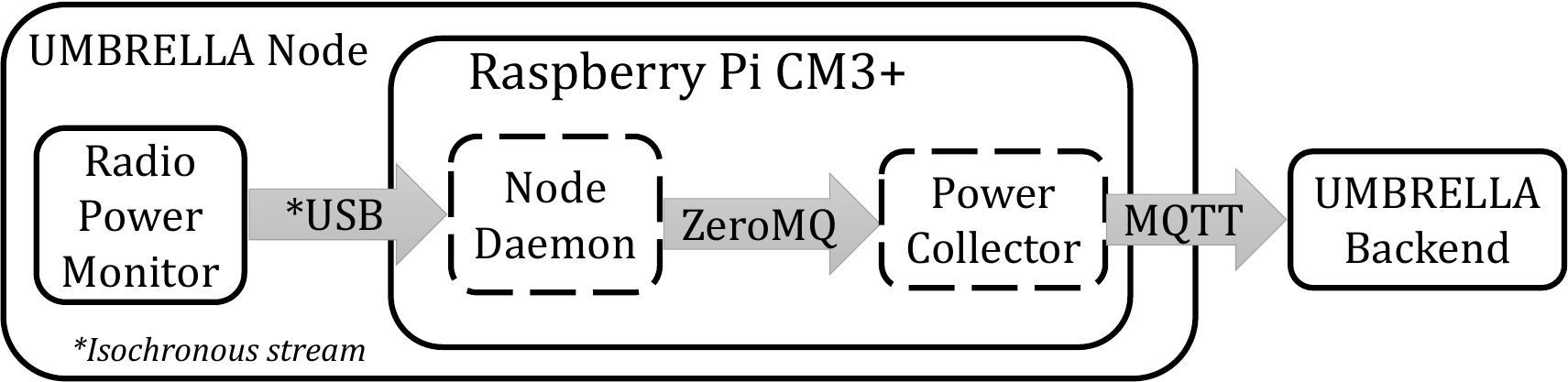}
    \caption{\textbf{A diagram showing the flow of power measurement data from the UMBRELLA node through to the backend services. Arrows indicate an interface through which data flows.}}
    \label{fig:power_measurement}
\end{figure}

\subsubsection{UMBRELLA Radio Experiment}
Radio experiment containers are provided by default by the UMBRELLA platform. These container images are paired with the custom radio firmware, uploaded on the nRF52840 or the CC1310 radios, and allow end users to access serial logs and the power profile of the devices. The data flow and system are presented in Fig.~\ref{fig:power_measurement}. More information on this functionality can be found in Sec.~\ref{subsec:wireless_testbed}.

\begin{figure*}[t]
    \centering
    \includegraphics[width=\textwidth]{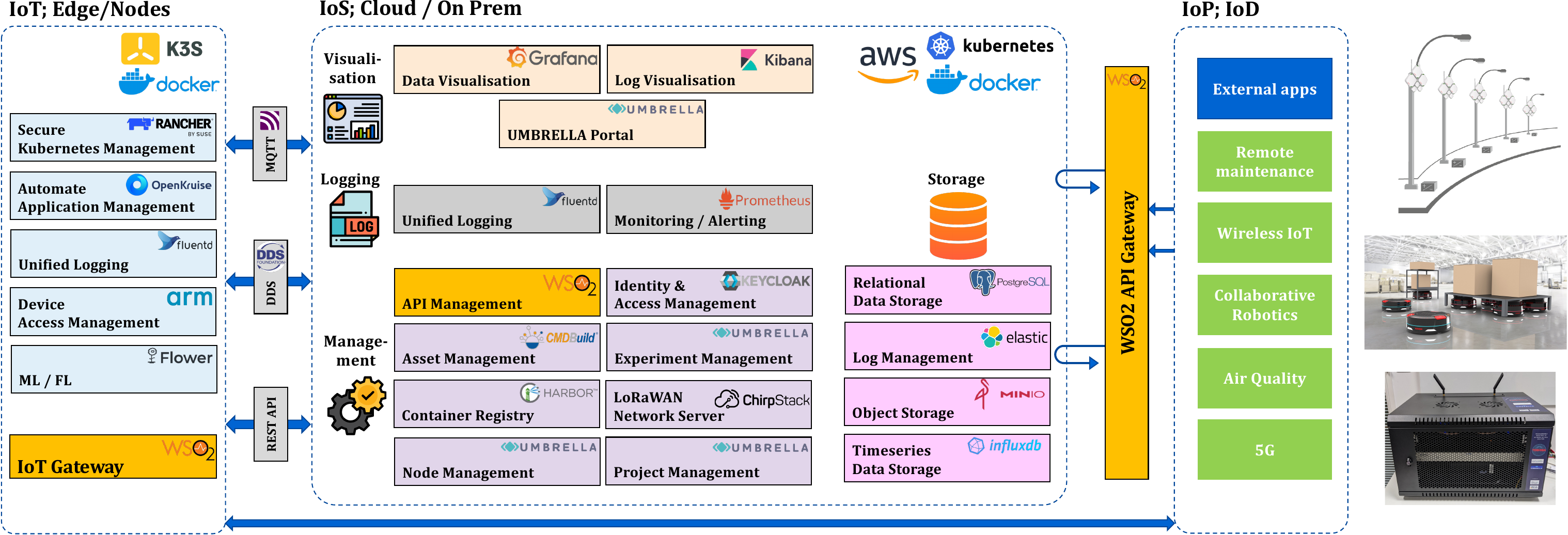}
    \caption{\textbf{UMBRELLA platform overview. Different services, either open-source or custom-built, ensure the seamless functionality of the system and that the key requirements are met.}}
    \label{fig:platform}
\end{figure*}

\subsection {Platform Overview}\label{subsec:platform}
The UMBRELLA platform employs a cloud-native, containerised architecture, facilitating unified management and secure operation across various testbeds and use cases. This architecture integrates edge and cloud computing and storage resources efficiently. A high-level overview of the platform design can be seen in Fig.~\ref{fig:platform}.

At the heart of UMBRELLA's architecture are three fundamental pillars:
\begin{itemize}
    \item \textbf{The IoT or Edge/Node Pillar}: Focused on managing the edge computing aspects.
    \item \textbf{The Internet of Services (IoS) Pillar}: Central to the platform's service-oriented operations.
    \item \textbf{The Internet of People (IOP) / Internet of Data (IoD) Pillar}: Integrating user-centric and data-centric functionalities.
\end{itemize}

The platform combines \textit{open-source tools} with \textit{custom-developed applications} to form a comprehensive ecosystem capable of \textit{managing users, orchestrating experiments, and handling the vast amounts of data generated}.

Applications across various domains are containerised and deployed within numerous Kubernetes clusters. The \textit{network fabric} for the K3s node clusters is based on Flannel\footnote{Flannel Network Fabric: \url{https://github.com/flannel-io/flannel}} due to its lightweight operation, while the backend K8s clusters use Calico\footnote{Calico: \url{https://www.tigera.io/project-calico/}} for the advanced features supported. This setup allows for dynamic scaling based on demand and available resources. \textit{Communication between applications} is facilitated through MQTT and Data Distribution Service (DDS) \textit{messaging buses}, complemented by well-defined REST APIs. An MQTT broker runs centrally in the IoS domain, whereas DDS (primarily used for the robotic testbed) operates distributedly. Moreover, MQTT's latency has a higher standard deviation but is more scalable than DDS, designed to be robust and used for real-time applications.

In the IoT domain, \textit{access to node resources} is controlled via Kubernetes Network and Security policies and is governed by Rancher's management platform\footnote{Rancher Enterprise Kubernetes Management: \url{https://www.rancher.com/}}. The deployed containers run in separate namespaces and are isolated in an application-centric way, blocking access and communication to resources outside the provided namespace. \textit{Routine tasks on the nodes} (e.g., cleanup of old images, backups, report generation, etc.) are automated using the OpenKruise suite\footnote{OpenKruise: \url{https://openkruise.io/}}. These routine tasks run as custom resources, enhancing operational efficiency and enabling their periodic execution. \textit{Interacting with the provided radio interfaces} requires, by default, elevated user rights for the end user. This can lead to significant security breaches. Therefore, ARM's Smarter Device Manager\footnote{Smarter Device Manager: \url{https://gitlab.com/arm-research/smarter/smarter-device-manager}} is used to handle access to all devices. This allows access to the serial port talking to the different interfaces (e.g., nRF52840~\cite{nRF52840}, CC1310, etc.) without the need for privileged containers. 

Throughout experiment execution, pod and container logs are collected via Fluentd\footnote{Fluentd: \url{https://www.fluentd.org/}}. Similarly, other pod and container logs running in the IoS domain are stored for debugging and management purposes. All \textit{experiment and system logs} are saved in Elasticsearch\footnote{Elasticsearch: \url{https://www.elastic.co/}}. This approach handles the challenging task of aggregating and analysing all the data coming from multiple nodes and applications. Collected logs can be visualised via Kibana\footnote{Kibana: \url{https://www.elastic.co/kibana}} (only accessible by the UMBRELLA admin team) and end users can also download them as compressed zip files. Prometheus\footnote{Prometheus: \url{https://prometheus.io/}} monitoring tool is used for real-time monitoring and issue alerting. The \textit{healthiness} of all pods, as well as various custom metrics published from the executed applications, are stored in Prometheus and surveyed in real-time. Alerts are generated based on predefined rules and when certain conditions are met. This functionality ensures that issues can be proactively addressed before propagating or before the system faces resource pressure.

Keycloak\footnote{Keycloak: \url{https://www.keycloak.org/}} and LDAP are used to manage UMBRELLA's \textit{identity and access management}. This system ensures secure user verification and Role-Based Access Control (RBAC), crucial for resource and testbed access. Users, after registration, need to verify their email address and are assigned a number of default roles (as discussed in Sec.~\ref{subsec:terminology}). All developed (internal and external APIs) APIs are handled by WSO2 API Manager\footnote{WSO2 API Manager: \url{https://wso2.com/api-manager/}}. That enables the management of the entire API's lifecycle and, most importantly, their governance (e.g., traffic limiting, monitoring, etc.) and their security (e.g., authentication, injection of malicious payloads, etc.). IoT Gateways\footnote{WSO2 IoT: \url{https://wso2.com/iot/}} operating along with the WSO2 API Manager enable a decentralised API architecture, provide easier control and configuration and ensure the lightweight and prompt operation of the system. 

The users interact with UMBRELLA via a \textit{custom-built portal} that provides role-based access to the different resources and testbeds. Visualisation of the collected data is built upon \textit{dashboards} served with Grafana\footnote{Grafana: \url{https://grafana.com/}}. Several services are provided to ensure that the use case requirements are met. A project management framework allows users to collaborate on different projects and intuitively structure their experiments. An experiment management framework handles all user experiments' scheduling, deployment and orchestration, considering the available resources on all nodes. Finally, a Node management framework manages all devices connected to UMBRELLA. This service is responsible for the IoT device identity lifecycle and handles their authentication, authorisation, and PKI infrastructure associated with that.

As described, all user applications are containerised. The \textit{container images} used for experimentation are organised and stored in the Harbor Container Registry. Moreover, Harbor\footnote{Harbor Container Registry: \url{https://goharbor.io/}} integrated with \textit{vulnerability scanners} allows scanning all images for vulnerabilities before being deployed as experiments. Hosting a \textit{private container registry} enables UMBRELLA to overcome rate-limiting controls from other Cloud-based registries (e.g., DockerHub) or the associated cost. 

Finally, all the above applications store the \textit{collected user data across multiple databases}. InfluxDB\footnote{InfluxDB: \url{https://www.influxdata.com/}} is used for all the time-series data from the experiments (e.g., air quality data), PostgreSQL\footnote{PostgreSQL: \url{https://www.postgresql.org/}} is used for all user data and their relations (e.g., experiments conducted, project details, etc.) and Minio\footnote{Min.io: \url{https://min.io/}} stores all firmware uploaded by the users and all compressed logs.

\begin{figure}[t]
    \centering
    \includegraphics[width=\columnwidth]{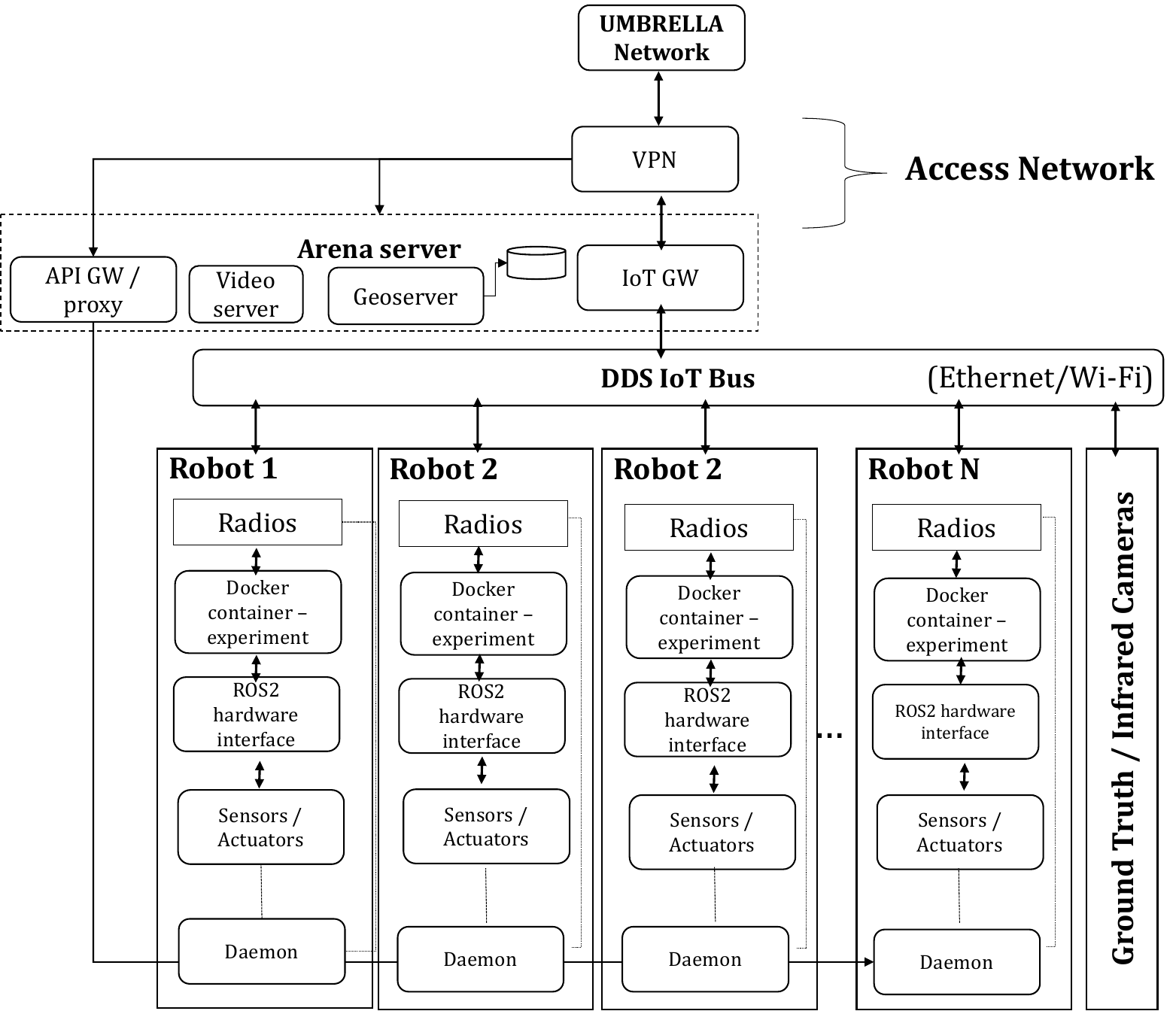}

    \caption{\textbf{Robotic platform architecture.}}
    \label{fig:robotarena}
\end{figure}

\begin{figure}[t]
    \centering
    \includegraphics[width=\columnwidth]{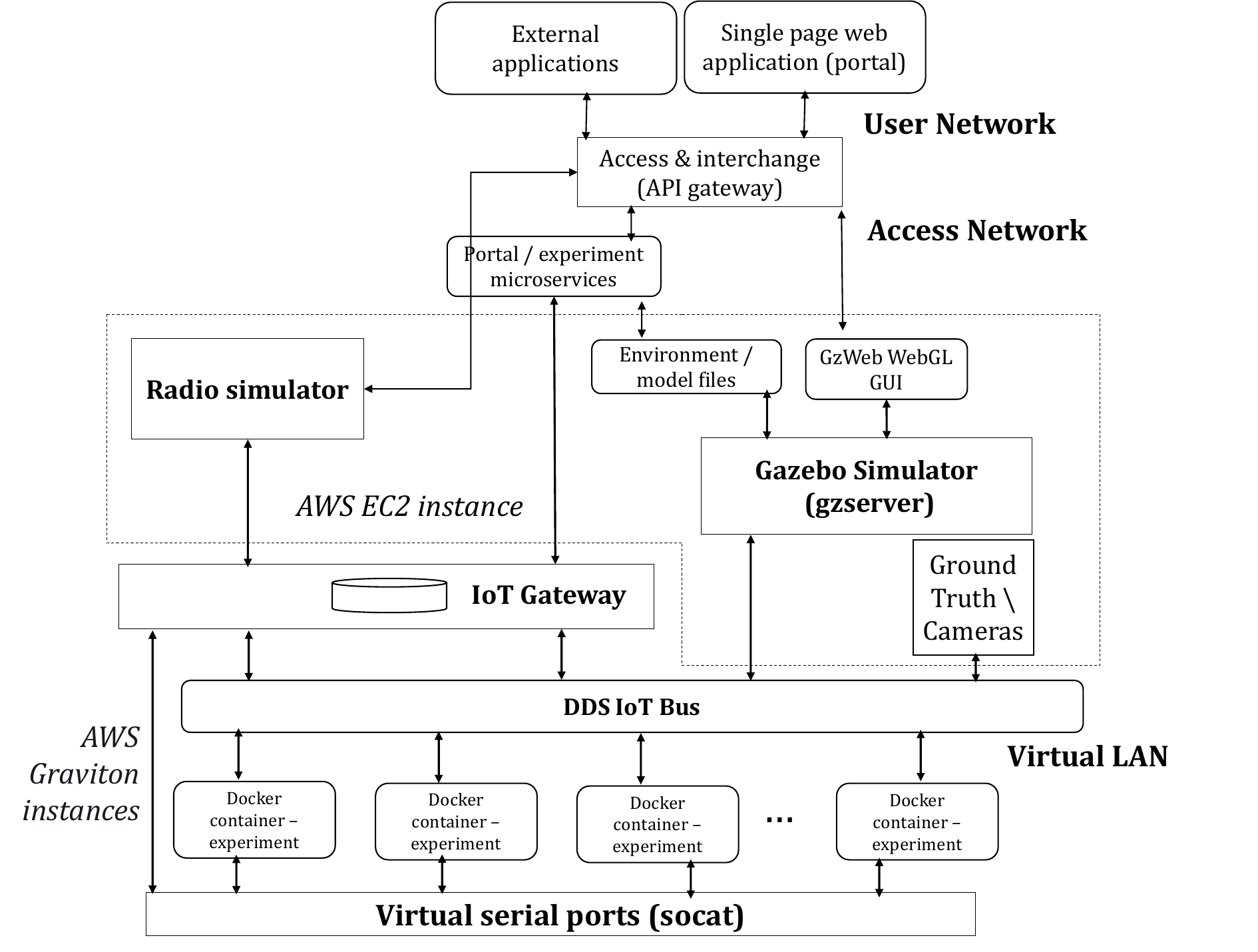}

    \caption{\textbf{Robot digital twin platform architecture.}}
    \label{fig:digitaltwin}
\end{figure}

The Robotic platform is an extension of the UMBRELLA platform and provides the functionality required for the robotic testbed. Its architecture is shown in Fig.~\ref{fig:robotarena}. It supports visualisation via spatial map overlays (using Geoserver) or 3D digital twin models using Gazebo\footnote{Gazebo simulator: \url{https://gazebosim.org/home}} in the single-page web application. The robotics platform and the corresponding digital twin platform (Fig.~\ref{fig:digitaltwin}) permit experimentation with real robots in the arena after prior validation using the digital twin environment. 

The robots, designed explicitly for evaluating collaborative intralogistics use cases, are called \textit{Distributed Organisation and Transport Systems (DOTS)} and are further detailed in~\cite{Jones2022DOTSAO}. The \textit{digital twin} environment is binary compatible with the DOTS as it utilises AWS Graviton2 ARM64 VM instances within the same Kubernetes cluster. It also supports radio simulation models for evaluating robot communication performance using onboard Bluetooth radios. In this manner, the GPU capabilities of the AWS node can be used to provide high-performance simulations without requiring additional external or offline services. 

The DDS IoT Bus supports all the interactions between sensors and actuators and between robots, as well as the ground truth and the IoT Gateway, which provides access to the data from the portal and other services. A robot daemon (similar to the one described in Sec.
~\ref{subsubsec:node_daemon}) also supports monitoring of the robots and deployment of binary firmware to the radios. The robot Kubernetes clusters (based on K3s) utilise Weave net\footnote{Weave Net: \url{https://www.weave.works/oss/net/}} Container Network Interface (CNI) as it supports multicast DDS operation, necessary for the DDS topic exchange and node discovery. In this manner, the physical arena and the digital twin environment are mimicked so that the controllers do not need modification.

Finally, a video server is used to \textit{access the arena's camera feeds}, and the Geoserver supports the heatmap overlays of the ground truth data. The arena camera feeds and the heatmap ground truth are available within the portal to permit remote observation of experiments.

\subsection{Potential Users for UMBRELLA}
UMBRELLA targets potential \textit{applications and users from different verticals} (or industrial domains). This focuses on (but is not limited to) Smart Cities (local authority services -- e.g., environment, street lights, waste), energy, logistics, transport, retail, robotics \& automation, manufacturing, healthcare, etc. Horizontally, it enables technology testing, including sensing, communications, edge computing, data analytics, and AI. Six different user types can use UMBRELLA:

\noindent \textbf{Hardware Designer}: Professionals engaged in creating new sensor or radio hardware components. A designer can easily integrate a new hardware component designed with the UMBRELLA nodes using common interfaces such as USB, I2C, SPI, etc. This allows for real-world testing and refinement, ensuring the hardware's effectiveness in diverse settings.

\noindent \textbf{Software Developer}: Coders and programmers focused on creating software algorithms or wireless protocols. Software engineers can remotely upload software algorithms and wireless protocol to all or a subset of UMBRELLA nodes and evaluate the efficiency and latency of their protocols in real time, leveraging the extensive sensing and radio technologies of UMBRELLA.

\noindent \textbf{Platform User}: Researchers and innovators who wish to test or refine their prototypes in real-world scenarios. Users can develop entire satellite nodes and connect them to UMBRELLA using one of the selected radio technologies. This allows them to gather crucial data remotely, using the UMBRELLA platform's web portal, leading to iterative improvements.

\noindent \textbf{Data User}: Professionals engaged in data-intensive fields, particularly those emphasising AI and ML. Data scientists can deploy ML models and applications to the edge infrastructure of UMBRELLA and evaluate distributed, federated, or personalised ML algorithms using locally collected data or datasets from the Internet.

\noindent \textbf{Application/Solution Developer}: Developers focusing on creating full-fledged applications or solutions, often incorporating custom hardware components. The users can utilise the UMBRELLA infrastructure to prototype applications at scale rapidly, evaluate their operation across different hardware architectures, and integrate custom hardware. KPIs such as seamless integration and performance can be evaluated on the fly and iteratively improved.

\noindent \textbf{Security Experts}: Professionals specialising in cybersecurity, assessing and ensuring the safety of digital systems, networks, and software against potential threats. Cybersecurity analysts can utilise UMBRELLA to evaluate the robustness of their security protocols, test security protocols over several wireless interfaces, demonstrate distributed virtual private network implementations or simulate cyber-attacks, e.g., container escape approaches.

\subsection{Implementation and Deployment} 
The UMBRELLA ecosystem was developed across a period of $\sim$2.5 years. The hardware and network deployment were split into three phases, these being: \textit{Phase 1}: design, development and lab testing; \textit{Phase 2}: six months field trial and server and network installation; \textit{Phase 3}: full batch manufacturing and roll outs.  

The backend and platform were developed over two phases: \textit{Phase 1}: Initial experimental functionality provided and initial portal design; \textit{Phase 2}: Integration with API management, enhanced security features, role-based access control and final portal design.  In this section, we briefly describe some of the key milestones during these phases and the necessary amendments in the initial requirements specifications.

\begin{figure}[t]
    \centering
    \includegraphics[width=\columnwidth]{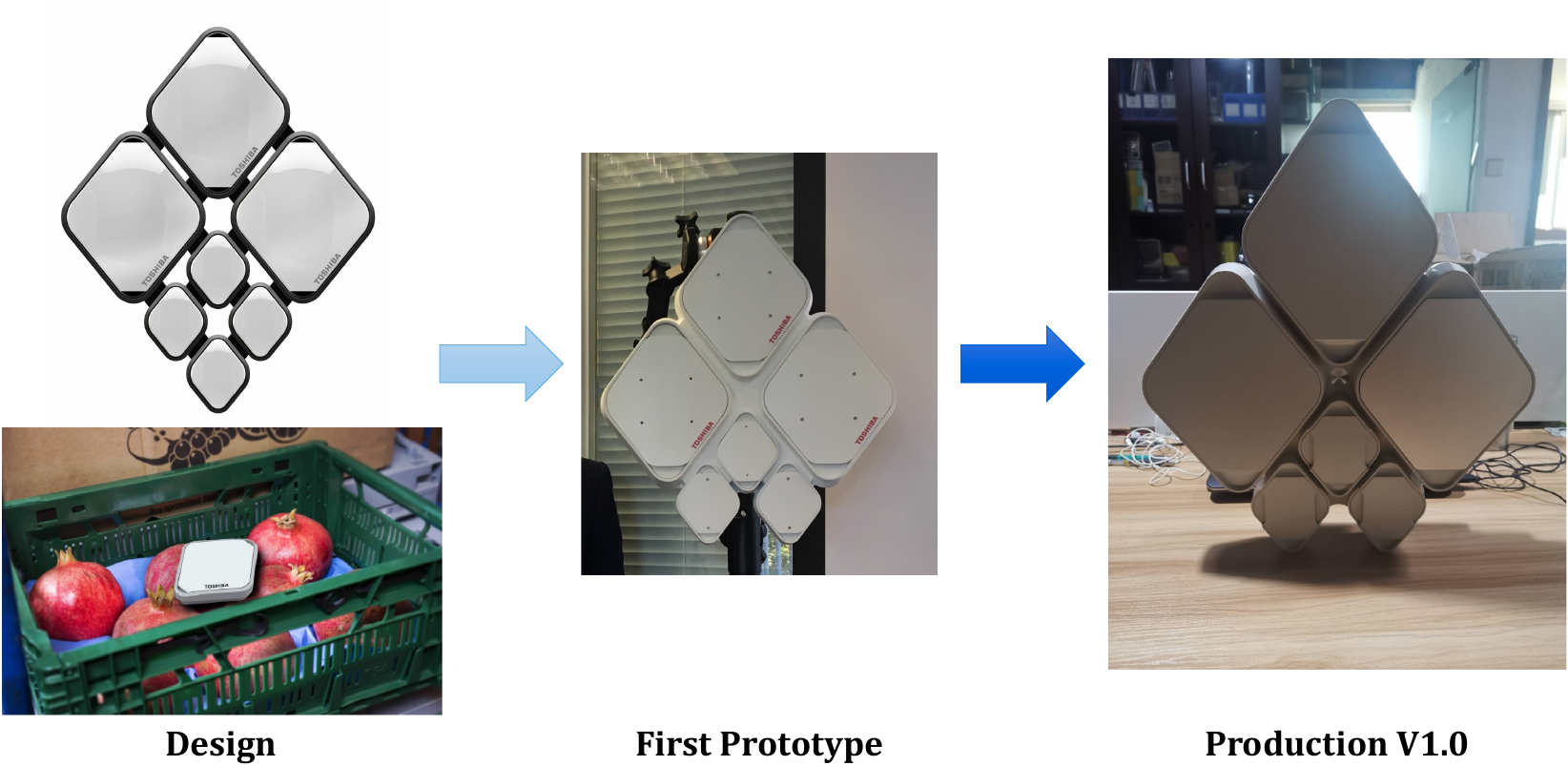}
    \caption{\textbf{The look of the UMBRELLA node changing through the different phases.}}
    \label{fig:umbrellanodelook}
\end{figure}

\begin{figure}[t]
    \centering
    \begin{subfigure}[b]{\columnwidth}
        \centering
        \includegraphics[width=\textwidth]{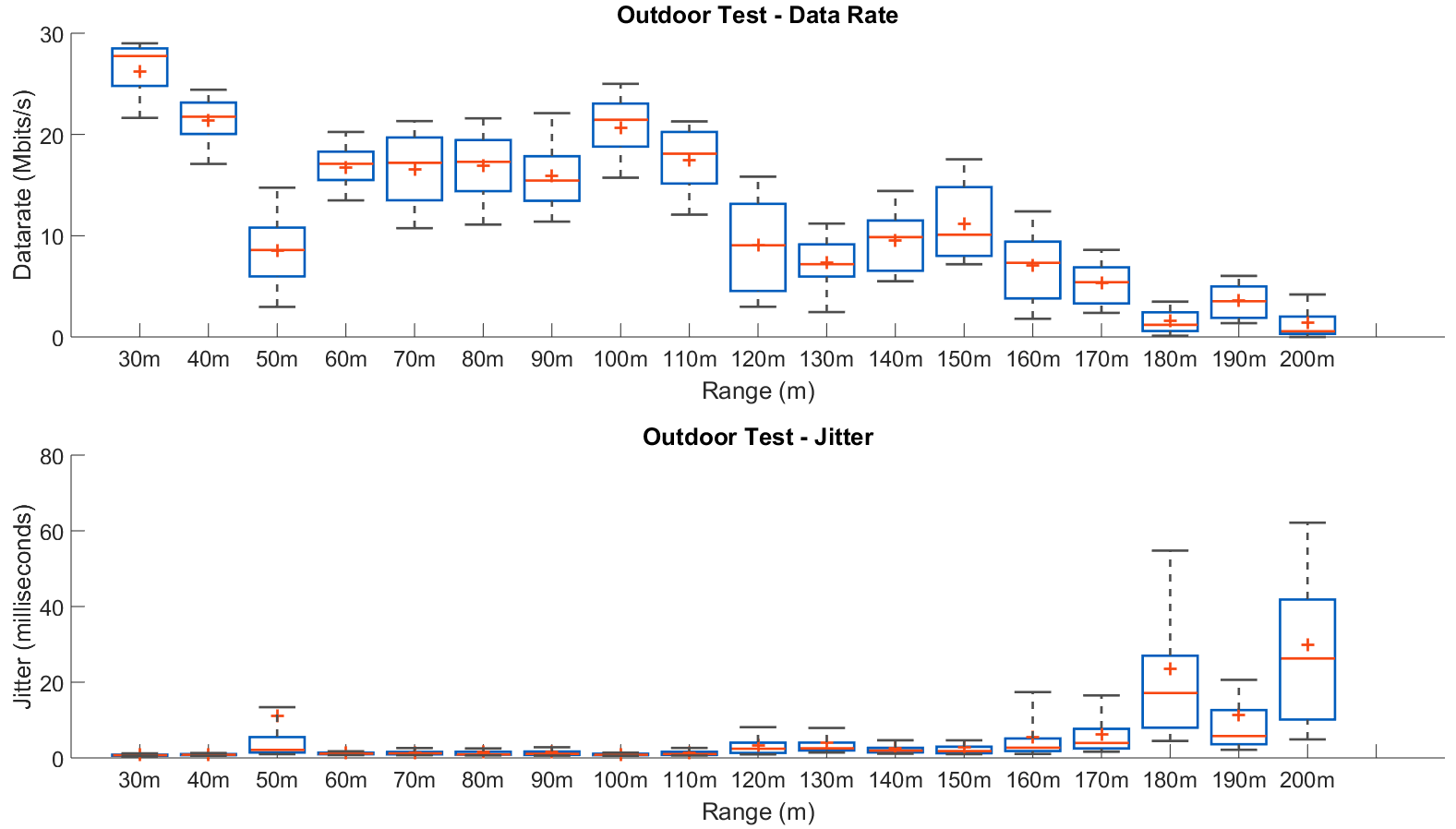}
        \caption{\textbf{UMBRELLA Wi-Fi connectivity test results.}}
        \label{fig:rangetest1}
    \end{subfigure}
    \vspace{3mm}
    \begin{subfigure}[b]{\columnwidth}
        \centering
        \includegraphics[width=\textwidth]{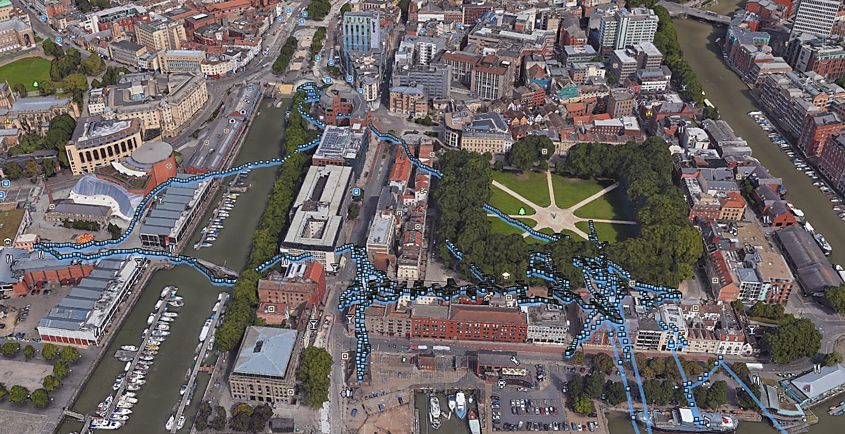}
        \caption{\textbf{Locations tracked while conducting experiments in the city of Bristol, United Kingdom.}}
        \label{fig:rangetest2}
    \end{subfigure}
    \caption{\textbf{UMBRELLA node Wi-Fi range testing in the city of Bristol, UK.}}
    \label{fig:Deployment2}
\end{figure}  

\subsubsection {Hardware and Network deployment}
Starting with the hardware and network design and deployment, we have:

\noindent \textbf{Phase 1}: The \textit{design of the UMBRELLA node} was driven by the modularity requirement and inspired by the aesthetic bee hive shape. The original concept considered each pod and the enclosed hardware as a standalone unit to accommodate new applications and plug-and-play functionality. For example, a sensor pod with a radio HAT can be used for logistic tracking applications, as shown in Fig.~\ref{fig:umbrellanodelook}. During the design phase, the enclosure design shifted substantially due to practical considerations such as weatherproofing for outdoor deployment and structural stability and safety as the unit would be deployed on public infrastructure. The final version is shown in Fig.~\ref{fig:umbrellanodelook}, keeping the same shape but having all pods injection moulded together. The enclosed hardware can still operate as a standalone unit mounted inside a new casing. 

The \textit{backbone wireless connectivity} was also tested as part of this phase. The results dictated the distance separation between the nodes (when installed on the lighting columns) and the required fibre connections for the entire infrastructure. A summary of our results is shown in Fig.~\ref{fig:rangetest1}, where adequate data rate and jitter were observed for distances up to \SI{100}{\meter}. 

\begin{figure}[t]
    \centering
    \includegraphics[width=0.5\columnwidth]{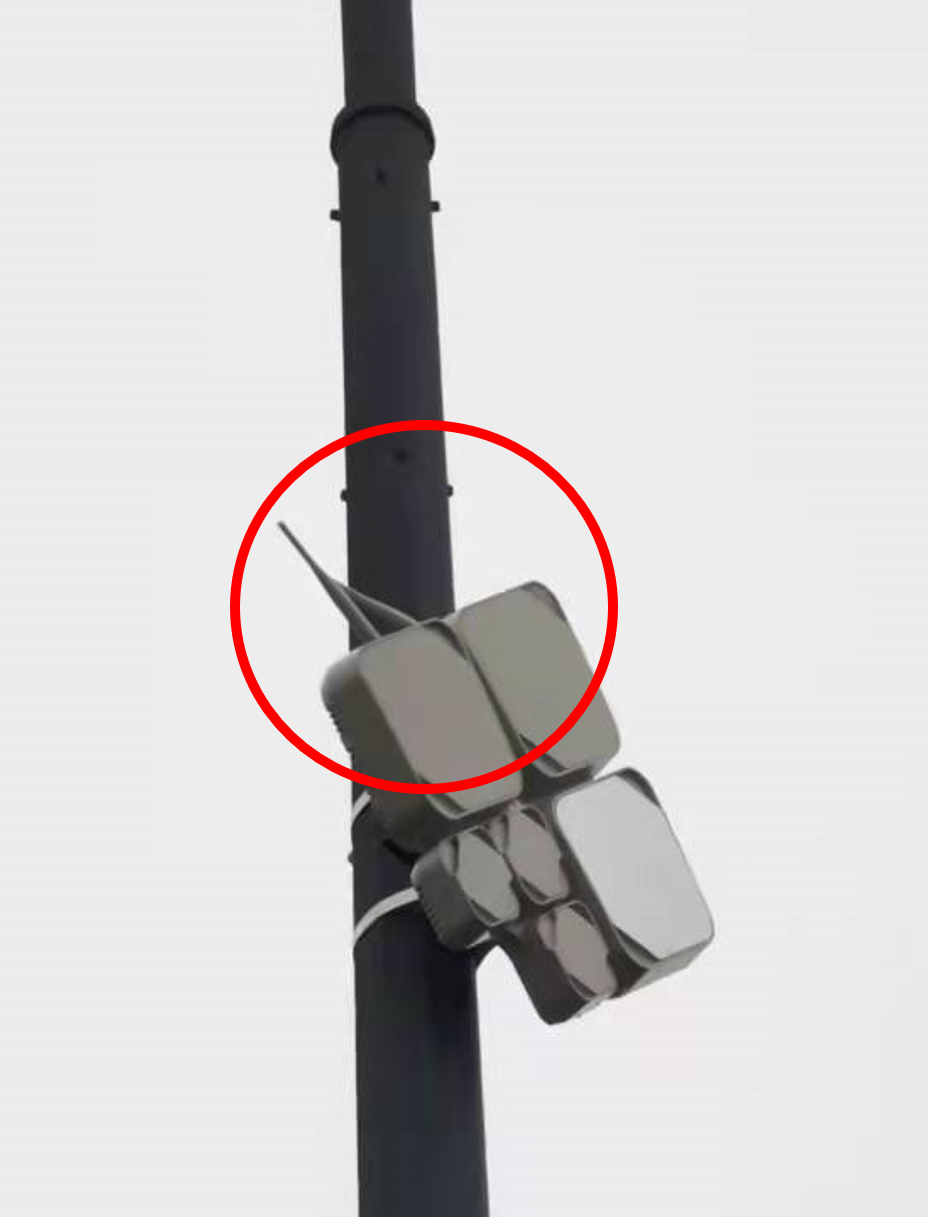}
    \caption{\textbf{Loose external antenna and misalignment after a few months of deployment, causing connectivity problems.}}
    \label{fig:loosen_antenna}
\end{figure}

\begin{figure}[t]
    \centering
    \includegraphics[width=\columnwidth]{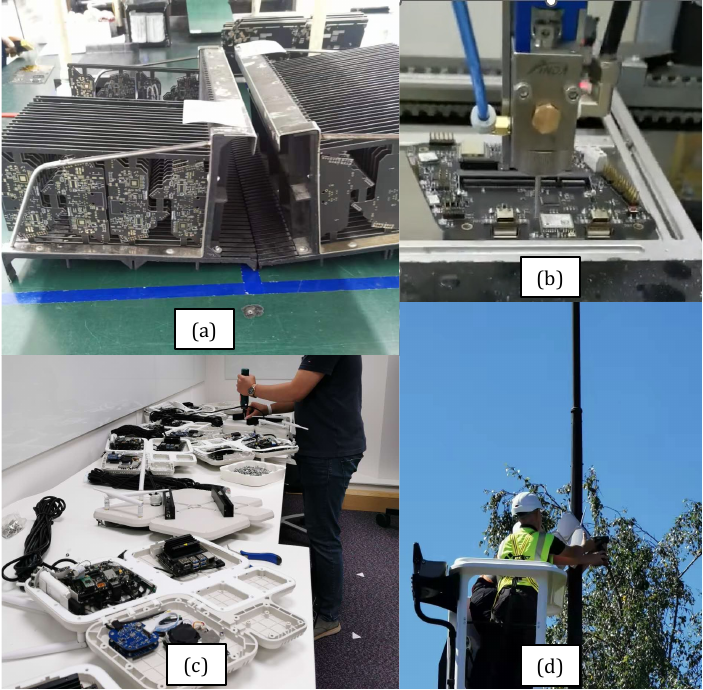}
    \caption{\textbf{a) UMBRELLA PCB manufacturing. b) Automated process to spray weatherproof coating. c) Node assembly. d) Node installation on lighting column.}}
    \label{fig:manufacturing}
\end{figure}

\begin{figure*}[t]
    \centering
    \includegraphics[width=0.8\textwidth]{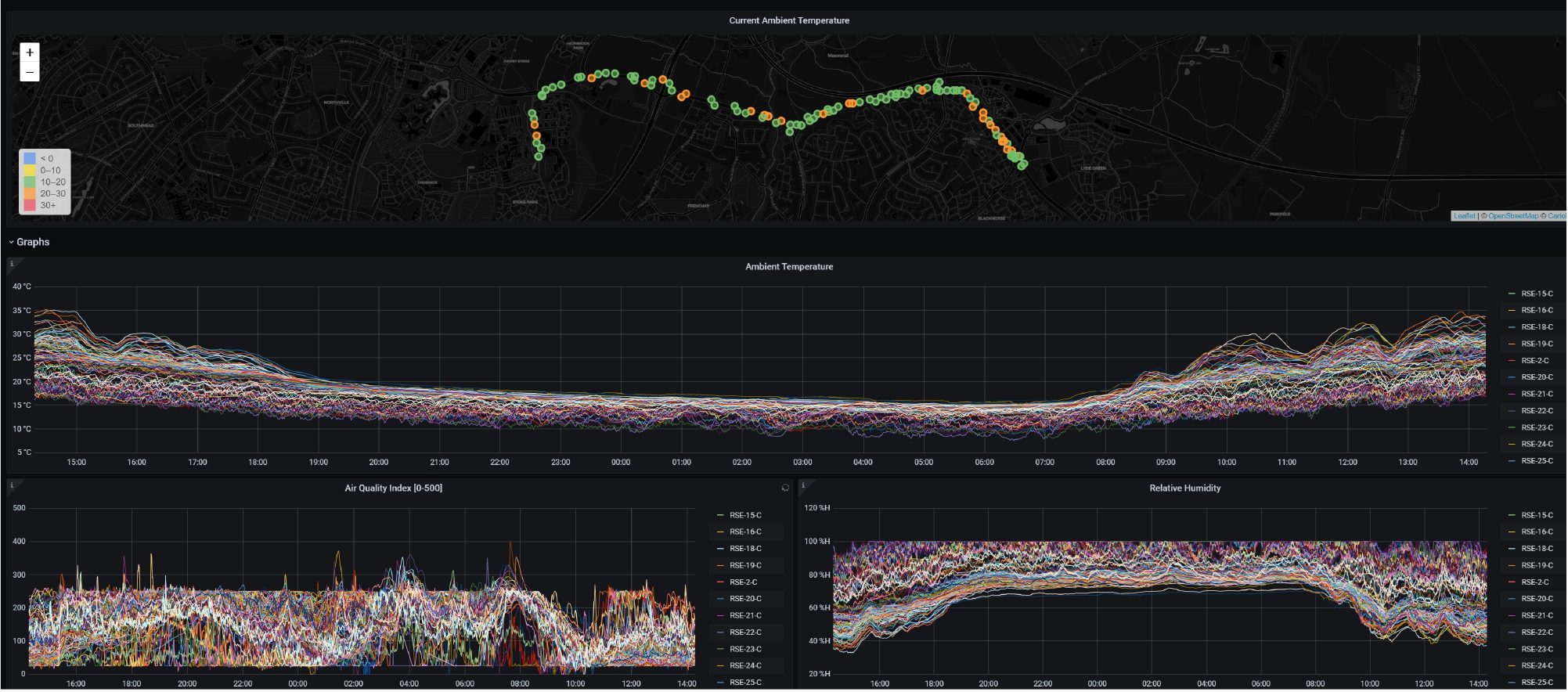}
    \caption{\textbf{Time series sensor data displayed on the UMBRELLA portal.}}
    \label{fig:sensor_data}
\end{figure*}

\quad

\noindent \textbf{Phase 2}: Before the \textit{large-scale manufacturing} and rollout, 20 UMBRELLA devices were initially produced and installed for a six-month field trial period. This revealed hardware and software defects addressed in the final phase. One example is shown in Fig.~\ref{fig:loosen_antenna}, where an issue with the antenna was identified. During this phase, the fibre network was also deployed, having a pair of fibre laid to each future UMBRELLA node. All fibres were patched at our network cabinet at BBSP (Fig.~\ref{fig:Deployment1}). The network is serviced by a redundant pair of servers running Proxox VM hypervisor and a collapsed core network topology, providing redundant links in the core layer that can tolerate the loss of a switch or uplink.

\quad

\noindent \textbf{Phase 3}: the UMBRELLA node went through the \textit{final simplification and refinement stages} called Design for Manufacturing (DFM) and Design for Assembly (DFA) to design parts, components for ease of manufacturing and assembly with the end goal of making a better product with low cost (Fig.~\ref{fig:manufacturing}). A \textit{self-testing code} was programmed once the nodes were produced and powered on. The \textit{final installation} was then carried out with all nodes being deployed. The field trial nodes from Phase 2 were replaced with the manufacturing batch.

\subsubsection {Backend and Platform Implementation}

Following, we have the backend and platform implementation split into two phases:

\noindent \textbf{Phase 1}: A \textit{simple portal} was implemented, allowing users to \textit{create and verify their accounts, create projects and run generic experiments} -- deployed on the field trial nodes but without allowing wireless and sensing capabilities. The basic functionality of the backend was implemented, \textit{integrating} the required \textit{databases, the container registry, and the node management capabilities}. At this stage, a highly available backend was not yet implemented. \textit{Initial versions of the use case applications} were developed and deployed for testing and debugging. Internal and external users and stakeholders thoroughly tested the functionality and captured a second round of functional requirements.

\quad

\noindent \textbf{Phase 2}: The new set of requirements was implemented as part of the second version of the platform. The enhancements provided included a \textit{fine-grained control} of the lifecycle of the available APIs, \textit{role-based access control} of the users, and \textit{scaling up system components} to meet demand and safeguard functionality. \textit{Security measures} were implemented, introducing cloud-native security and network policies. Vulnerability scanning and IDS mechanisms were integrated. All use cases were extended, providing their current functionality and the portal was finalised. 

\textbf{The UMBRELLA network was fully functional in Sep. 2021. Since then, more than 6 million sensing samples are being collected daily via UMBRELLA,} as shown in Fig.~ \ref{fig:sensor_data}.

\section{UMBRELLA Use Cases and Applications}\label{sec:use_cases}
In the previous section, we briefly described the main functionality of UMBRELLA and the core components of the hardware, the software, and the backend. Our incentive is to motivate researchers to demonstrate their own solutions on top of UMBRELLA,  with the potential applications being limited only by the users' creativity.  We envision UMBRELLA to be used for large-scale pilots~\cite{umbrellawireless2022}, wireless protocol evaluations~\cite{protocolEvaluation}, cybersecurity proof-of-concepts investigations~\cite{cybersecurityExamples}, ML-enabled Smart Cities applications~\cite{anand2023}, and more. In this section, we will describe in more detail some core use cases demonstrated, reflecting on the UMBRELLA's capabilities. We encourage the reader to check our Case Studies\footnote{UMBRELLA Case Studies: https://www.umbrellaiot.com/use-cases/} on the UMBRELLA's website for more examples of how the UMBRELLA ecosystem has been utilised in the past.

\subsection {Large Scale Wireless Testbed}\label{subsec:wireless_testbed}
UMBRELLA incorporates several wireless technologies for short- and long-range communications accessible by the end user. The functionality provided is two-fold. At first, UMBRELLA can be used for \textit{over-the-air} large-scale wireless experimentation using Bluetooth. An end-user can reprogram the available interfaces to evaluate new protocols or applications across a large geographical area. Secondly, an end-user can create and run LoRaWAN applications. A user could either display data and log files directly on our portal or utilise UMBRELLA as a backbone for connection to the collaborative ecosystem of The Things Network (TTN)\footnote{The Things Network (TTN): https://www.thethingsnetwork.org}.

Starting with the \textit{over-the-air} experimentation, the unique feature of UMBRELLA, compared to another similar testbed (e.g., FitLab-IoT~\cite{fitlab_iot} D-Cube~\cite{dcube}, etc.), is that it allows not only upload of firmware binaries but also containerised applications that can directly interact with the wireless interfaces (Sec.~\ref{subsec:terminology}). This provides a more diverse and complex set of experiments and algorithms that can be evaluated on UMBRELLA.

The Bluetooth interfaces available on all UMBRELLA nodes (Fig.~\ref{fig:umbrella_network}) are a Nordic Semiconductor nRF52840~\cite{nRF52840} and a Texas Instruments CC1310~\cite{CC1310}. Both interfaces are connected to a dipole antenna with \SI{5}{\dBi} gain. In Fig.~\ref{fig:cad}, we see three antennas. The left one is for the \SI{2.4}{\giga\hertz} nRF52840, the middle one is for the sub-\SI{}{\giga\hertz} CC1310, and the one on the right is for the \SI{2.4}{\giga\hertz} Wi-Fi interface. Between the two interfaces and their antenna exists a Skyworks RF Front-End Module~\cite{skyworksFEM}, integrating a Low Noise Amplifier (LNA) and Power Amplifier (PA). This results in \SI{22}{\dB} of TX power gain and increases RX sensitivity up to \SI{6}{\dB}, approximately doubling the range of a typical IoT device~\cite{skyworksFEM}.  

\begin{figure}[t]
    \centering
    \includegraphics[width=1\columnwidth]{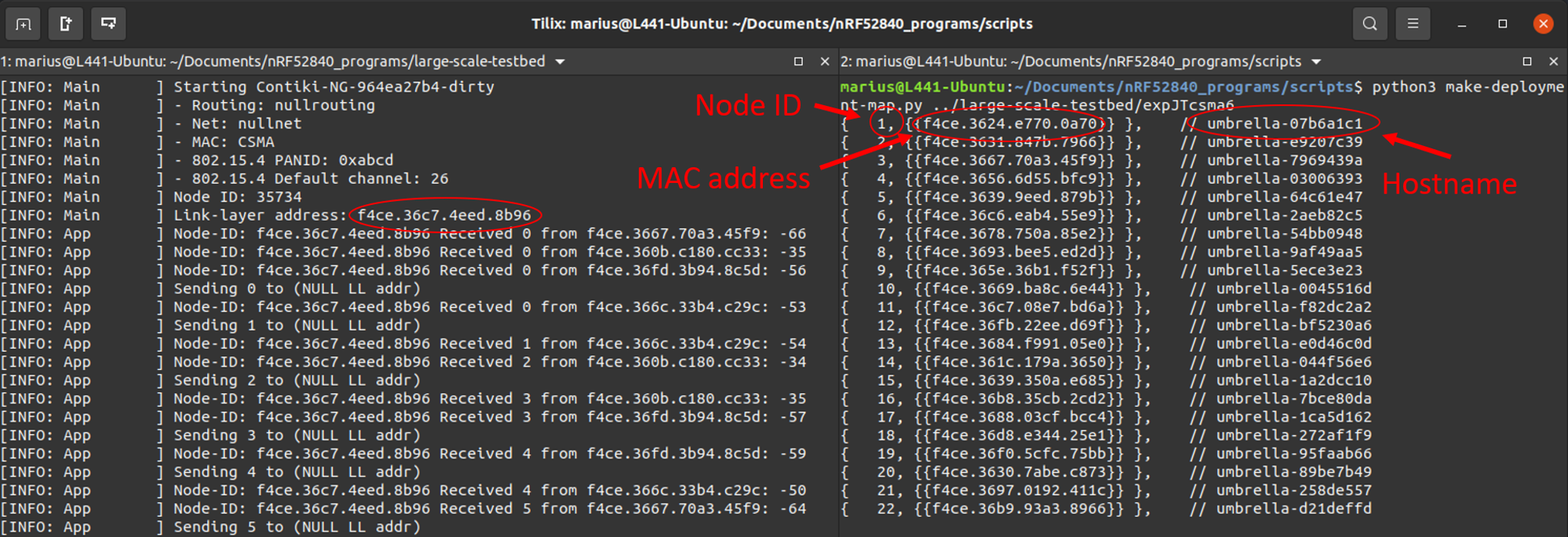}
    \caption{\textbf{An example of the logs collected when running a wireless experiment.}}
    \label{fig:logs}
\end{figure}

\begin{figure}[t]
    \centering
    \includegraphics[width=0.85\columnwidth]{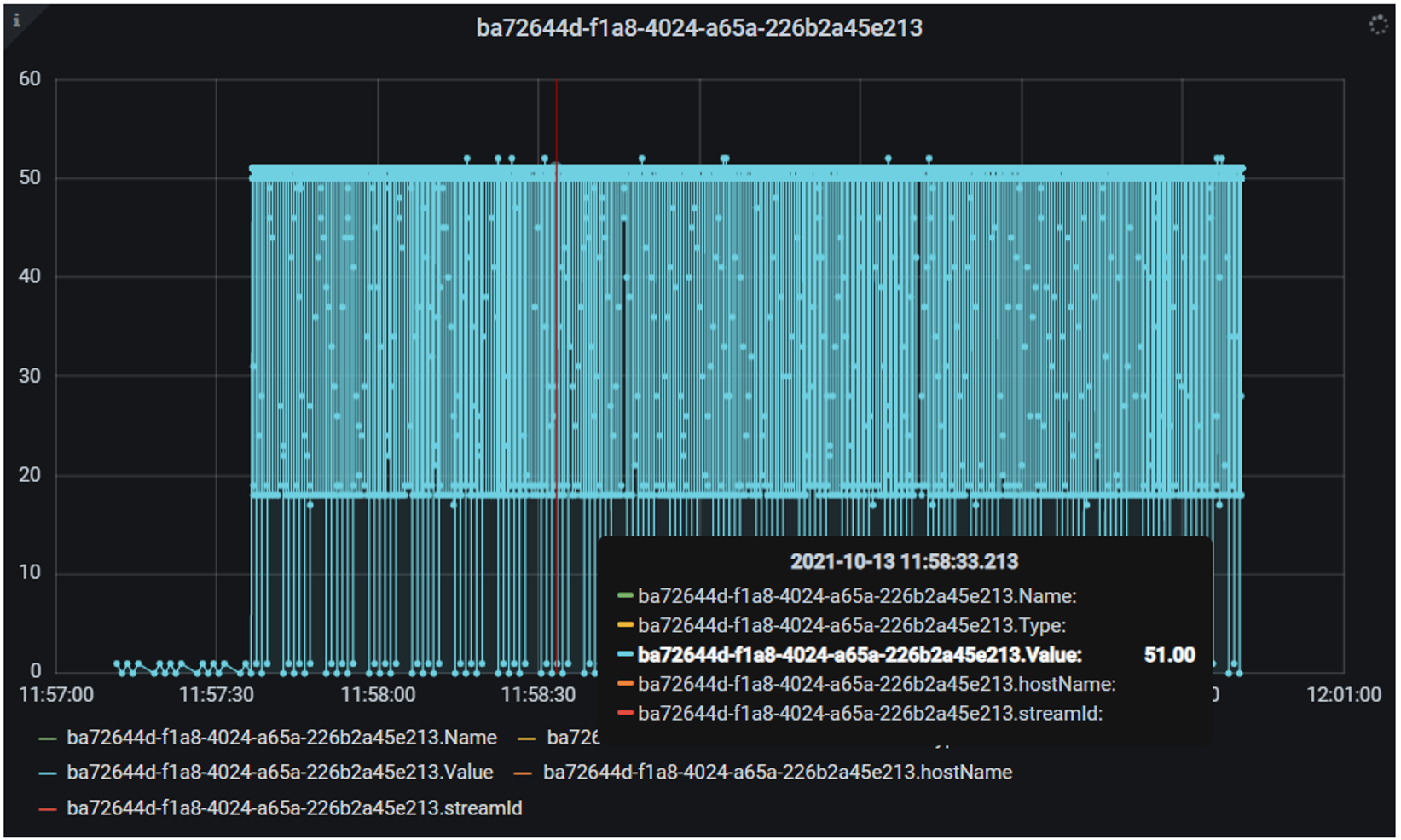}
    \caption{\textbf{An example of the real-time power profiling of Nordic nRF52840.}}
    \label{fig:power}
\end{figure}

\begin{figure}[t]
    \centering
    \includegraphics[width=1\columnwidth]{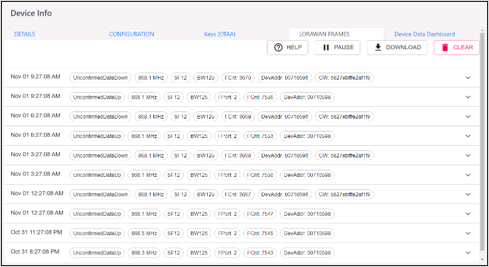}
    \caption{\textbf{An example of a LoRaWAN application running on our portal.}}
    \label{fig:lora_frames}
\end{figure}

As described in Sec.~\ref{subsec:terminology}, a user can run \textit{experiments} flashing a firmware binary on either of the wireless interfaces (Fig.~\ref{fig:combined}). When the experiment is about to start, the backend loads the binaries to the chosen wireless interfaces, deploys the required container images, and initiates the experiment. As described, users can define device ``configurations'' for subsets of nodes and have a fine-grained control of the experiment. For example, a subset of nodes can run as data producers and a subset as data consumers. The different ``roles'' can be assigned based on the different binaries flashed to the interface through the appropriate configuration. 

Moreover, users can upload supporting containerised applications and interact directly with the wireless interfaces (as described in Sec.~\ref{subsec:node_software}). For example, one could require access to the serial logs (the stdout output of the serial interface) of the binary deployed or require a more elaborate data processing algorithm before the data are sent to the backend for storage. The UMBRELLA portal, by default, provides access to two pre-existing base images that a user can utilise (Fig.~\ref{fig:configuration_screen}), one that gives access to the serial logs (Fig.~\ref{fig:logs}) and another that displays in real time the power profile of the interface (Fig.~\ref{fig:power}).

For LoRaWAN experimentation, users can create their own applications and run them over the UMBRELLA network. Two device types can be found in UMBRELLA. One acts as a LoRaWAN gateway equipped with a RakWireless RAK2247 interface~\cite{rakwireless}, and one acts as a LoRaWAN receiver using a Hope RFM95W interface~\cite{hoperf}. UMBRELLA has eleven LoRaWAN gateway nodes deployed (purple nodes in Fig.~\ref{fig:umbrella_network}).

\begin{figure*}[ht]
    \centering
    \begin{subfigure}[b]{0.3\textwidth}
        \centering
        \includegraphics[width=\textwidth]{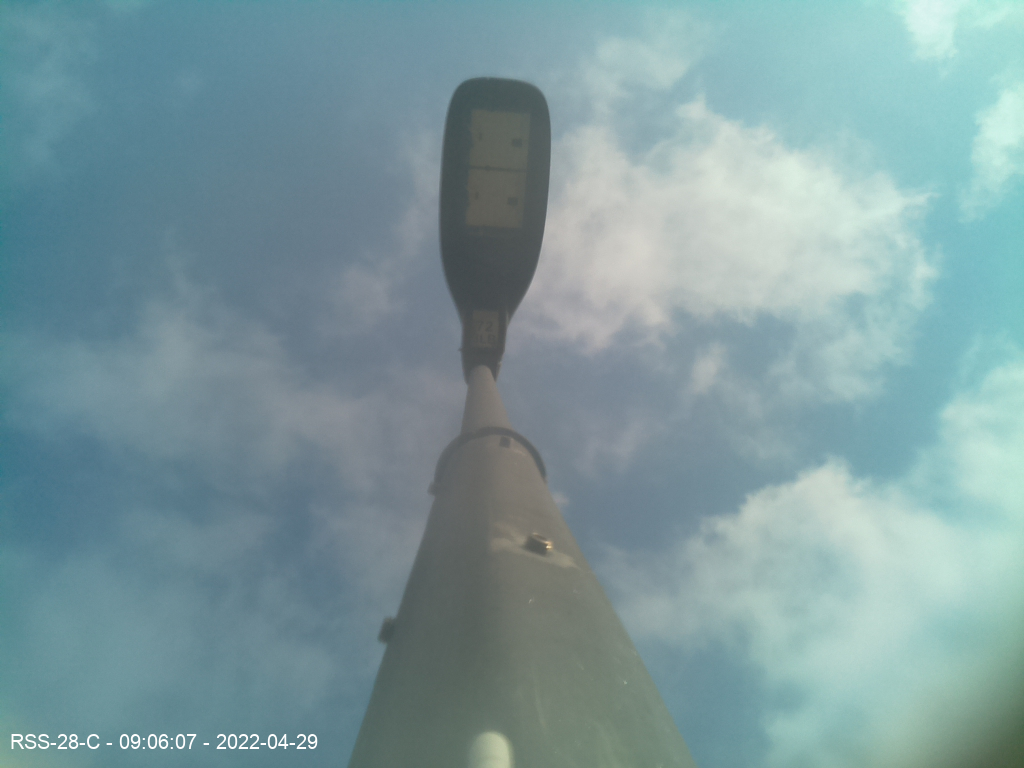}
        \caption{\textbf{Node Type 1}}\label{fig:node0}
    \end{subfigure}
    \vspace{1mm}
    \begin{subfigure}[b]{0.3\textwidth}
        \centering
        \includegraphics[width=\textwidth]{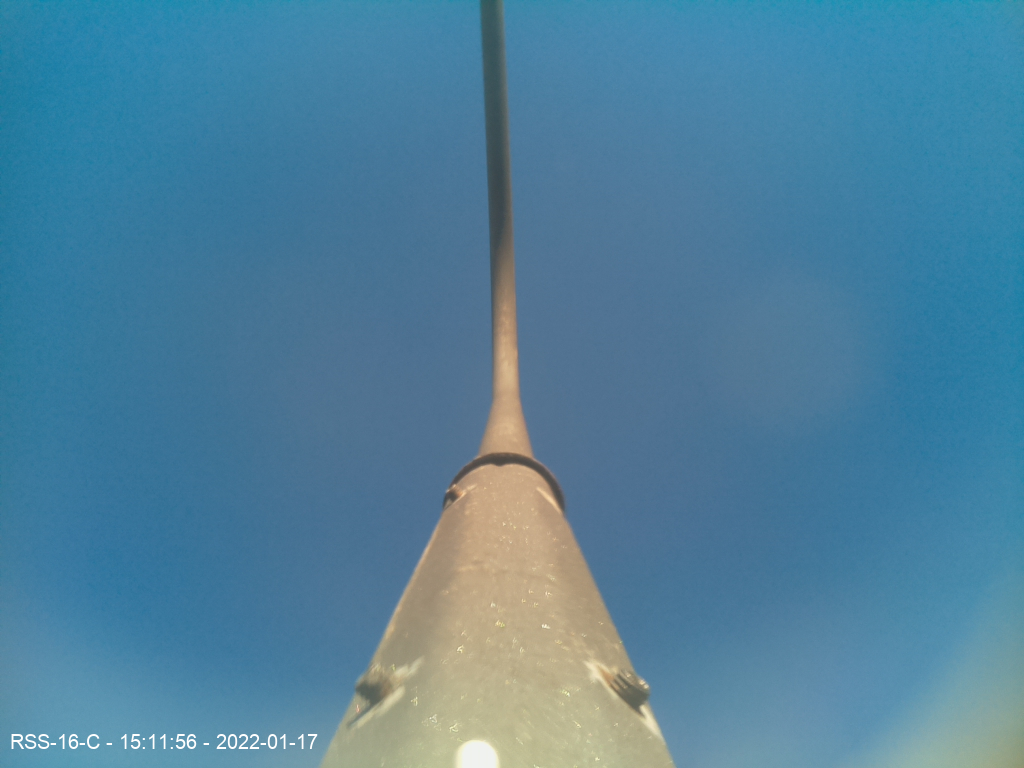}
        \caption{\textbf{Node Type 2}}\label{fig:node1}
    \end{subfigure}
    \begin{subfigure}[b]{0.3\textwidth}
        \centering
        \includegraphics[width=\textwidth]{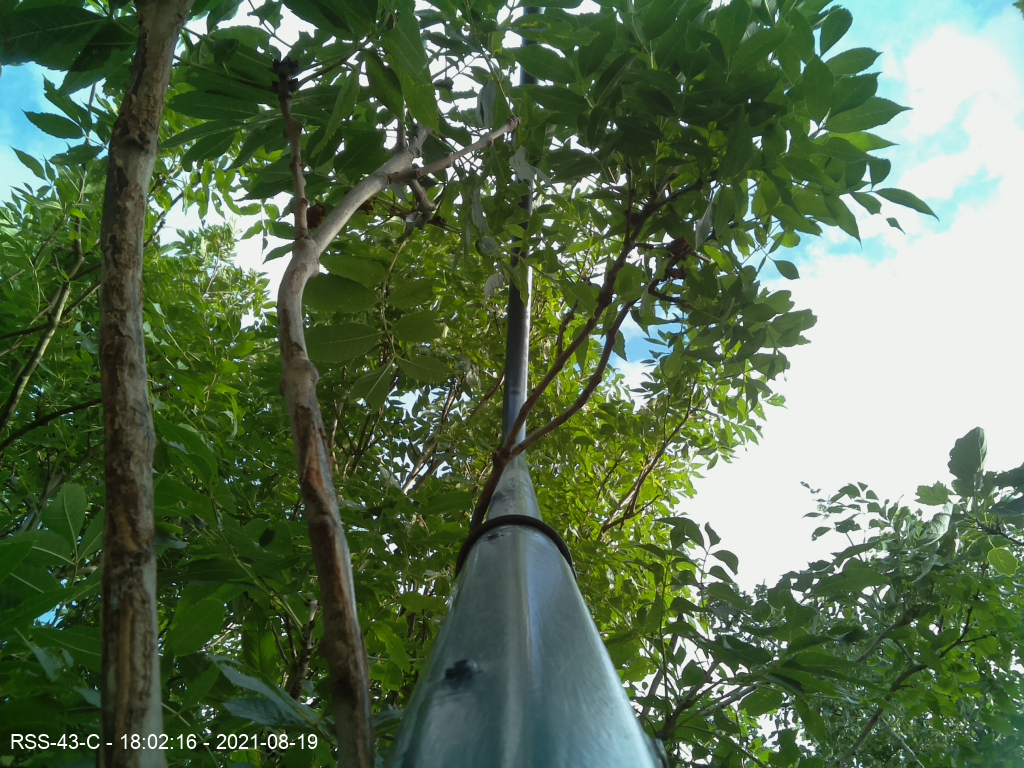}
        \caption{\textbf{Node Type 3}}\label{fig:node2}
    \end{subfigure}\\ %\hspace{-10.5cm}
    \begin{subfigure}[b]{1\textwidth}
        \centering  
        \includegraphics[height=3cm]{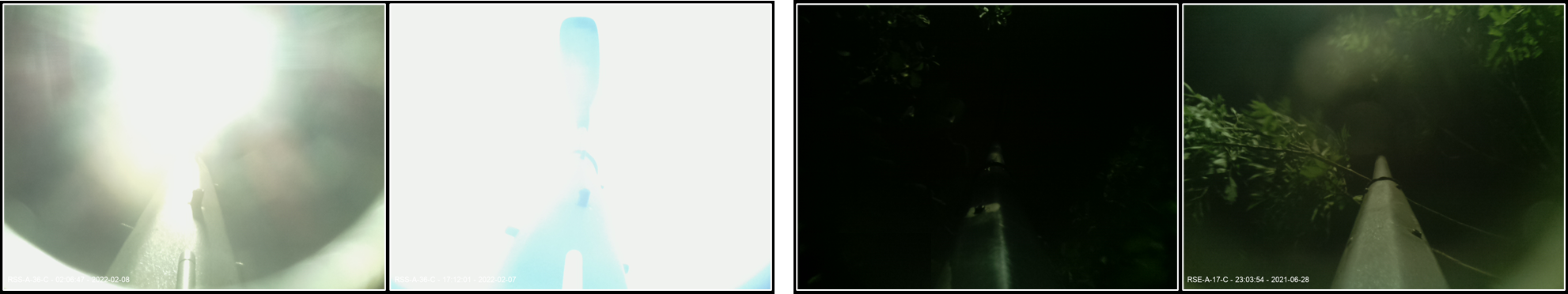}
        \caption{\textbf{Examples of the highly exposed photos used for the nighttime evaluation.}}
        \label{fig:nodeExposed}
    \end{subfigure}
    
    \caption{\textbf{(a) Node 1: Ideal lamppost in which the light is clearly visible; (b) Node 2: In this type, the light is not directly visible; (c) Node 3: In this type, the light is completely covered by vegetation or the camera has slipped. In (d), high-exposure examples are shown; on the left is a lamppost of type Node 1, and on the right is a lamppost of type Node 3.}}
\label{fig:example-of-lampposts}
\end{figure*}

To run a new application, a user needs to create the following: a service profile, which defines the features enabled and the rate of messages sent over the network; one or many device profiles that describe the capabilities associated with the devices; and finally, the LoRaWAN application. Users are not limited in terms of the applications they run. An example application can be seen in Fig.~\ref{fig:lora_frames}. This is a heartbeat application we developed for management and demonstration purposes. As our LoRaWAN implementation is based on ChirpStack, we refer the user to ChirpStack's documentation for more details on creating new applications.

\textbf{UMBRELLA has already been used for large-scale wireless studies}, e.g., in~\cite{umbrellawireless2022} where a large-scale firmware update scenario via a Bluetooth mesh network was investigated, and in~\cite{atomicencryption2022} that the encryption of the Synchronous Flooding protocol Atomic-SDN was demonstrated. Overall, the flexibility provided by the testbed allows a large number of experiments and protocols to be deployed and demonstrated on UMBRELLA. Users can design their own simple or complex experiments and demonstrate them in a realistic environment.

\subsection {Street Light Remote Maintenance}\label{sec:slrm}

This use case aims to monitor the functionality of the street lighting fixtures. Unexpected operation should alert the street lighting maintenance team. Most lights (except the ones that use a custom schedule) turn off 15 minutes after sunrise and turn on 15 minutes before sunset. The council's street lighting team runs periodic (approx. four weeks) manual checks to ensure normal operation. In most cases, they rely on the general public to report problems with the light fixtures. Once multiple fixtures experience unexpected behaviour (i.e., ON/OFF status outside expected times), the team performs corrective measures in batches (for cost-saving reasons) over a stretch of a road. Thus, the reporting requirements on our automatic lamppost monitoring application are in the order of days. 

\begin{figure}
    \centering
    \includegraphics[width=1\linewidth]{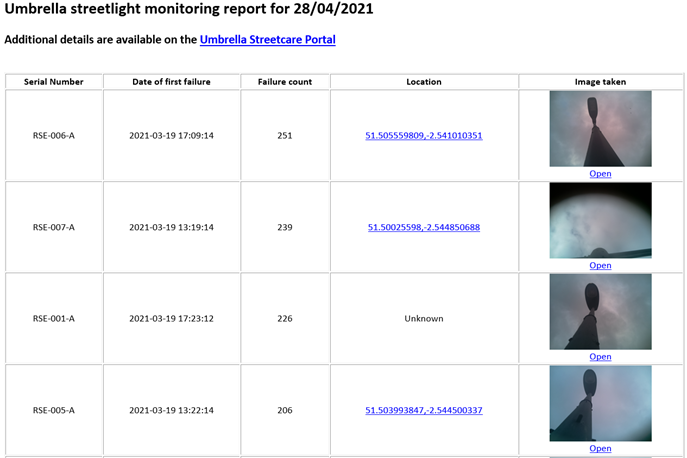}
    \caption{\textbf{An example of the report sent to the Street Care Council Team when one or several faulty lights are identified.}}
    \label{fig:example_report}
\end{figure}

All UMBRELLA nodes are equipped with an RPi Camera Module ver.1 (Fig.~\ref{fig:node_open}-5)~\cite{picamera}, facing the streetlight. Our approach for automating the reporting of fixture problems involves capturing images of the streetlights on several occasions during the day and night and detecting the ON/OFF status. More particularly, if the light is OFF at night or ON during the day (observed across multiple days), a report is generated and sent to the street lighting team as an email. 

Three types of streetlights can be found in the network (Fig.~\ref{fig:example-of-lampposts}), these being: a) the streetlight is within the field of view of the camera, b) the streetlight is outside of the field of view, c) the streetlight is covered with heavy vegetation. Our decision-making is based on two methods. Firstly, we have a simplistic computer vision mechanism. Adjusting the camera's exposure, we can identify the ON/OFF status of the streetlight and label the image accordingly. Fig.~\ref{fig:nodeExposed} shows two examples of the same photo taken with and without increased exposure. Secondly, an ML model classifies the normal image as ON/OFF and provides a confidence interval as well. We have identified that the simple computer vision approach works well at night, while the ML-based approach operates accurately during the day. Thus, our reports are based on this assumption. Complementary to this, \textbf{a dataset was collected and shared with the research community}~\cite{MAVROMATIS2022108658}. The dataset was automatically labelled based on the mechanisms mentioned above.

\begin{figure*}[t]
    \centering
    \begin{subfigure}[b]{0.5\columnwidth}
        \centering
        \includegraphics[width=\textwidth]{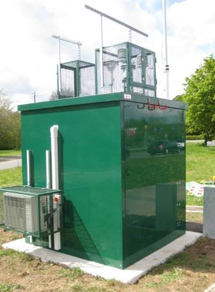}
        \caption{\textbf{Tier 1: High-cost air quality measurement station.}}
        \label{fig:aq1}
    \end{subfigure}
    \vspace{3mm}
    \begin{subfigure}[b]{0.67\columnwidth}
        \centering
        \includegraphics[width=\textwidth]{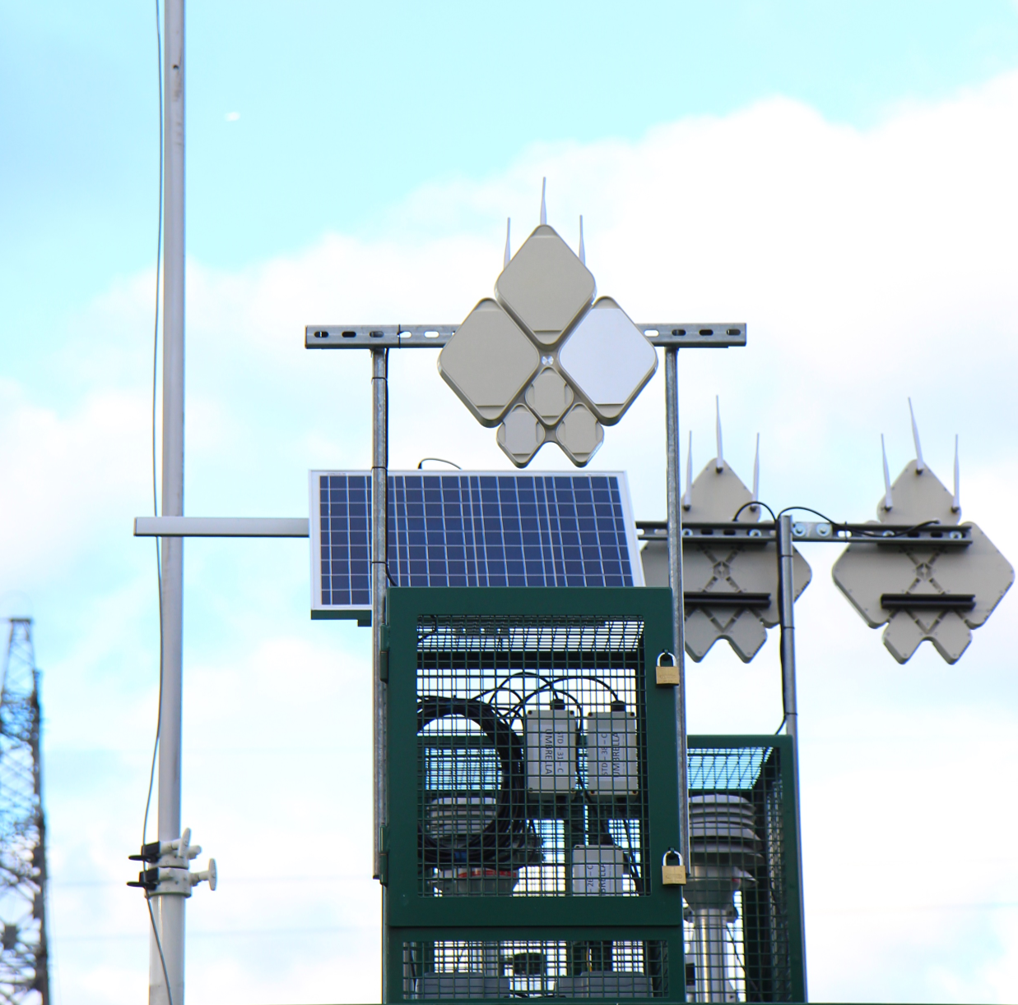}
        \caption{\textbf{Co-located UMBRELLA nodes installed on top of the Tier 1 unit.}}
        \label{fig:aq2}
    \end{subfigure}
    \vspace{1mm}
    \begin{subfigure}[b]{0.6\columnwidth}
        \centering
        \includegraphics[width=\textwidth]{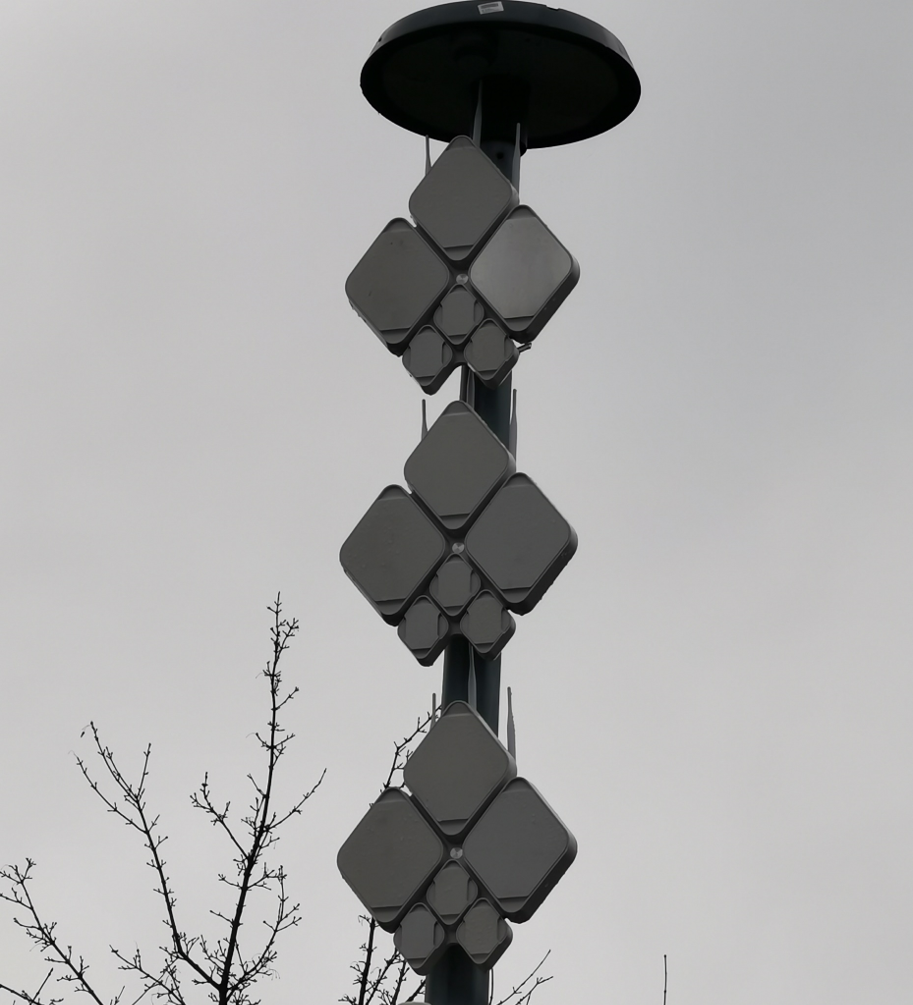}
        \caption{\textbf{Co-located UMBRELLA triplets installed on lighting column.}}
        \label{fig:aq3}
    \end{subfigure}\\
       \begin{subfigure}[b]{0.7\columnwidth}
        \centering
        \includegraphics[width=\textwidth]{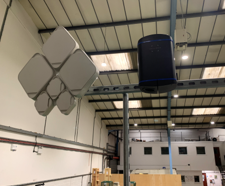}
        \caption{\textbf{Co-located UMBRELLA node and Tier 2 medium cost Zephyr node (8 Units).}}
        \label{fig:aq4}
    \end{subfigure}
    \vspace{1mm}
       \begin{subfigure}[b]{\columnwidth}
        \centering
        \includegraphics[width=\textwidth]{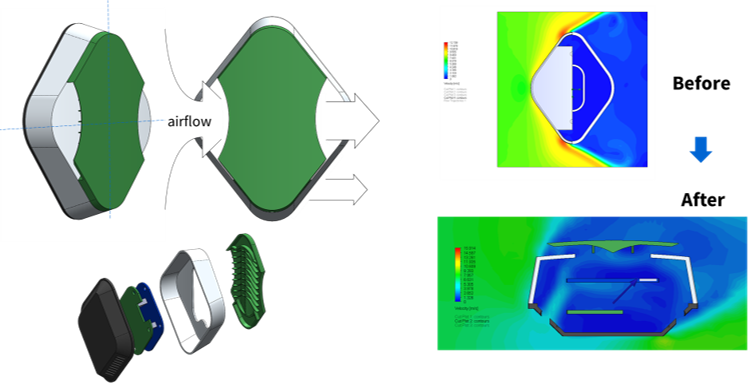}
        \caption{\textbf{Tier 3 - UMBRELLA node air quality pod design for enhanced air venting without water ingress.}}
        \label{fig:aq5}
    \end{subfigure}
    \caption{\textbf{UMBRELLA city traffic and air quality monitoring use case.}}
    \label{fig:Airquality}
 \end{figure*}

An image is collected every hour from each node available. All images are stored on our servers and are displayed on the Street Care portal designed (Fig.~\ref{fig:networks_to_choose}), providing status information for each lamppost in the network. Our UMBRELLA platform is responsible for providing a report for all the lights that are turned off periodically. This report is generated every Monday at 9.00am and sent to the Street Care Council team for further investigation. An example of this report can be seen in Fig.~\ref{fig:example_report}. The street care team also uses our portal to double-check for false positives (e.g., when the sun is directly above the street light). This implementation saves from unnecessary field trips and provides vital debugging information.

Detailed information related to the data collection and the system's operation are presented in~\cite{MAVROMATIS2022108658}. The publicly available dataset consists of $\sim$350k images captured using 140 UMBRELLA nodes over a period of six months. It covers all the different streetlight types presented earlier. The use case is still operational, resulting in $\sim$1.9M images already collected. This dataset can become available via a request to the UMBRELLA's admin team. Potential use cases based on the dataset could be around (but not limited to) assessing the status of the street and emergency lights in real-time, used for enhancing models of outdoor smart city CCTV deployments, real-time weather warning and monitoring systems or used for object recognition combined with datasets of images from other street furniture. Finally, a more \textbf{in-depth investigation for automating the detection of the status of the streetlights using ML was carried out in}~\cite{anand2023}.

\subsection {Three-tiered low-cost Air Quality sensing}\label{subsec:airquality}
Air pollution has been a worldwide issue. The increased level of pollutants in the air has long-term effects on both human health and nature. As A4174 is one of the main arteries of the greater Bristol area, road traffic is the area's primary source of air pollution. South Gloucestershire Council uses the UMBRELLA network to understand the effects of traffic on the local air quality. The data collected are analysed to design policies for combating air pollution without compromising the region's economy.

To meet the requirements of the above use case, UMBRELLA developed a cost-effective solution. As discussed in~\cite{low-cost-sensors}, low-cost air quality sensors are a good alternative for pollution hot spots, such as the road on which the UMBRELLA nodes are deployed. Their mean absolute error might be significant, but the relative measurements follow the trends of expensive air quality sensing stations, costing hundreds of thousands of pounds and covering limited areas. To ensure high-quality decision policy making, and enable the infancy of more diverse use cases from researchers around the world, UMBRELLA designed and deployed a three-tiered air quality monitoring system as shown in Fig.~\ref{fig:Airquality}, which includes: 
\begin{itemize}
    \item Tier 1: One unit of high-cost air quality monitoring station (Fig.~\ref{fig:aq1}), which provides reliable and verified data. The UMBRELLA sensors can be calibrated against the reference station for the individual pollutants of interest.
    \item Tier 2: Eight units of medium-cost sensors (Fig.~\ref{fig:aq4}). The Zephyr air quality sensors\footnote{Zephyr air quality monitor: \url{https://www.earthsense.co.uk/zephyr}}, widely used in many studies, provide accurate results and add a middle tier of robustness to the network. The Zephyr units have been co-located with the UMBRELLA nodes and deployed at main junctions and roundabouts. 
    \item Tier 3: 200+ UMBRELLA nodes equipped with various air quality sensors (as presented in Sec.~\ref{subsubsec:sensingpod}) are placed roadside on lampposts every 40 – 80 meters on either side of the $\sim$\SI{7.2}{\kilo\meter} of road section to ensure appropriate coverage. 
    \item Co-located triplets deployment: Multiple UMBRELLA nodes are purposely deployed at the same location as shown in Figs.~\ref{fig:aq2} and Figs.~\ref{fig:aq3} to improve sensing accuracy and to mitigate drifts. 
\end{itemize} 

The main technical benefits of such a system are: 1. Real-time collection of sensory data in large areas without the need to travel to and collect samples. 2. Improve overall sensing accuracy by correlating and calibrating the low-cost sensors using the three-tiered system. 3. Less frequent service cost of expensive sensing equipment such as replacing air quality filters. 

All the data collected are stored on the UMBRELLA backend and are displayed on various Grafana dashboards (Fig.~\ref{fig:sensor_data}), allowing visualisation of the data and the download in CSV format. Apart from the provided dashboards, a Sensor REST API has also been developed and is accessible by UMBRELLA users. The API gives access to all collected data and can be easily integrated into applications outside the UMBRELLA ecosystem. Access to the API can be requested from the UMBRELLA admin team. Since deployed, the UMBRELLA air quality monitoring system has been used by academic air quality researchers~\cite{jim} and local council officers.

\subsection {Robot Arena Testbed}\label{subsec:robotarena}
The robotics experimentation testbed consists of a physical arena (\SI{5}{}$\times$\SI{5}{\meter}) with up to 20 DOTS robots (see Tab.~\ref{table:robotmainspec}) and a digital twin environment that can simulate the robots and their environment for both validation and comparisons. In order to achieve the comparison and validation, sub-mm accurate ground truth data is provided by an Optitrack(TM) optical tracking system in the real arena environment and a ground truth publisher plugin for the Gazebo-based simulation environment. Both these approaches generate ROS2 topics for distributing the odometry data relating to each robot. In this manner, playback/replay and direct comparisons are possible between the arena and digital twin environments. 

Visualisation is also provided in the 3D Gazebo web view (gzWeb) and the 2D view (for the arena) within the UMBRELLA portal. Accessing the ground truth data is possible via a REST API provided. The sensors provided on the robot (see Tab.~\ref{table:robotsensorspec}) are accessible through the corresponding ROS2 topics. There are two types of radios provided (see Tab.~{\ref{table:robotradiospec}}). The two Bluetooth radios are accessible through the corresponding serial port devices. All information about the robots and the interfaces can be found on our Wiki page.

\begin{table}[t]
\renewcommand{\arraystretch}{1.1}
    \caption{\textbf{Robot main capabilities.}} 
    \centering
    
    \begin{tabular}{r||l}
    \textbf{Specification} & \textbf{Value} \\
    \hline \hline
    
    Diameter & 250mm \\
    \hline
    Height & 120mm lifter down and 200mm up \\
    \hline
    Speed & 2 metres per second \\
    \hline
    Acceleration & 4 metres per second squared \\
    \hline
    Max load & 2kg \\
    \hline
    Battery capacity & 100Wh \\
    \hline
    Battery life & 6 hours to 12 hours \\
    \hline
    Battery charge time & 3 hours \\
    \hline
    Drive & 3 Holonomic servomotor omniwheels \\
    \hline
    Processors & RockPi4B 4GB 4x Raspberry Pi Zero \\
    
    \end{tabular}\label{table:robotmainspec}
\end{table}

\begin{table}[t]
\renewcommand{\arraystretch}{1.1}
    \caption{\textbf{Robot sensor specifications}} 
    \centering
    
    \begin{tabular}{r||c c}
    \textbf{Sensor} & \textbf{Type} & \textbf{Rate}  \\
    \hline \hline
    Cameras x5 & OV5647 5MP & < 60Hz \\
    \hline
    Microphones x2 & IM69D130 & 44.1kHz  \\
    \hline
    Laser time-of-flight distance x16 & VL53L1X & 50Hz  \\
    \hline
    6DoF IMU & LSM6DSM & 1kHz  \\
    \hline
    Magnetometer & LIS2MDL & 1kHz \\
    \hline
    Barometer & LPS22HD & 1Hz  \\
    \hline
    Temp/humidity & Si7021-A20 & 1Hz  \\
    \hline
    Strain gauge ADC x3 & MAX11210 & 1kHz  \\
    \hline
    General purpose 8 channel ADC & MAX11615 & 1kHz \\
    \hline
    Power monitor x3 & INA219 & 100Hz  \\
    \hline
    PSU status & STM32L031 & 100Hz  \\
    \hline
    Battery status & STM32L031 & 10Hz  \\
    
    \end{tabular}\label{table:robotsensorspec}
\end{table}

\begin{table}[t]
\renewcommand{\arraystretch}{1.1}
    \caption{Robot radio specifications} 
    \centering
    
    \begin{tabular}{r||c}
    \textbf{Technology} & \textbf{Radio} \\
    \hline \hline
    Bluetooth & 2x Nordic nRF52840 radios \\
    \hline
    Wi-Fi & IEEE 802.11ac compliant onboard radio \\
    
    \end{tabular}\label{table:robotradiospec}
\end{table}

\begin{figure}[h]
    \centering
    \includegraphics[width=1\columnwidth]{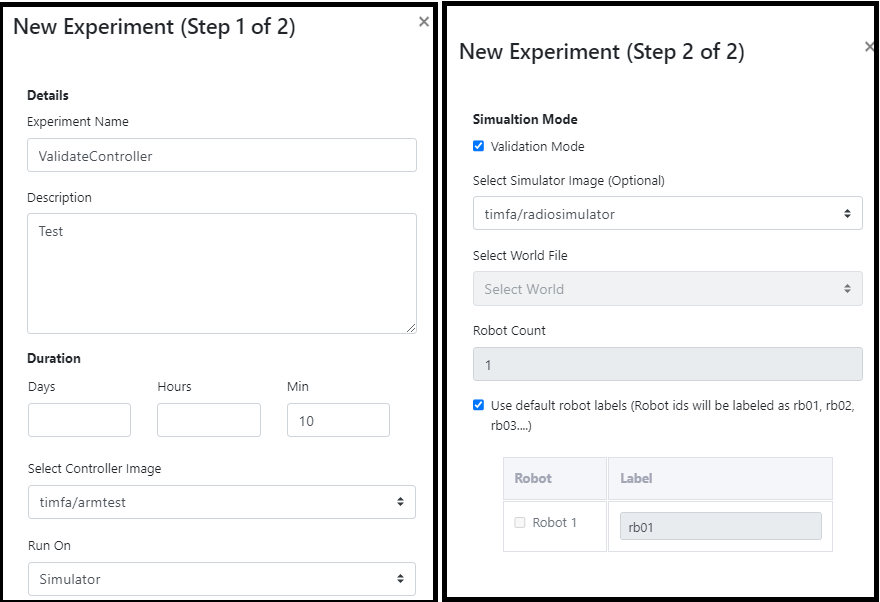}
    \caption{\textbf{Robot experiment configuration selection steps.}}
    \label{fig:robotexperiment}
\end{figure}

\begin{figure}[h]
    \centering
    \includegraphics[width=1\columnwidth]{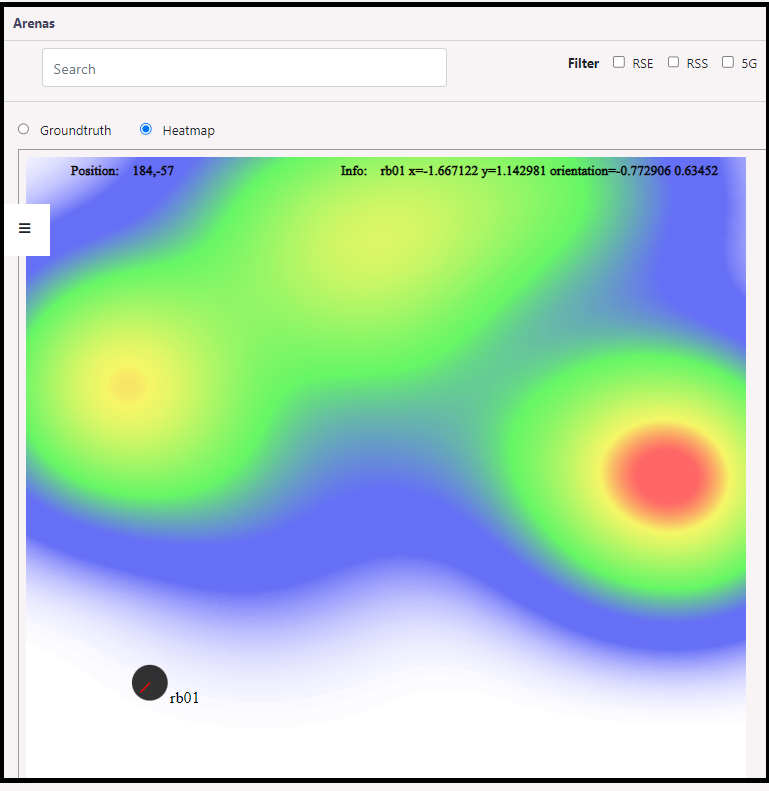}
    \caption{\textbf{Robot arena heatmap ground truth view.}}
    \label{fig:robotarenaview}
\end{figure}

The robotic experiments follow the same steps described in Sec.~\ref{subsec:terminology}. Moreover, the user can define the robot's friendly name and the radios to be used during the experiments (Fig. \ref{fig:robotexperiment}). All experiments must be validated on the robot simulator before being used in the robot arena. The validation mode uses a predefined arena world (corresponding to the real arena) to ensure collision-free operation. User-defined worlds (in SDF format\footnote{SDF Format: \url{http://sdformat.org/spec}}) can also be uploaded for evaluation of larger or more complex scenarios. Radio simulators are also provided to evaluate custom protocol stacks supporting swarm (robot to robot) interaction. The radio simulator instances in the digital twin environment receive ground truth updates via an exposed REST API that is polled at specified intervals. 

In order to monitor the experiment within the arena environment, a video feed is provided for the experiment owner. The status can also be monitored using the ground truth view shown in Fig. ~\ref{fig:robotarenaview}. This view shows the position and orientation of the robots as well as the heatmap of all positions during the experiment.

\begin{figure}[t]
    \centering
\includegraphics[scale=0.07]{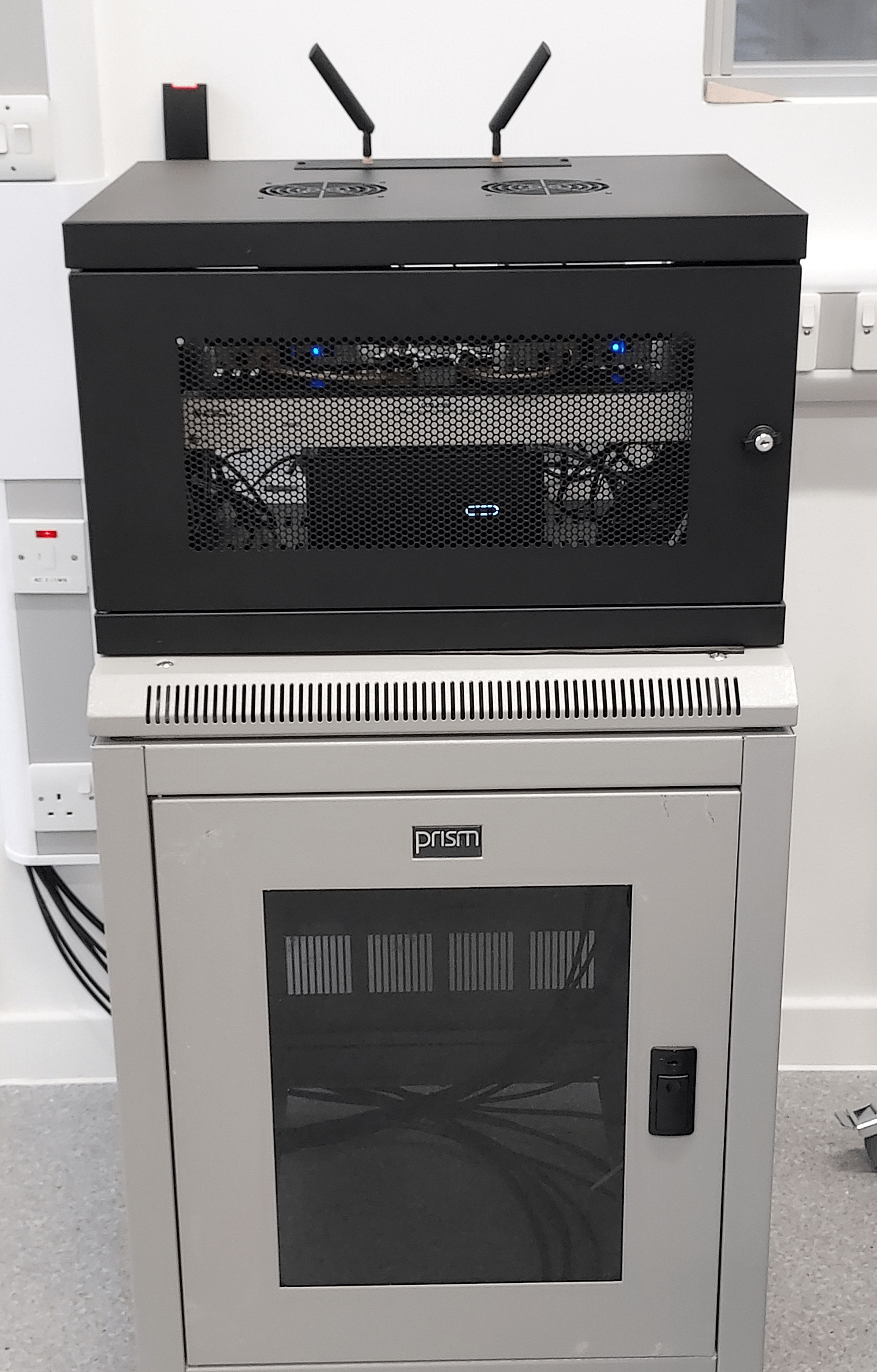}
    \caption{\textbf{One of the 5G network-in-a-box solutions deployed in UWE (RIF).}}
    \label{5g_node}
\end{figure}

The robot arena is assisted by a technician for health and safety reasons, which prevents unattended operation. This necessitates that the experiments be placed into a queuing system and activated only when the arena technician is present. The technician can also remotely initiate the experiments via a developed admin portal. \textbf{Further information regarding the DOTS robots and the robotics testbed platform can be found in~\cite{Jones2022DOTSAO} and~\cite{UMBRELLA_collaborative}, respectively}.

\subsection{Private 5G Capabilities}\label{subsec:private5G}

Private 5G~\cite{Pvt_5G} unlocks the potential of 5G technology for IIoT applications with additional benefits of dedicated coverage, intrinsic control, and exclusive capacity, thereby providing dependable connectivity. The UMBRELLA testbed provides the capability of testing and validating private 5G~\cite{Pvt_5G} solutions and use cases. This capability is based on a 5G network-in-a-box solution. 

The initial prototype implementation of this network-in-a-box solution is based on an open-source 5G software stack from the OpenAirInterface (OAI) community\footnote{\url{https://openairinterface.org/}} and general-purpose hardware for the Radio Access Network (RAN) and core network. It operates in the 5G non-standalone (NSA) mode where the 5G base station (gNB) uses a 4G base station (eNB) as an anchor, and it is connected to a 4G core network, also known as Evolved Packet Core (EPC). The gNB and eNB front-ends are based on Ettus B210 software radio platforms~\cite{ettus}. 

The network-in-a-box solution supports Commercial-Off-The-Shelf (COTS) handsets and dongles in different sub-\SI{6}{\GHz} frequency bands, some of which are dedicated to private networks. Further details regarding the \textbf{implementation and performance of the 5G network-in-a-box solution are available in~\cite{5G_Netsoft}}. 

As part of the UMBRELLA testbed, two 5G network-in-a-box solutions have been deployed in key locations at the UWE campus. One is co-located with the robotics testbed and provides the capability to test 5G connectivity for cooperative robotics applications and architectures. The second has been deployed in the Robotics Innovation Facility (RIF). It provides the capability to test teleoperation-centric use cases over 5 G. Both solutions provide indoor 5G coverage/connectivity testing capabilities.

\subsection{MLOps/FL on UMBRELLA}\label{sec:mlops}
Machine Learning Operations (MLOps) refers to the practices that aim to deploy and maintain ML models efficiently and reliably in production. MLOps combines ML, development operations (DevOps), and data engineering methodologies. Some aspects of the MLOps framework have been tested in UMBRELLA, particularly focusing on edge-based FL. The aim was to effectively use the GPU-enabled UMBRELLA nodes to function as FL clients performing local training.  

\begin{figure}[t]
    \centering
    \includegraphics[width=1\columnwidth]{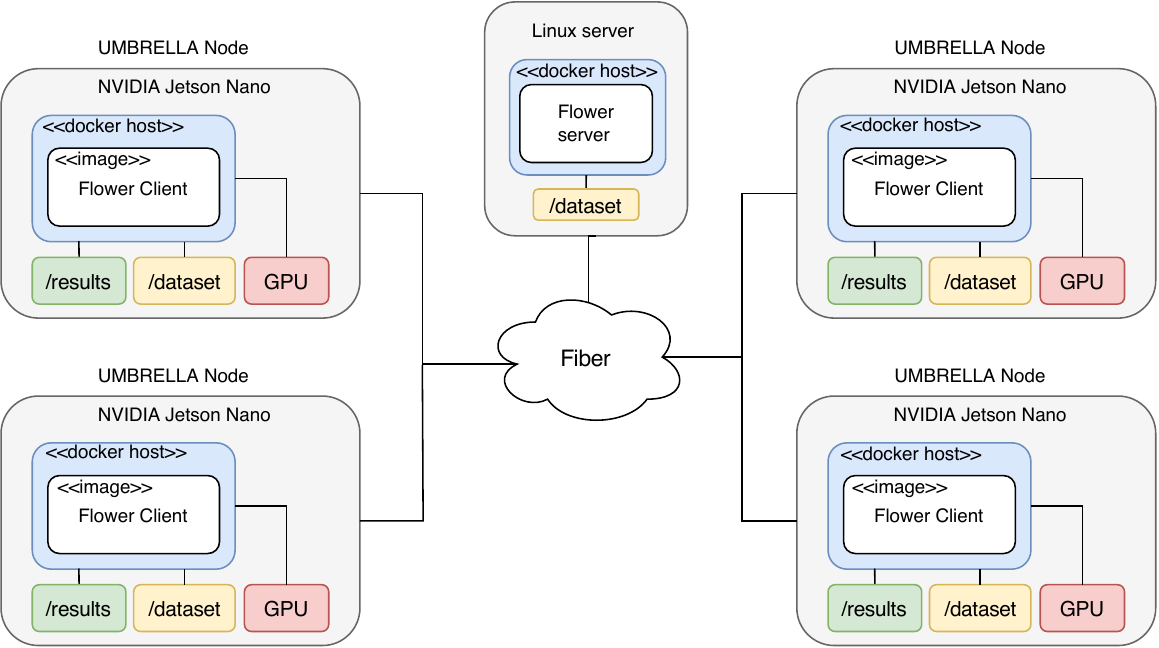}
    \caption{\textbf{K3S and Docker containerisation system architecture for proof of concept FL simulations using UMBRELLA IoT network devices.}}
    \label{fig:fl_mlops}
\end{figure}

For this use case, our FL platform was extended to function within the Docker containerisation and Kubernetes orchestration systems. Together, these two functionalities cover the deployment and infrastructure management aspects of an MLOps framework. This allows our FL platform to be more hardware-independent and scalable, allowing hundreds of clients to be ``live" training and updating a global model aggregated in the central parameter server. These features are illustrated in Fig.~\ref{fig:fl_mlops}, showing an example of four connected devices from our UMBRELLA IoT network. Each UMBRELLA device/node is equipped with a GPU (as detailed in Sec.~\ref{sec:edge_module}), which in turn loads a pre-built Docker container with all relevant Flower~\cite{flower} and PyTorch~\cite{pytorch} libraries, as well as an (initial) dataset for FL training. Each FL container then interfaces with a central parameter server (at the UMBRELLA servers) via the Flower framework to conduct FL training at scale. 

For our initial testing, FL was used within the context of the street lights remote maintenance use case (see Sec.~\ref{sec:slrm}). Images of the lamppost lights (as shown in Fig.~\ref{fig:example-of-lampposts}) from each UMBRELLA node are used for training local FL models. Some \textbf{results associated with this study have been reported} in~\cite{anand2023}, where we benchmark the performance of our privacy-preserving FL-trained models, achieving over 98\% accuracy for correctly detecting the status of the street lights. \textbf{Some more results and use cases are described in~\cite{umbrella_past_present_future}}.

\section{Spin-offs and UMBRELLA-enabled Projects}\label{sec:spinoffs}
UMBRELLA enabled the delivery of a number of projects throughout the years. Utilising either the existing infrastructure, the designed hardware, or the backend platform, spin-off projects have leveraged the existing functionality and extended UMBRELLA's capabilities or demonstrated unique use cases. Below is a list of the most prominent projects that benefited from the UMBRELLA ecosystem.

\textbf{LoRD}\footnote{SemanticLCA: https://www.list.lu/en/environment/project/semanticlca/} is a subcontracted project from EPSRC SemanticLCA (running until Dec. 2023). Two UMBRELLA nodes were deployed at the School of Engineering of Cardiff University as part of the project. Both nodes served as LoRaWAN gateways. Several custom LoRaWAN-enabled devices monitored the indoor air quality of the building. The data collected are used by the University's researchers for developing energy models for the building. The implementation of the project is presented at~\cite{Lord}.

\textbf{SYNERGIA}\footnote{SYNERGIA: https://synergia.blogs.bristol.ac.uk/about/} is an Innovate UK-funded project concluded in Dec. 2022. SYNERGIA utilised the hardware produced in UMBRELLA and the backend platform. The project demonstrated novel secure-by-design, end-to-core capabilities for IoT platforms and resource-constrained devices. It focused on secure configuration management and control of IoT devices, a distributed IDS at the edge, extended the cloud-native capabilities of UMBRELLA with security features and provided solutions on drift detection for IoT data~\cite{le3d}. 

\textbf{Green Lamp Post}\footnote{Green Lamp Post: https://tinyurl.com/grlamppost}, completed in Nov. 2021, is a project that utilises hardware produced in the UMBRELLA. The project aimed to combine the sensing elements of UMBRELLA with decorative floral displays attached to lamp posts. The project was conducted in collaboration with Data Communications Company (DCC) and Scotscape. The hardware features a UMBRELLA endpoint with a customised GLP HAT that allows for control of watering pumps. A modified UMBRELLA software stack was deployed on a stock Raspberry Pi 3b+ to enable communication with DCC's backend.

\textbf{BEACON-5G}\footnote{BEACON-5G: https://www.beacon-5g.com/} (finished in Sep. 2023) enhanced UMBRELLA's 5G capabilities via a 5G standalone (SA) deployment. The 5G SA designed and deployed is based on Open RAN (O-RAN) architecture. Having only the 5G radio unit (RU) and antenna installed at a busy traffic junction (Hambrook junction at Bristol, UK) reduces the installation's footprint. The rest of the RAN and core network components are located at BBSP, connected via a \SI{4.3}{\kilo\meter} fibre link. BEACON-5G shares UMBRELLA's infrastructure for the backbone connectivity. Moreover, extending UMBRELLA backend functionality developing a novel 5G API marketplace~\cite{marketplace5G} that enables the integration and monetisation of different O-RAN Digital Twins (DTs) and allowing API-centric monetisation policies, such as pay-per-use, performance-based criteria, service level agreements, as well as flat subscription fees. Further details about the multi-vendor 5G O-RAN system, the field deployment, and a designed Smart City use-case are available in~\cite{BEACON-5G_conf}. 

\textbf{Goal-oriented Communications for Robotics}:
The UMBRELLA robotics testbed will be used for technology trial activities as part of the newly funded EU Horizon SNS 6G-GOALS project (starting in Jan./Feb. 2024). The 6G-GOALS project will lay the theoretical, algorithmic, and operational foundations of a novel \emph{goal-oriented communication paradigm}, underpinned by semantic awareness in networks~\cite{Semcom_Open_RAN} and AI-based architectures. 

\textbf{MILO} delivered a minimal-infrastructure robotic solution for remote delivery of supplies in deployed arenas in response to crises, such as the Nightingale Hospital. It was funded by Innovate UK to address COVID-19 disruption. The UMBRELLA radio hardware and communication technologies enabled infrastructure-less communication among MILO robots.

\section{Lessons Learnt and Critical Reflections}\label{sec:lessons_learned}

UMBRELLA was a massive collaborative effort to create a sustainable, multi-technology, real-world testbed aiming to address IoT innovation challenges faced by many sectors and users. Most of the UMBRELLA system's design, development, and deployment tasks occurred during the COVID-19 lockdown period, a journey filled with challenges and learnings. While the UMBRELLA project has been successful in many respects in delivering the original vision, it has also presented numerous challenges that provided us valuable learning throughout the journey. We hope some of the learnings of both technical and non-technical nature will help anyone who wishes to undertake a similar endeavour to meet the objectives through better consideration of some of the challenges mentioned.

\subsection{Design Challenges}
Building a holistic IoT testbed to cater to the diverse needs of stakeholders ranging from local authorities, hardware developers, and solution developers to researchers by incorporating multiple technologies requires a complex process. The challenge has been made harder by the requirement for the testbed to be useful for the next ten years. Since the development of IoT technology is happening rapidly, developing a system that will not be obsolete within a few years needs to make design choices, especially on what hardware to use, how it will be deployed, how it will be maintained, etc.

\begin{enumerate}[wide, labelwidth=!, labelindent=9pt, itemindent=0pt]

    \item \textbf{Accessibility and Future-Proofing}: Initial key and ambitious requirements may not be feasible due to practical limitations. Adaptability in the design and accommodating rapid changes without deviating from the end goal are really important for future-proofing a system. \\
\textit{Reflections}: While modularity and a plug-and-play approach for the UMBRELLA pods was one of the key initial requirements, practical limitations like, e.g., the installation height that could lead to falling pods, the need for road closures to access and install the nodes, and the difficulty of weatherproofing exposed USB interfaces, played a pivotal role in the final node design. The injection moulded casing was converted from ``pods'' to a single-piece casing, but the electronics and hardware were designed with the initial plug-and-play approach in mind. This, even though it does not allow for easy replacement on the existing testbed, enables the futureproofing of UMBRELLA, allowing the deployment of smaller pods (with a new casing) in other use cases.

\vspace{0.2em}

\item \textbf{Balance between Scope and Specialisation}: While a broad scope can cater to a wider audience, it can also dilute the effectiveness of the testbed for specialised tasks and struggle to attract certain user groups. A clear understanding of primary stakeholders and use cases is crucial to finding the first user group for the testbed. \\
\textit{Reflections}: As the primary stakeholders were local authorities, there has been a great interest in environmental sensing and smart infrastructure solutions. Thus, UMBRELLA focused on integrating sensors and IoT devices catering to these needs, leading to high utility and satisfaction among the initial users. Discussing further with non-key shareholders and broader audiences from the academia and the industry revealed needs that could also be accommodated on the UMBRELLA platform with minor modifications in the initial architecture design. The adaptability of a design is of paramount importance, and UMBRELLA delivered a platform that is use case agnostic, making the system a great playground for multiple communities.

\vspace{0.2em}

\item \textbf{Bottom-up design risks}: While the technology-driven design is essential and considering many potential use cases can lead to innovation, it can also result in an overly complex system. Co-developing example use-case trials to demonstrate the use of the testbed not only reduces the risk of major design oversights but also acts as a vital tool to help future users.\\
\textit{Reflections}: The way functional and technical requirements are captured must be very well thought out, precise, clearly communicated, and cross-validated. For example, working closely with the local authorities that specialise in environmental monitoring led to a highly tailored system that is very effective. On the other hand, due to miscommunication, a large subset of the lamps on targeted lampposts are outside of the field-of-view of the UMBRELLA camera, as the final rollout involving three different types of lampposts which was not identified during the initial field testing period. A workaround had to be provided to ensure the delivery of the streetlight monitoring use case, but with proper communication, a different camera model with a wider lens could have been used.

\vspace{0.2em}

\item \textbf{Regulatory and Compliance related risks}: Continuous dialogue with local communities, authorities, and other stakeholders can smoothen the deployment and operation of the system. Oversight to meet certain straightforward requirements could lead to significant costs and delays in operationalising the system. In the worst case, such requirements could force part or the whole system unusable for a prolonged period until such conflicts are resolved. The opposite is also true – making hard decision choices to avoid compliance risks could mean a system's inability to provide certain services without major updates later. Considering a simple technological solution such as software control to address such issues could easily overcome such constraints to make the system future-proof. \\
\textit{Reflections}: As UMBRELLA was built primarily during COVID-19, some design decisions were severely impacted. An example is the position of the camera, which is pointing upwards. In the initial design, cameras were positioned in parallel with the road plane. However, the design was changed to avoid a lengthy approval process that had significantly delayed the nodes' manufacturing. This, unfortunately, limited potential applications such as traffic monitoring that could have been implemented without the need for node modifications.

\end{enumerate}

\qquad

\subsection{Development \& Deployment Challenges}
While the testbed is expected to operate at high Technology Readiness (TRL) and Commercially Readiness (CRL) levels so that potential users can use the services offered by the testbed reliably, the need to use state-of-the-art technologies often creates a conflicting requirement. 
\begin{enumerate}[wide, labelwidth=!, labelindent=9pt, itemindent=0pt]

    \item \textbf{Balancing product and research conflicts}: Managing expectations between research and innovation teams who are responsible for defining the scope and the specification, and the product teams who are tasked with the development of the system is crucial, which at times could manifest as a human resource management challenge. Working with multi-skilled teams and organisations enhances the breadth of expertise. However, clear communication, defined roles, and aligned goals are essential to prevent misunderstandings and redundancies.\\
    \textit{Reflections}: Even though UMBRELLA is envisaged to drive innovation and research, one of the key goals was a stable production system. Thus, during its development, some state-of-the-art solutions, even though were considered, they did not leave the drawing board to ensure stable operation. An example was a serverless deployment of the robotic digital twins that could reduce cloud costs in the long term and allow the number of experiments to scale. Serverless across hybrid cloud/on-prem environment was still in its infancy back in 2019, thus, a simpler approach was designed, based on a simplistic Continuous Integration and Continuous Delivery (CI/CD) system that allows execution to three existing VMs on AWS (thus limiting the parallel experiments to three).

    \vspace{0.2em}
    
    \item \textbf{Technical Enablers are only part of the success}: While sensing, actuation, and remote programmability are crucial, user experience, interface design, and training tools are equally important for user adoption and satisfaction.\\
    \textit{Reflections}: UMBRELLA was built with ease-of-operation in mind. As the majority of the users are non-technical in nature, use cases and data are accessible via GUIs. This ensures that even non-technical stakeholders can easily interact with the system via a web-based portal that is easily accessible.

    \vspace{0.2em}

    \item \textbf{Prepare for Evolution}: The technology and standards related to IoT and industry trends are changing rapidly, which means that today's cutting-edge solution might be obsolete tomorrow. Modularity, reconfigurability, and re-programmability will prevent major redevelopment costs later on. In practice, technical teams who design and develop testbeds will rely on delivery partners not only for deployment but also for operation and maintenance. Such partnerships should be designed into the work programme from early on to maximise the service delivery capability and keep evolving the system.\\
    \textit{Reflections}: The supporting ecosystem built around UMBRELLA, involving both technical and non-technical teams, allows for the ongoing operation and maintenance of the system. The modularity and reconfigurability that are tightly integrated with the system design, both in the hardware and software domains, allow for interchangeable sensor and wireless modules, easy upgrades, and repairs. UMBRELLA development teams followed all the recommended design practices, well established in software and hardware communities, thus ensuring that individual changes would not impact the system.

    \vspace{0.2em}

    \item \textbf{Selecting the right location for the testbed}: Firstly, the physical deployment site can significantly impact the performance of IoT systems - factors like connectivity, interference, and environmental conditions are often considered during the design stage. Further, physical location, together with the relevance of the IoT installations and accessibility, will dictate not only the system's suitability for applications but also to attract potential users.\\
    \textit{Reflections}: The location of the various testbeds provided was very carefully considered. As UMBRELLA connects five innovation hubs in the UK's South Gloucestershire region, it ensures community engagement and enables securing funding for ongoing operation and maintenance. Long-term support mechanisms and partnerships with local stakeholders can sustain the project for longer, providing more community benefits.

    \vspace{0.2em}
    
    \item \textbf{Specialised Users vs. Potential Audience}: While niche users could familiarise themselves with the testbed system quickly, the broader audience requires more involved support, including tailored training sessions, comprehensive documentation, and responsive support to bridge this gap.\\
    \textit{Reflections}: As some of the key stakeholders are local authorities and people with non-strong software development backgrounds, the ease of use was prioritised for the interfaces designed. For example, remote terminal interfaces and direct access to the nodes, even though they are features found in other specialised wireless testbeds (e.g., in FitLab) and are usually expected by certain communities, were not finally provided to avoid an overly complex implementation. This has not limited the functionality provided, but users coming from these communities may find the use of UMBRELLA unorthodox and will need to adapt to the way that experiments are executed. 

\end{enumerate}

\qquad

\subsection{Cross-sector collaboration challenges}
Developing a digital collaboration platform for local government to bolster regional innovation, education, and job opportunities using advanced technologies for future scenarios presents several challenges. 
\begin{enumerate}[wide, labelwidth=!, labelindent=9pt, itemindent=0pt]

    \item \textbf{Bridging the Technical-Operational Gap}: Engaging senior government leaders in a platform that does not immediately cater to their current operational issues can be difficult. \\
    \textit{Reflections}: Convincing senior stakeholders, particularly non-technical, about how an IoT platform could streamline a government operation, was rather tricky initially. The ongoing engagement with the stakeholders and the demonstration of real-world examples that IoT solutions can benefit citizens across the globe (using examples from other councils/countries and the research community) pivoted various beliefs and demonstrated the benefit of UMBRELLA to the community. Currently, being an integral part of the council's daily operations with regard to air quality and street monitoring, as well as offering continuous support with major IIoT infrastructure for other innovative projects (those shown in Sec.~\ref{sec:spinoffs}), makes it a success story that can drive similar innovation and deployments in other places as well.

    \vspace{0.2em}
    
    \item \textbf{Complexity of Cutting-Edge Technologies}: Implementing and maintaining state-of-the-art technologies in a governmental context, where rapid technological changes might not be the norm, can be overwhelming to the stakeholders. \\
    \textit{Reflections}: As described earlier, even though the delivery of an underlying platform could be rather complex for a non-technical audience, the ease of use provided ensures that this complexity does not leak to the general audience. Moreover, the modular design of the node and the cloud-native implementation based on APIs allow changes to be made that enhance the functionality or the security of the system without directly affecting the end-users in most cases.
    
    \vspace{0.2em}
    
    \item \textbf{Aligning Long-Term Vision with Immediate Needs}: While the platform is geared towards future use cases, ensuring it aligns with the present-day agenda and immediate requirements of the government is crucial to the project's successful operation. \\
    \textit{Reflections}: As UMBRELLA demonstrates features already being used by the key stakeholders, it is aligned with their immediate needs. The functionality provided to evaluate new protocols and use cases shows that UMBRELLA can be adaptable to future Smart City applications as an ecosystem. This provides a good balance between a long-term vision and an implementation that brings immediate benefits to the community.

\end{enumerate}

\qquad

\subsection{Operational challenges}
Many testbeds are developed as projects expecting them to define a viable operational and business model so that the testbed could be sustainable. Given the wider scope, dedicated support teams, both technical and non-technical, become essential to achieve this. 
\begin{enumerate}[wide, labelwidth=!, labelindent=9pt, itemindent=0pt]

    \item \textit{Business Models for Sustainability}: While initial funding might cover setup costs, a clear business model is necessary for long-term sustainability, and interim support mechanisms are essential until a sustainable operation and business model can be realistically applied. As the system scales and evolves, regular updates and maintenance become critical. Ignoring this can lead to system failures and dissatisfied users. \\
    \textit{Reflections}: The established partnerships with local businesses and governmental entities ensure the sustainability of the testbed. However, as UMBRELLA is not-for-profit, the cost of the current operation should be kept within limits, and modifications have been made over time to reduce the operational cost. One example is a backup 4G link on each node that acts as a backup ingress point to the nodes if the main backbone connection is down. This link was recently discontinued due to the increased data plan from the ISP. 
    
    \vspace{0.2em}
    
    \item \textit{Parallel Development Environments}: Developing pre-production and staging environments, these being identical replicas of the real system (to a certain extent), is very critical. Comprehensive testing procedures in the pre-production environment can be conducted to ensure no service disruption would take place before pushing system updates or configurations to the actual production environment.  \\
    \textit{Reflections}: The core UMBRELLA development team initially worked on pop-up small-scale testbeds built on demand to test individual features and were dismantled later. However, as the project progressed and the complexity of the implementation increased, the need for E2E testing became a necessity. Therefore, a replica system was built as a staging and development environment. Even though that increased the maintenance efforts within the team, it ensured that all solutions were tested in an E2E fashion before being deployed in the production system.

\end{enumerate}

\section{Conclusion and Future plans}\label{sec:conclusion}
UMBRELLA has successfully established significant and practical benefits to the region by creating a world-leading Centre of Excellence for research and innovation into the future of IoT. The ecosystem built offers a real-world IoT ecosystem, enabling technology evaluation at scale. UMBRELLA combines a mixture of testbeds with a wide range of hardware, from multi-sensor nodes to robots and 5G infrastructure with a unified backend platform. The unique approach of holistic testing of solutions in an SoS fashion and a real-world deployment, puts UMBRELLA at the forefront compared to other existing testbed solutions. Such a design has demonstrated UMBRELLA's capability to effectively handle and test complex interconnected systems and scenarios, providing a unique playground for IoT innovation that is not confined within a lab environment. The openness, heterogeneity, and tools provided make UMBRELLA uniquely suited for Smart City, Robotics, Wireless, and Edge AI research. This diverse integration has not only broadened the scope of IoT applications but also provided a fertile ground for multidisciplinary research and innovation. The lessons learned from UMBRELLA's journey offer invaluable insights for future IoT initiatives, emphasising the importance of adaptability, user-centric design, and cross-domain collaboration. As it was reflected, there is a fine balance between the technical innovation and the system's stability, the technical expertise of the target audience, and the alignment of a long-term vision with the immediate needs of a community. The sustainability of testbeds like UMBRELLA requires continuous evolution aligned with user needs and a supporting ecosystem covering the maintenance and operational costs.  Overall, operational for over two years now, UMBRELLA has supported various research studies and commercial trials. Moving forward, new wireless technologies like 6G capabilities and AI areas like embodied intelligence will be incorporated. The lessons from UMBRELLA's development can guide future testbed initiatives to maximise impact. With its multi-domain capabilities and ecosystem for collaboration, UMBRELLA aims to continue accelerating technology innovation and translating emerging research into real-world progress. Since its deployment, the UMBRELLA ecosystem has been nationally and internationally recognised, receiving numerous awards. UMBRELLA won the \textbf{Connected Britain, the Industrial Innovation Award 2022}, and was in the finalist for \textbf{Smart City Expo Awards 2021}, \textbf{IoT Global Awards 2021}, and \textbf{UK Local Government Chronicle Awards 2022}. This recognition serves as a testament that the UMBRELLA project not only achieved its goals but also set new standards for IoT research. We envision UMBRELLA continuing to accelerate IoT innovations with the unique ecosystem provided. Finally, its contributions extend beyond technological advancements, fostering a collaborative and innovative spirit that will inspire and guide future projects in the IoT landscape.

\appendices

\section*{Acknowledgment}
UMBRELLA is funded in part by the West of England Local Enterprise Partnership (LEP) Local Growth Fund, administered by the West of England Combined Authority (WECA), in part by the South Gloucestershire Council, and in part by Toshiba Europe Ltd. and Bristol Research Innovation Laboratory (BRIL).

We would like to thank all project partners: \textit{Bristol and Bath Science Park, Bristol Robotic Laboratory, Kasden Electronics, Kinneir Dufort, National Composite Centre, Select Electrics Limited, Toshiba Software (India) Pvt. Ltd, University of Bristol, University of West England}. The entire UMBRELLA development team would like to express our sincere gratitude to all our colleagues and ex-team members, these being: \textit{Pietro Carnelli, Michael Baddeley, Adrian Sanchez-Mompo, Theo Spyridopoulos, Vijay Kumar, Simon Jones, Fanyu Meng, Mahendra Tailor}, for helping us develop the presented use cases; \textit{Ajith Sahadevan, John Ramsden, Victor Nwaesei} for helping with the DevOps and testing; \textit{Adina Nistor} for the CRF project coordination;  \textit{John Hebditch, Nita Patel}, that were part of the Project Management and Council Coordination team; and finally, \textit{Thomas Bierton} for the business insights provided around UMBRELLA. They all played a pivotal role in the development and delivery of the project.

\bibliographystyle{IEEEtran}
\bibliography{bib.bib,IEEEabrv}

\begin{IEEEbiography}[{\includegraphics[width=1in,height=1.25in,clip,keepaspectratio]{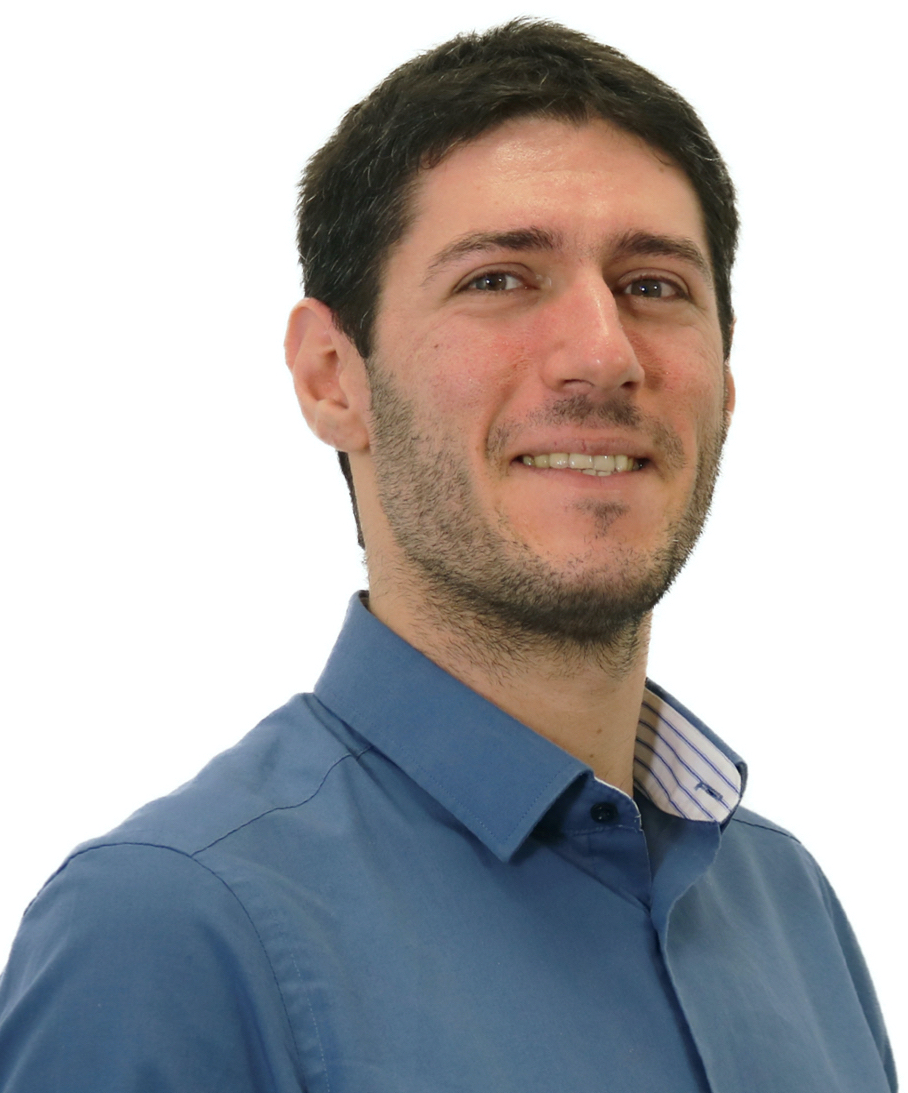}}]{Ioannis Mavromatis} is a Lead 5G/Future Networks Technologist at Digital Catapult, London, UK. He has extensive experience in 5G-and-beyond technologies, cloud-native computing, testbed deployments, wireless networking, software architecture and development. Dr Mavromatis received his PhD in ``5G Connected and Automated Vehicles'' in 2018 from the University of Bristol. He was the lead backend architect of the award-winning UMBRELLA framework, and, in the past, while working at Bristol Research and Innovation Laboratory of Toshiba Europe Ltd. and the University of Bristol, he was involved in several publicly and privately funded projects (SYNERGIA, CAVShield, BEACON-5G, FLOURISH, VENTURER, etc.). His research interests span the areas of 5G-and-beyond Communications, Cloud-native Computing, Cybersecurity, Machine Learning \& Federated Learning, and Sustainability. Dr Mavromatis received the IEEE Popularity Award from IEEE VNC 2018 and the IEEE Best Paper Award from VTC-Spring 2019.
\end{IEEEbiography}

\begin{IEEEbiography}[{\includegraphics[width=1in,height=1.25in,clip,keepaspectratio]{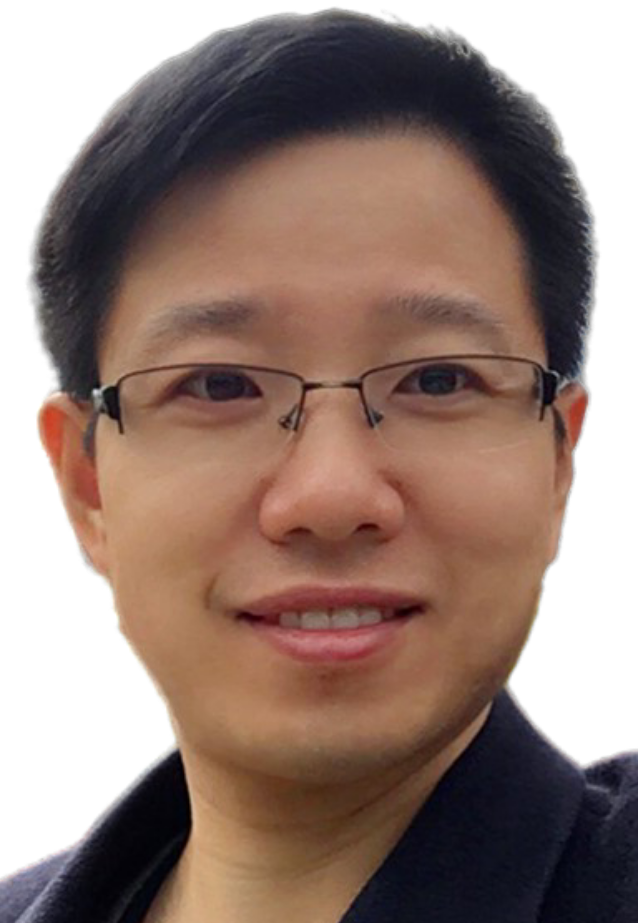}}]{Yichao Jin} is an industrial automation and wireless communication expert with over 15 years of industrial research and engineering experience. He leads the Digital System Integration team in Toshiba Bristol Research and Innovation laboratory, focusing on Building and logistic automation systems. He has authored more than 40 publications and 15 patents. His research interests including highly reliable industrial wireless communication, smart IoT Sensing technologies, multi-agent robotics, and digital twins technologies etc. He is a Fellow of IET (FIET) and Chartered Engineer (CEng), and has been involved in various of standardization activities with IETF, IEC and ETSI. He is the Technical Lead for the UMBRELLA project. 
\end{IEEEbiography}

\begin{IEEEbiography}[{\includegraphics[width=1in,height=1.25in,clip,keepaspectratio]{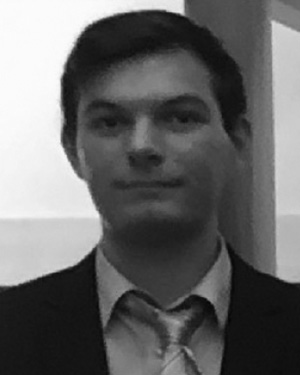}}]{Aleksandar Stanoev} is currently a Senior R\&D Engineer at Nordic Semiconductor. He received his MEng degree (Hons.) from the University of Bristol, Bristol, U.K., in 2018. In the past, he worked as a Senior Research Engineer with the Bristol Research and Innovation Laboratory, from Toshiba Europe Ltd. He is also a maintainer of the Contiki-NG Open-Source IoT Operating System. His current research interests include low-power wireless MAC protocols, large-scale smart city deployments and their design challenges, and embedded IoT hardware design.
\end{IEEEbiography}

\begin{IEEEbiography}[{\includegraphics[width=1in,height=1.25in,clip,keepaspectratio]{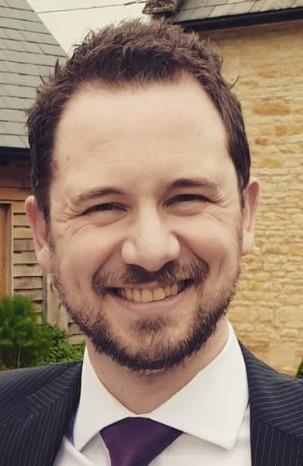}}]{Anthony Portelli} is currently a Director of Technology at Turtle Beach working on Strategic Product Categories. In the past, he worked as a Senior Research Engineer at Bristol Research and Innovation Laboratory of Toshiba Europe Ltd. His experience ranges from the development of electronics, robotics, and software engineering. He was the Lead Hardware architect and designer of the UMBRELLA nodes. His research interests include novel Human-Computer Interaction (HCI), Brain-Computer Interfacing (BCI), Robotics and Computer Games.
\end{IEEEbiography}

\begin{IEEEbiography}[{\includegraphics[width=1in,height=1.25in,clip,keepaspectratio]{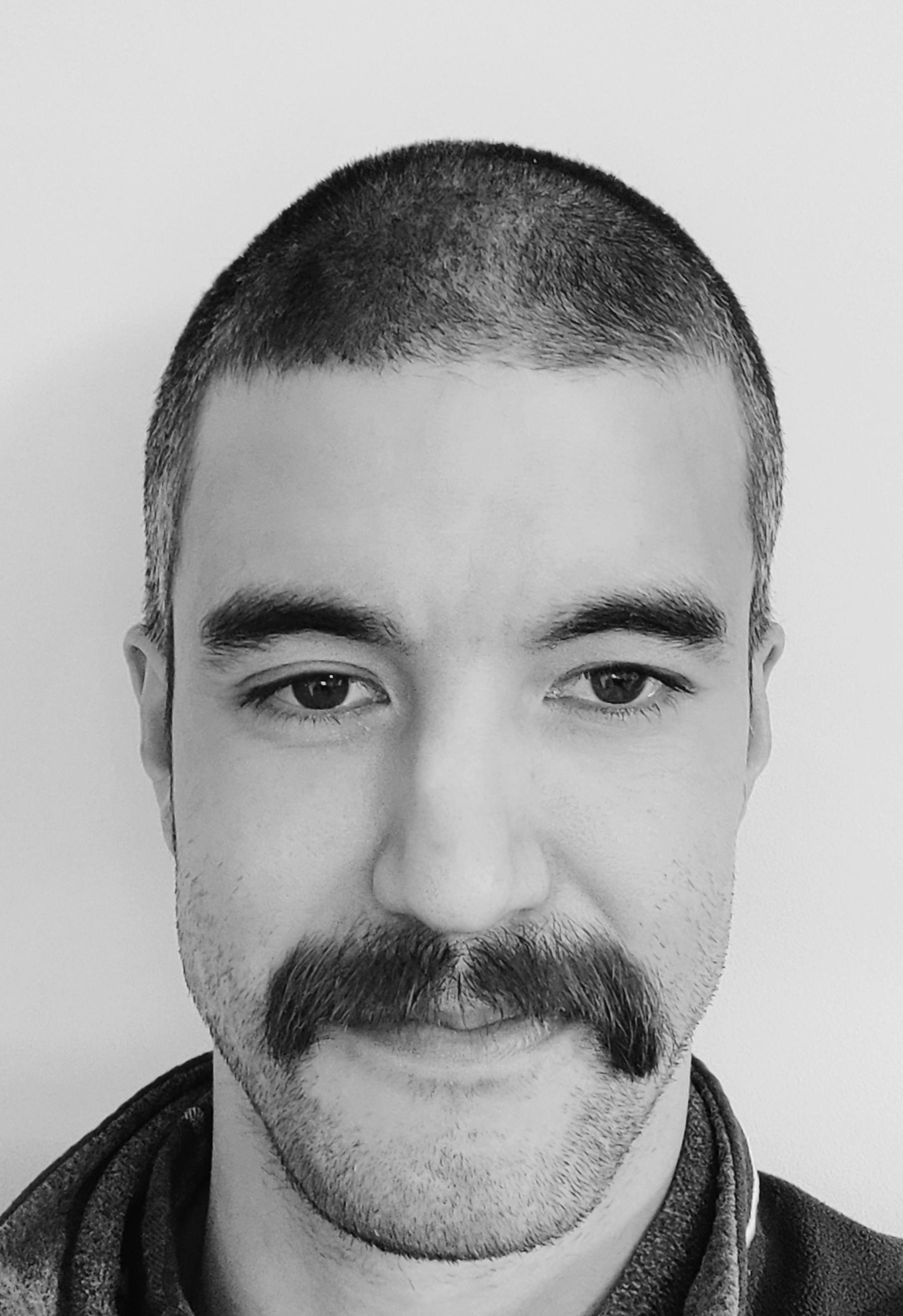}}]{Ingram Weeks} received an MEng degree in Mechanical Engineering and also went on to complete a PhD in Tribology at Cardiff University. He later worked at Dyson in Test Rig Automation. He joined Toshiba in 2020 as a Research Engineer, where he enjoys delivering software for Linux and Embedded Systems and building mechanical assemblies for robotics projects.
\end{IEEEbiography}

\begin{IEEEbiography}[{\includegraphics[width=1in,height=1.25in,clip,keepaspectratio]{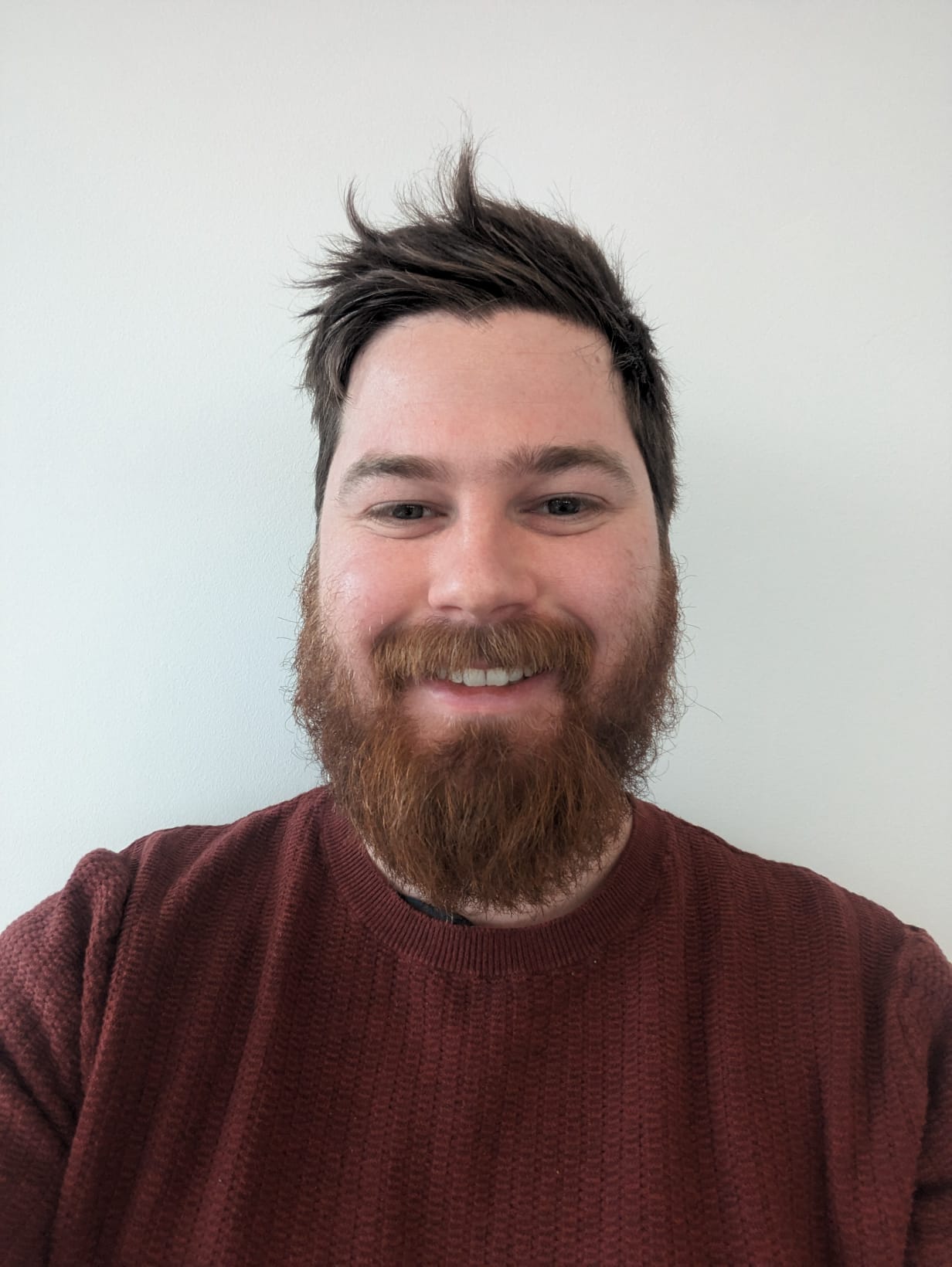}}]{Ben Holden} is currently a Firmware Engineer at Q5D Technologies Ltd. In the past, he worked as a Senior Research Engineer at Bristol Research and Innovation Laboratory of Toshiba Europe Ltd. He holds a BSc in Robotics from the University of West of England. He joined Toshiba in 2020 as an embedded software developer and 5G engineer for the UMBRELLA project. Since then, his research interests have shifted towards robotics, hyper-scalable digital twins and sim-to-real industrial applications.
\end{IEEEbiography}

\begin{IEEEbiography}[{\includegraphics[width=1in,height=1.25in,clip,keepaspectratio]{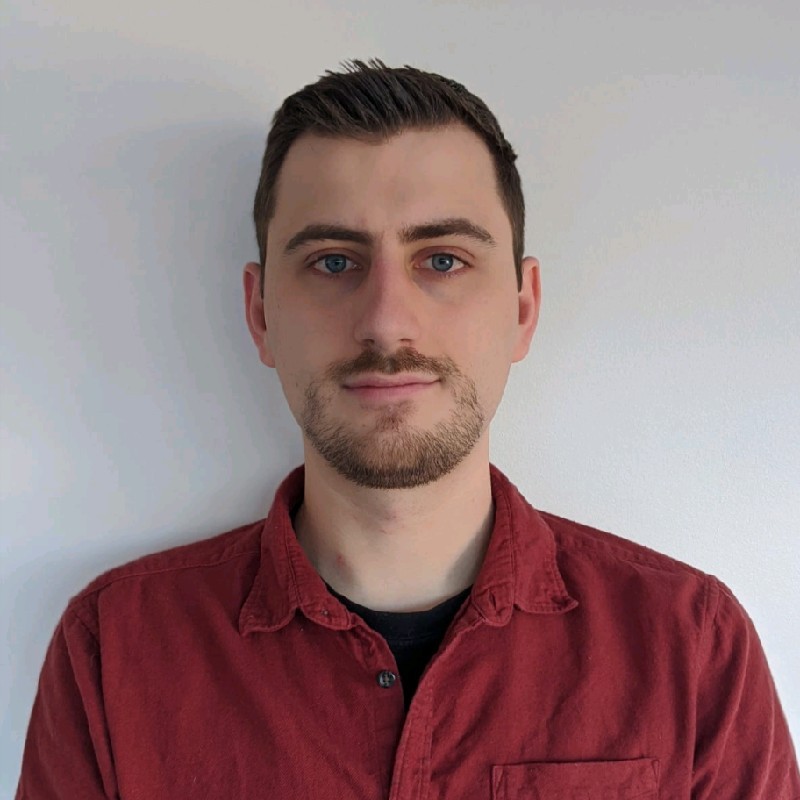}}]{Eliot Glasspole} is currently a Principal Software Engineer at Amiosec Ltd. In the past, he worked as a Research Engineer at Bristol Research and Innovation Laboratory of Toshiba Europe Ltd. He holds a BSc in Electronic Engineering from the University of West of England. He joined Toshiba in 2019 to work as a software and electronics engineer for the UMBRELLA project.
\end{IEEEbiography}

\begin{IEEEbiography}[{\includegraphics[width=1in,height=1.25in,clip,keepaspectratio]{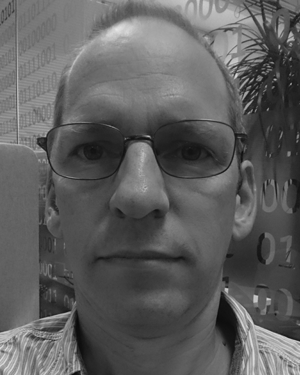}}]{Tim Farnham} (M’92) received the B.Eng. degree from Manchester University and the Ph.D. degree from DeMontfort University, Leicester, U.K. He is currently a Chief Research Fellow at Toshiba Europe Ltd. His recent research activities have included cognitive radio and radio environment mapping techniques for heterogeneous wireless networking and assisting radio location. Evaluation of such techniques has been performed within various scenarios, such as video distribution in visitor attractions (Innovate U.K. AIYP Project), indoor home media distribution and small-cell networks (EU FP7 ARAGORN and FARAMIR projects), and sensor networks used within water distribution networks (EU FP7 ICeWater Project). He has also contributed to relevant standards, such as IEEE 1900.4 and 1905.1 as well as the IEC 61360 common data dictionary standard that provides the necessary data ontologies for sharing product data.
\end{IEEEbiography}

\begin{IEEEbiography}[{\includegraphics[width=1in,clip,keepaspectratio]{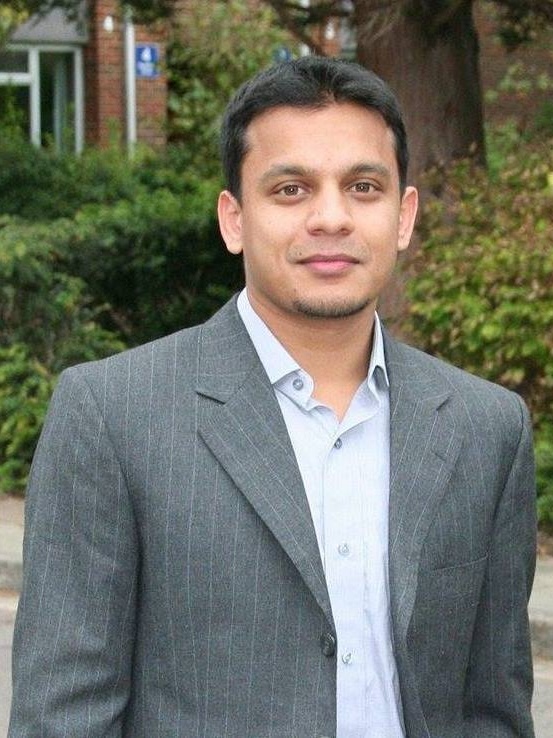}}]{Aftab Khan} is the Distributed AI Programme Leader at the Bristol Research and Innovation Laboratory, Toshiba Europe Ltd., U.K. He received his PhD in Machine Learning from the University of Surrey, U.K. (2013). His research agenda is mainly focused on distributed machine learning, AI-driven cyber security, computational behaviour analysis and pattern recognition. He has been involved in several EU and EPSRC projects (REPLICATE, SiDE, TEDDI, ACASVA) as well as industry-led Innovate UK projects (SYNERGIA, CAVShield).
\end{IEEEbiography}

\begin{IEEEbiography}[{\includegraphics[width=1in,height=1.25in,clip,keepaspectratio]{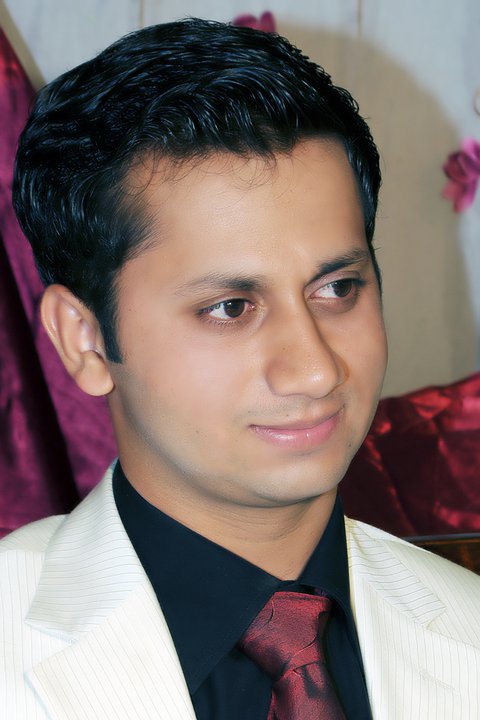}}]{Usman Raza} has over 15 years of experience as an engineering manager, product lead, and applied researcher in wireless networking, machine learning, indoor/outdoor positioning systems, and smart city IoT applications. He currently leads an agile software team at Waymap, a London-based start-up company that develops iOS/Android applications to estimate the location of smartphone users using inertial sensors, camera, GNSS, and mapping data. Previously, he held multiple roles as a systems researcher at Toshiba Europe Limited UK, Bruno Kessler Foundation Italy, and the University of Trento Italy. He is an author of 15+ patents and 40+ research articles in reputed IEEE \& ACM journals and conferences, four of which received best paper awards. He received the Australian Government’s Endeavour Research Fellowship, an offer of the U.S. Department of State’s Fulbright scholarship, the IEEE Communication Society's Heinrich Hertz Award, the Mark Weiser Best Paper Award, and the National Management Foundation’s Gold Medal. He holds a PhD degree from the University of Trento, an MS from Lahore University of Management Sciences, and a BS from the National University of Computer \& Emerging Sciences in Computer Science
\end{IEEEbiography}

\begin{IEEEbiography}[{\includegraphics[width=1in,height=1.25in,clip,keepaspectratio]{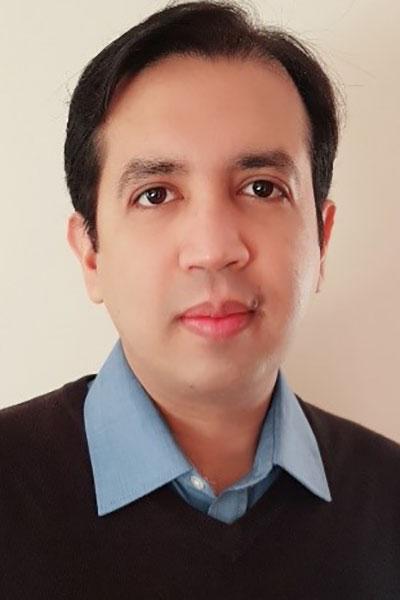}}]{Adnan Aijaz} (M’14–SM’18) studied telecommunications engineering at the King’s College London, U.K., where he received a Ph.D. degree in 2014 for research in wireless networks. He is currently the Programme Leader for Beyond 5G at the Bristol Research and Innovation Laboratory, Toshiba Research Europe Ltd., U.K. His recent research interests include 5G/6G wireless systems, Open RAN, time-sensitive networking, high-altitude platforms, robotics, and autonomous systems.
\end{IEEEbiography}

\begin{IEEEbiography}[{\includegraphics[width=1in,height=1.25in,clip,keepaspectratio]{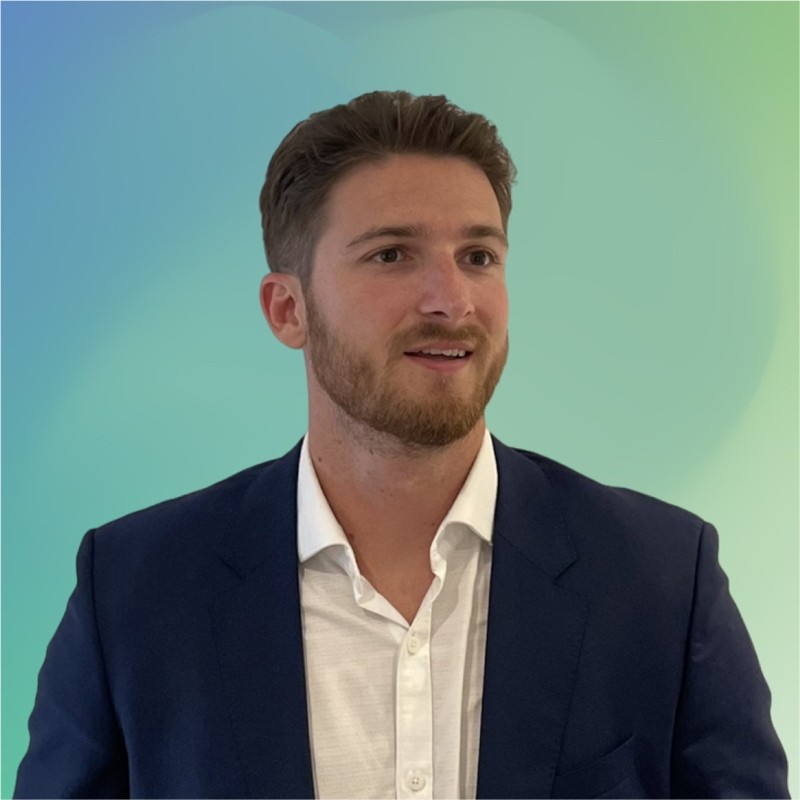}}]{Thomas Bierton} works as an Innovation Analyst in the Business and Strategy Development function of Toshiba’s Bristol Research and Innovation Lab. He has an MBA from the University of the West of England where he focused on Strategy, Operations and Digital Transformation. Upon joining Toshiba in 2019, he worked on developing the strategy for the UMBRELLA project and its marketing, including the website, launch event, and promotional materials. Outside of UMBRELLA, he enjoys working on researching business models and new markets for new technologies in development.
\end{IEEEbiography}

\begin{IEEEbiography}[{\includegraphics[width=1in,height=1.25in,clip,keepaspectratio]{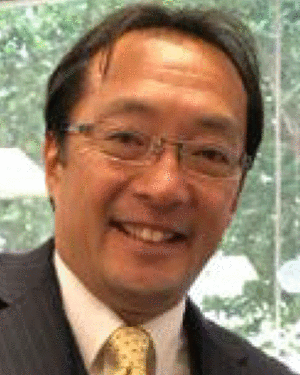}}]{Ichiro Seto} (Member, IEEE) received the B.E. and M.E. degrees from Keio University, Tokyo, Japan, in 1991 and 1993, respectively. In 1993, he joined the Corporate Research and Development Center, Toshiba Corporation, Tokyo, where he engaged in R\&D of optical and wireless communication systems. During 2005–2006, he was a Visitor Researcher with UC Berkeley Wireless Research Center, U.S. During 2017–2022, he belonged to Toshiba Europe, Bristol Research and Innovation Laboratory, U.K. He has contributed to business start-ups for distributed antenna systems in cellular and wireless LSI for broadband. He has filed 50 or more patents. He was the recipient of the 2018 National Invention Award from the Japan Institute of Invention and Innovation.
\end{IEEEbiography}

\begin{IEEEbiography}[{\includegraphics[width=1in,height=1.25in,clip,keepaspectratio]{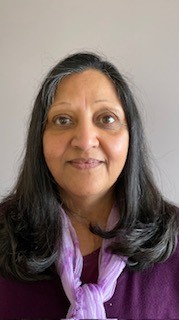}}]{Nita Patel} serves as the Digital Strategy \& Technology lead at South Gloucestershire Council. Her background is rich in telecommunications, blending longer term strategic planning, technical know-how, information technical and strategic leadership.  In her current role, she leverages this diverse experience to develop digital and technology strategies at South Gloucestershire Council focusing on how digital and technology advancement can support public sector to address societal challenges that supports citizens, business and academia in the regional, improving lives and supporting business growth and skills development.
\end{IEEEbiography}

\begin{IEEEbiography}[{\includegraphics[width=1in,height=1.25in,clip,keepaspectratio]{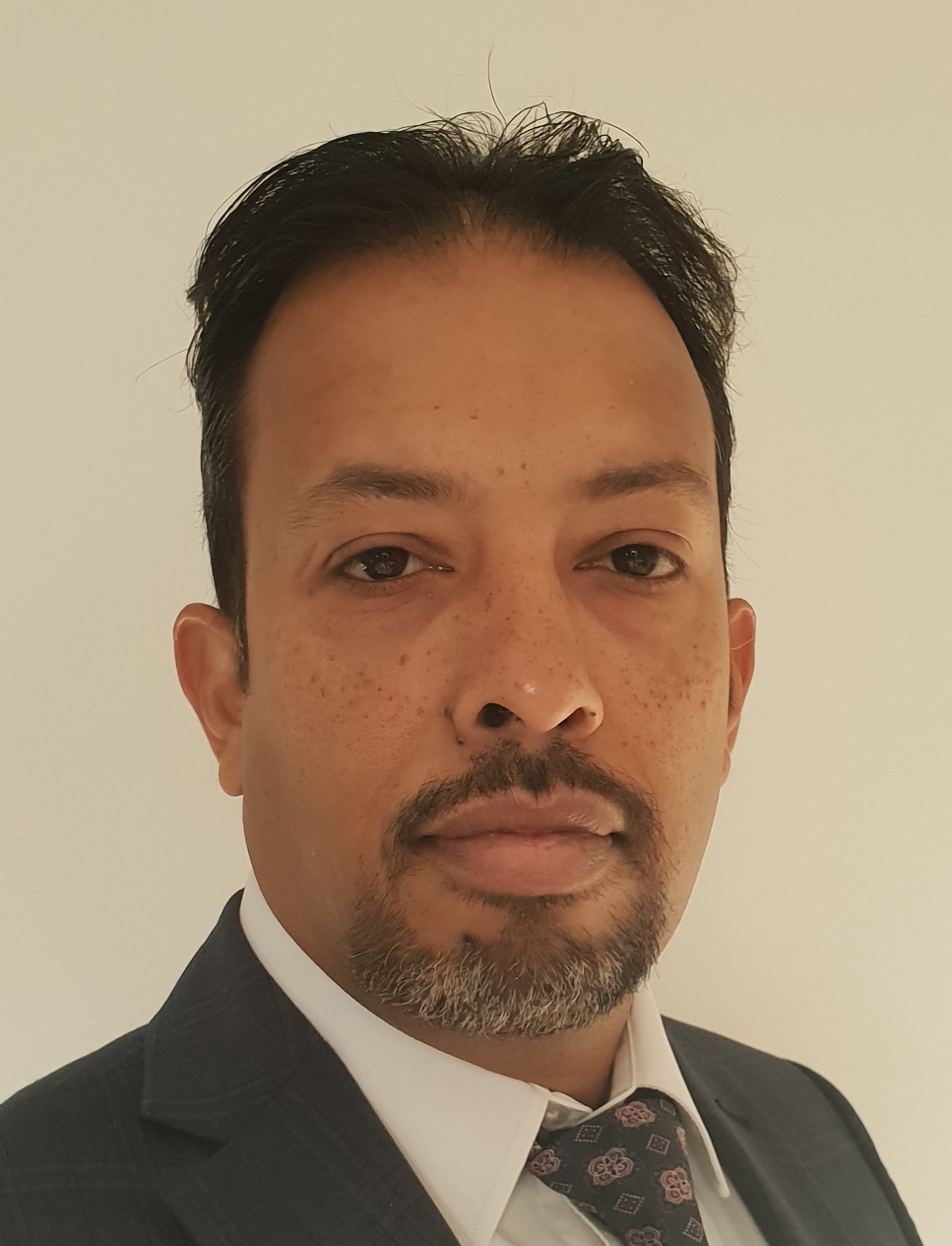}}]{Mahesh Sooriyabandara} received his BSc Eng degree from the University of Peradeniya, Sri Lanka, and a PhD degree from the University of Aberdeen in the United Kingdom. In 2004, he joined the Bristol Research \& Innovation Lab of Toshiba Europe Limited, where he is currently the Managing Director. His research interests include industrial IoT networks, smart energy networks and intelligent and distributed systems. He is a Fellow of IET, a senior member of the IEEE and ACM and an honorary visiting professor at the School of Engineering, Cardiff University.
\end{IEEEbiography}

\EOD

\end{document}